\documentclass[12pt, a4paper]{article}
\pdfoutput=1
\usepackage[utf8]{inputenc}
\usepackage[T1]{fontenc}
\usepackage{lmodern, textcomp}
\usepackage[american]{babel}
\usepackage{csquotes}
\usepackage{bbold}

\usepackage{eurosym}
\usepackage[nottoc]{tocbibind}
\usepackage{authblk}
\usepackage{fancyhdr}
\usepackage{xspace}
\usepackage{bm}

\usepackage{graphicx}
\usepackage{subcaption}
\usepackage{float}
\usepackage{makecell}
\usepackage{multirow}
\usepackage{multicol} 
\usepackage{marginnote}
\usepackage[multiple]{footmisc}

\usepackage[usenames, dvipsnames, svgnames, table]{xcolor}

\usepackage[backend=bibtex8, citestyle=numeric-comp, maxbibnames=99,
			sorting=none, sortcase=false, sortcites=true, giveninits=true]{biblatex}

\usepackage{amsmath, amsfonts, amssymb}
\usepackage{mathtools}
\usepackage{mathrsfs}
\usepackage{cancel}
\usepackage{braket}
\usepackage{tensor}
\usepackage{slashed}
\usepackage{siunitx}
\usepackage{listings}

\usepackage{stmaryrd}
\usepackage{caption}
\usepackage{relsize,exscale}
\usepackage{array}
\usepackage[toc,page]{appendix}

\usepackage{enumerate}

\DeclareSymbolFont{largesymbols}{OMX}{cmex}{m}{n}

\setcellgapes{1pt}
\makegapedcells
\newcolumntype{R}[1]{>{\raggedleft\arraybackslash }b{#1}}
\newcolumntype{L}[1]{>{\raggedright\arraybackslash }b{#1}}
\newcolumntype{C}[1]{>{\centering\arraybackslash }b{#1}}

\newcommand{\Tr}{\mathrm{Tr}}

\newtheorem{definition}{Definition}

\newtheorem{proposition}{Proposition}
\newtheorem{remark}{Remark}
\newtheorem{claim}{Claim}

\newcommand{\cZ}{{\mathcal Z}}

\newcommand{\beq}{\begin{equation}}
\newcommand{\eeq}{\end{equation}}
\newcommand{\bea}{\begin{eqnarray}}
\newcommand{\eea}{\end{eqnarray}}
\definecolor{mygray}{gray}{0.3}

\newcommand{\bes}{\begin{eqnarray}}
\newcommand{\ees}{\end{eqnarray}}

\newcommand\restr[2]{{
  \left.\kern-\nulldelimiterspace 
  #1 
  \vphantom{\big|} 
  \right|_{#2} 
  }}

\usepackage{multicol}
\usepackage{amssymb}

\def\XXint#1#2#3{{\setbox0=\hbox{$#1{#2#3}{\int}$}
     \vcenter{\hbox{$#2#3$}}\kern-.5\wd0}}

\usepackage[heightrounded, top=4cm, bottom=4cm, left=3cm, right=3cm, headheight=14pt]{geometry}

\usepackage[pdftex, breaklinks=true, linktocpage,
	colorlinks=true, urlcolor=blue, linkcolor=blue, citecolor=red
]{hyperref}

\newcommand{\email}[1]{\href{mailto:#1}{\nolinkurl{#1}}}
\newcommand{\emailfoot}[1]{\thanks{\email{#1}}}

\usepackage{marginnote}
\newcounter{draftcommentcnt}
\NewDocumentCommand{\draftcomment}{s O{red} m}{%
	\def\margnote{\IfBooleanTF{#1}{\marginnote}{\marginpar}}%
	\stepcounter{draftcommentcnt}%
	\textcolor{#2}{#3}%
	\margnote{\textcolor{#2}{$\Leftarrow$ \arabic{draftcommentcnt}}}%
}

\fancypagestyle{preprint}{
	\fancyhf{}

	\fancyhead[R]{\textsf{\preprint}}
}

\numberwithin{equation}{section}

\hypersetup{
	pdftitle={Frequency regulator for quantum $p$-spin renormalization group in the large $N$ limit},
}

\title{The quantum $p$-spin renormalization group in the large $N$ limit as a benchmark for functional renormalization group}

\author[1]{Vincent Lahoche\emailfoot{vincent.lahoche@cea.fr}}
\author[1,2]{Dine Ousmane Samary\emailfoot{dine.ousmanesamary@cipma.uac.bj}}
\author[1]{Parham Radpay\emailfoot{parham.radpay@cea.fr}}

\affil[1]{%
	Université Paris Saclay, \textsc{Cea}, Palaiseau, F-91191, France
}
\affil[2]{%
	Faculté des Sciences et Techniques (ICMPA-UNESCO Chair)
	\protect\\
	Université d'Abomey-Calavi, 072 BP 50, Bénin
}

\begin{document}
\maketitle

\hrule

\hrule
\begin{abstract}
To gain a deeper understanding of the glassy phase in $p$-spin quantum models, this paper examines the dynamics of the $N$-vector $\bm{x} \in \mathbb{R}^N$ through the framework of renormalization group theory. First, we focus on perturbation theory, which is more suitable than nonperturbative techniques due to the specific temporal non-locality of the model after disorder integration. We compute the one-loop $\beta$-function and explore the structure of its fixed points.
Next, we develop the nonperturbative renormalization group approach based on the standard Wetterich-Morris formalism, using two approximation schemes to address the model's non-locality. We investigate the vertex expansion in the symmetric phase and assess the reliability of the approximations for the fixed-point solutions.
Finally, we extend our analysis beyond the symmetric phase by using an expansion around the vacuum of the local potential. Our numerical investigations particularly focus on the cases $p = 2$ and $p=3$.
\end{abstract}

%
\hrule

\hrule
\newpage
\pdfbookmark[1]{\contentsname}{toc}
\tableofcontents
\bigskip

\newpage

\section{Introduction}

Over the past six decades, significant progress has been made in the understanding of glassy systems \cite{talagrand2010mean,talagrand2010mean2,talagrand2003spin,castellani2005spin}. Moreover, these developments span a broad range of disciplines, lying at the intersection of physics, mathematics, economics and more recently, data science \cite{nishimori2001statistical,fan2023searching,ghio2023sampling,mezard2009information,bouchaud2023application}. Spin glass models are idealized representations of realistic metals and alloys, replicating their glassy behavior in the low-temperature regime due to randomly distributed impurities \cite{mezard1987spin}.
\medskip

One of the most important features of the glassy phase is the complexity of its energy landscape, which is characterized by an exponential number of metastable states. Computationally, this means that finding the ground state of the landscape becomes an NP-hard problem. Non-equilibrium effects at low temperatures also mark the slow dynamics of the spin glass phase, and some solvable models, such as the spherical Sherrington-Kirkpatrick model \cite{panchenko2013sherrington}, exhibit this phenomenon. This complex energy landscape arises from frustration—random impurities create effective couplings between spins that cannot all be satisfied simultaneously along a closed loop, resulting in negative products of couplings along the loop. Moreover, the disorder is quenched, meaning that the typical time evolution for the “spin” is much longer than for the disorder (impurities' configuration) \cite{cugliandolo2002dynamics}. 
\medskip

Spin glass models can often be tackled using classical tools, mainly mean-field techniques from statistical physics. This is justified in cases where the transition temperature and corresponding energy scale are far from those of quantum fluctuations \cite{biroli2001quantum,cugliandolo2001imaginary,baldwin2017clustering,wu1991classical}. Quantum transitions typically occur at zero temperature, where quantum fluctuations disrupt long-range order. However, external parameters can be tuned in spin glass systems such that the glassy transition temperature approaches the scale of quantum energy fluctuations. This allows quantum fluctuations to influence dynamics through the complex energy landscape significantly. These effects are particularly relevant to quantum annealing (as opposed to thermal annealing), which explores how quantum particles escape metastable states via quantum tunnelling.
\medskip

There are various approaches to studying quantum spin glasses. One common approach is the quantum analogue of the Heisenberg model, where classical three-dimensional spin vectors are replaced by Pauli operators coupled through a quenched coupling constant -- see for instance \cite{baldwin2017clustering,baldwin2018quantum,bapst2013quantum} and references therein. In this approach, quantum fluctuations arise because of the non-commutativity between the Ising term and some transverse magnetic field coupling. Another important model is the Sachdev-Ye-Kitaev (SYK) model, originally introduced in condensed matter physics to describe the behavior of non-Fermi liquids \cite{sachdev2024quantum,rosenhaus2019introduction,gurau2018prescription}. The SYK model has an intriguing connection with quantum gravity, though it is not a true spin glass because "annealed = quenched" and is "maximally chaotic." A popular alternative is the quantum spherical $p$-spin glass model, which describes a quantum particle moving through an $N$-dimensional random energy landscape, considered for instance in \cite{biroli2001quantum,cugliandolo2001imaginary}, and is the model we focus on in this paper. Such a model is expected for instance to make contact with the low-temperature physics of some alloy like $\text{LiHo}_x\text{Y}_{1-x} \text{F}_4$ \cite{cugliandolo2001imaginary,ancona2008quantum}. 
\medskip

Three major approaches dominate the investigation of glassy behavior in spin glass models. The first is the static approach, based on replica symmetry breaking \cite{mezard1987spin}. The second is the dynamical approach, focused on dynamical transitions and aging effects systems that have longer relaxation times than younger ones \cite{cugliandolo2002dynamics,de2006random}. Finally, the complexity and entropy crisis approach is based on the Thouless, Anderson, and Palmer (TAP) equation \cite{castellani2005spin,nishimori2001statistical}. These approaches are considered complementary, each providing insight into different aspects of the system. For example, the $p=2$ spherical spin glass behaves like a ferromagnet in disguise (no replica symmetry breaking and no true spin glass phase) in the static approach, while the dynamical approach reveals non-trivial out-of-equilibrium processes with aging effects. In this paper, we consider a quantum dynamical process and focus on the late-time effects of disorder on the quantum behavior of a particle.
\medskip

A powerful tool for investigating the low-energy behavior of a system is the Wilsonian renormalization group (WRG). WRG constructs a series of effective Hamiltonians by successively eliminating high-energy degrees of freedom, while preserving large-scale physical properties, such as fixed points and critical exponents. Although WRG is not typically used in spin glass studies, which mainly rely on mean-field theory, there are notable exceptions, including both perturbative and nonperturbative approaches -- see for instance \cite{kovacs2011renormalization,Tarjus,Tarjus2,castellana2010renormalization,banavar1988heisenberg,balog2018criticality,parisi2001renormalization}. See also \cite{dupuis2024mott,daviet2021chaos} for recent applications to disorder systems, implying a formalism close enough to ours. 
\medskip

In this paper, we apply WRG to the quantum problem of a particle moving through a rough landscape. We use both perturbative expansions and more sophisticated nonperturbative approximations, focusing on the Wetterich-Morris framework. Future works will delve into more advanced topics, including investigations using the 2PI formalism \cite{benedetti20182pi,blaizot2021functional}. Besides, it is fully connected and mean field theory is expected to provide a satisfactory result, our goal is not to compete with it, but rather to propose a kind of alternative, a complementary vision of the nature of glass transitions for example, adding a characterization to those already existing. The proximity with models admitting an analytical solution is to be able to make a precise comparison and to build reliable approximations of the RG, which can then be used with a good degree of confidence to investigate problems more challenging for analytical approaches. This is the case for example of problems involving "structured" disorders \cite{mezard2024spin}, in this specific case, of the detection of quantum signals for instance.
\medskip

This work moreover follows a recent bibliographical line \cite{lahoche2024largetimeeffectivekinetics,lahoche2024Ward,achitouv2024timetranslationinvariancesymmetrybreaking} of the same authors, aiming to consider RG as a tool to characterize the underlying physics of quantum glassy systems. These past investigations considered the "$2+p$"-quantum spin glass models, mixing a matrix-like disorder materialized by a Wigner matrix \cite{potters2020first} with a tensor-like disorder as soon as $p>2$. One of the originality of our work was to consider a coarse-graining over the Wigner spectrum. The corresponding RG flow was characterized by essentially three features:

\begin{enumerate}
\item Despite the absence of background space, the scaling behavior of the system is reminiscent of a $4$ dimensional (non-local) field theory in the deep infrared. 
\item For intermediate scales, the canonical dimension depends on scale, and all the explicit dependencies of the flow equation on the scaling parameter cannot be removed. 
\item Finite scale singularities appear for strong enough tensorial disorder.
\end{enumerate}

The last point is reminiscent of phenomena observed for other disorder models in the literature \cite{gredat2014finite,dupuis2020bose,dupuis2024mott}, and should be connected by the concept of Larkin scale. From our investigations, the mechanism at the origin of these phenomena is the following: As the disorder becomes strong enough some metastable states dominate the flow, corresponding to interactions forbidden by the large $N$ perturbation theory from which the ansatz for the effective average action is done \cite{gredat2014finite}. In \cite{lahoche2024Ward}, we provided some numerical evidence confirming this assumption for a specific choice of local interactions that couple replica, which is not expected from perturbation theory and formal large $N$ gap equation. Moreover, in \cite{achitouv2024timetranslationinvariancesymmetrybreaking}, we provided some evidences for the same mechanism for interactions breaking explicitly time reversal symmetry. 
The physics underlying these observations is at this stage the following: The RG we constructed indeed maps the system with the same system for a smaller variance for the matrix like disorder. In case where the matrix like disorder dominates, the system is close to a ferromagnet, and exhibit a second order phase transition where a single component of the spins (the projection along the smaller eigenvalue of the disorder) has a macroscopic occupation number below some finite critical temperature \cite{lahoche2024largetimeeffectivekinetics,de2006random}. As the strength of the matrix like disorder go down, the tensorial like disorder increases, and induces a glassy behavior whose singularities are symptomatic. 
\medskip

In this paper, we focus on the $p$-spin model, and we consider a coarse-graining over time. Coarse-graining over time has been considered for analogue quantum mechanical systems in the past \cite{aoki2002non}, and one of the mean difficulty in that case is that all interactions are relevant, and relevance increases with the number of fields involved in the coupling. This why we considered different approximation schemes to improve the reliability of our conclusions. In particular, we also considered perturbation theory and a large $N$ technique coming from tensorial field theory and called effective vertex expansion \cite{Lahoche:2020pjo,Lahoche_2020b,Lahoche:2018oeo}, allowing us to keep into account all the local sector in the infrared regime. Our conclusions are essentially the same as in our previous work. Once again we observe finite (time scale) singularity as the disorder is large enough, and metastable states correspond to interactions forbidden by perturbation theory. Taking into account some of them is the construction of the effective average action moreover cancel some of these divergences. Finally, note that similar results were obtained for the counterpart classical system \cite{Lahoche:2021tyc}, quantum fluctuation being replaced by thermal fluctuations in that case.

\paragraph{Outline.} In detail, the paper is organized as follows: In Section \ref{sec1}, we introduce the model and conventions, especially the replica trick used to compute the quenched disorder average.
In Section \ref{sec2}, we introduce Litim’s regularization and outline the general strategy for computing perturbative $\beta$-functions. In Section \ref{sec3}, we explicitly compute and investigate the one-loop $\beta$-functions for low-rank models ($p=2,3$).
In Section \ref{sec4}, we explore nonperturbative aspects, focusing on the vertex expansion in the symmetric phase and evaluating the robustness of the fixed-point solution for large truncations.
In Section \ref{sectionbeyond}, we extend the flow beyond the symmetric phase using an expansion around the running vacuum of the local potential and including interactions corresponding to metastable states along the flow, canceling finite scale singularities we observed above. In section \ref{sectionenhanced}, we consider the same kind of improved truncation in the symmetric phase, using vertex expansion. Finally, in Section \ref{sec6}, we conclude and outline future research directions. 
\medskip

\section{The model and conventions}\label{sec1}

In this section, we introduce the model and the conventions used throughout the paper. Additionally, we present the standard replica trick to compute the quenched average of the generating functional of connected correlation functions, which corresponds to the free energy.

\subsection{Definition of the model path integral}
\begin{figure}
\begin{center}
\includegraphics[scale=0.4]{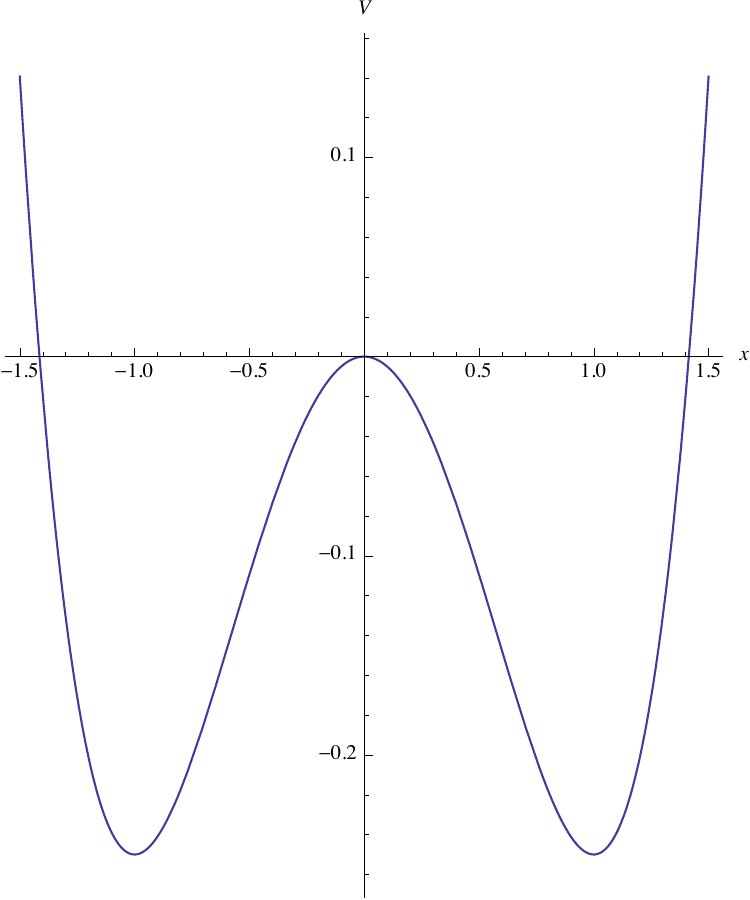}\qquad\qquad \includegraphics[scale=0.4]{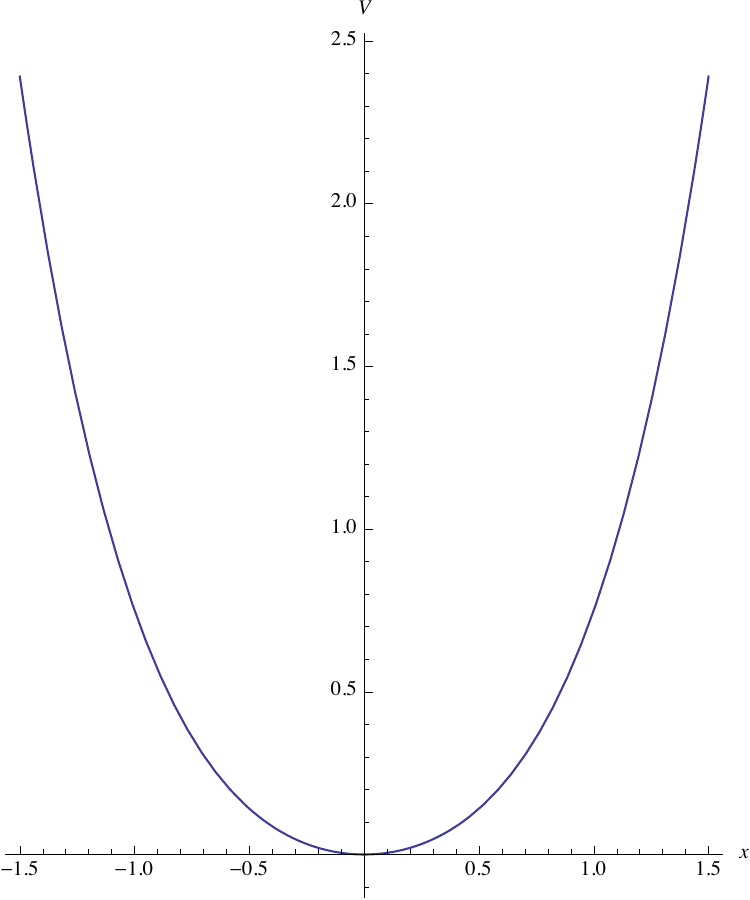}
\end{center}
\caption{Typical shape of the potential $(N=1)$ for $\mu_1<0$ (on left) and $\mu_1>0$ (on right).}\label{figpot}
\end{figure}

We consider the dynamics of an $N$-dimensional (non-relativistic) quantum particle moving through a rugged landscape represented by a rank-$p$ Gaussian random tensor, $\bm{J}$. The components $J_{i_1,\cdots,i_p}$ of the Gaussian tensor are independent random variables with zero mean, and their variance is defined as follows:
\begin{equation}
\overline{J_{i_1\cdots i_p}J_{i_1^\prime \cdots i_p^\prime}}=\left(\frac{\lambda p!}{N^{p-1}}\right) \,\prod_{\ell=1}^p \delta_{i_\ell i_\ell^\prime}\,.\label{averageJ}
\end{equation}
We use the notation $\overline{X}$ for the average. The classical variable $\bm{x}\in \mathbb{R}^N$ corresponds to the classical position of the particle, which is described in the point of view of quantum theory by the wave function $\Psi(\bm{x},t)$, satisfying the Schrodinger equation:
\begin{equation}
i \hbar \frac{\partial }{\partial t}\Psi(\bm{x},t)= \hat{H}_{\text{SG}}\, \Psi(\bm{x},t)\,.\label{schro}
\end{equation}
The Hamiltonian $ \hat{H}_{\text{SG}}$ is furthermore assumed to be of the form:
\begin{equation}
 \hat{H}_{\text{SG}}= \frac{\hat{\bm{p}}^2}{2m}+\sum_{i_1<i_2\cdots <i_p}\, J_{i_1i_2\cdots i_p} \, \hat{x}_{i_1} \hat{x}_{i_2}\cdots  \hat{x}_{i_p}+V(\hat{\bm{x}}^2)\,.
\end{equation}
Note that as for the Hamiltonian, we denoted the momentum and position components operators with a hat notation, to distinguish them with their classical analogue. In particular, they satisfy the non vanished canonical  commutation relations $[\hat{p}_k,\hat{x}_j]=i \hbar \delta_{kj}$.
The potential $V(\bm{x}^2)$, depending only on the square length $\bm{x}^2$ of the classical coordinate avoid large $\bm{x}$ configurations in the classical  equation of motion:
\begin{equation}
m \frac{d^2 x_i}{dt^2}= -\frac{\partial}{\partial x_i} \left(\sum_{i_1 < i_2\cdots < i_p}\, J_{i_1i_2\cdots i_p} \, {x}_{i_1} {x}_{i_2}\cdots {x}_{i_p}+V({\textbf{x}}^2)\right)\,.\label{classicalequation}
\end{equation}
There are essentially two strategies to avoid large $\bm x$ configurations:
\begin{enumerate}
\item The hard constraint such that $V({\bm{x}}^2):=\mu \left({\bm{x}}^2-N\right)$, where $\mu$ is some \textit{Lagrange multiplier} enforcing the spherical constraint:
\begin{equation}
{\bm{x}}^2(t)=N\,,
\end{equation}
along the dynamics (i.e. $\forall\,t$).
\item The soft constraint, with a confining potential:
$
V({\bm{x}}^2)=\frac{1}{2}\mu_1 {\bm{x}}^2+\frac{\mu_2}{4 N}\, ({\bm{x}}^2)^2\,,
$
such that for $\mu_1<0$ and $\mu_2>0$, there are two stables minima (see Figure \ref{figpot}) and the constraint writes as:
\begin{equation}
{\bm{x}}^2_0(t)=-\frac{\mu_1}{\mu_2}\,N\,,
\end{equation}
enforcing dynamically the spherical constraint.
\end{enumerate}
In the rest of this manuscript, we mainly focus on the second approach, which allows exploring a richer phase space than the first one. It is moreover closer to the formalism used in recent papers in the WRG literature \cite{lahoche2022functional,Lahoche:2021tyc,erbin2023functional} aiming to investigate classical $p$-spin dynamics -- see also \cite{lahoche2023low,van2010second} for a classical treatment of the late time $p=2$ spin dynamics with confining potential.
Note that in our investigation, we give up the Schrodinger formalism \eqref{schro} and consider the Feynman path integral approach. This framework is based on the  definition of the partition function $\cZ[\bm{J},\bm{L}]$ (see \cite{Zinn-Justin:1989rgp,ZinnJustinBook2,peskin2018introduction})
\begin{equation}
\cZ[\bm{J},\bm{L}]:=\int \, [d x]\, e^{\frac{i}{\hbar}S_{\text{cl}}[\bm{x}]+i\int_{-\infty}^\infty\,dt\,\sum_{k=1}^N L_k(t)x_k(t)}\,,\label{pathintegralZ}
\end{equation}
where the \textit{classical action} $S_{\text{cl}}$ is:
\begin{equation}
S_{\text{cl}}[\bm{x}]:=\int_{-\infty}^{+\infty} dt \left(\frac{1}{2}\dot{\bm{x}}^2-\sum_{i_1 < \cdots < i_p}\, J_{i_1i_2\cdots i_p} \, {x}_{i_1}(t)\cdots {x}_{i_p}(t)-V({\bm{x}}^2(t))\right)\,,
\end{equation}
for:
\begin{equation}
V({\bm{x}}^2):=N \sum_{\ell=1}^{\ell_{\text{max}}}\, \frac{\mu_\ell}{(2\ell) N^\ell} (\bm{x}^2)^\ell\,.\label{potentiallocal}
\end{equation}
Note that we use the notation ‘‘$[d x]$'' for the path integral measure, to distinguish it from the standard Lebesgue measure ‘‘$d x$''. As well known in quantum field theory (QFT), the path integral \eqref{pathintegralZ} allows to compute vacuum-vacuum expectation value of fields correlations at different times \cite{peskin2018introduction}:
\begin{equation}
\langle \Omega \vert \hat{x}_{i_1}(t_1)\hat{x}_{i_2}(t_2)\cdots \hat{x}_{i_n}(t_n) \vert \rangle = \frac{(-i)^{n}}{\cZ[0]} \frac{\partial^n \cZ[J]}{\partial L_{i_1}(t_1)\partial L_{i_2}(t_2)\cdots \partial L_{i_n}(t_n)}\,\Bigg\vert_{J=0}\,.
\end{equation}
We will mainly focus on the Euclidean version of the path integral, assuming Wick rotation holds along the field configuration domain. This formally corresponds to the transformation $t\to -i\tau$ (imaginary time), and avoid the difficulty coming from the mass pole $\omega^2=\pm \mu_1$ in the original theory, and the additional singularities coming from Goldstone modes dynamics. Hence, the path integral we consider finally is the following:
\begin{equation}
\cZ_{\text{E}}[\bm{J},\bm{L}]:=\int \, [d x]\, e^{-\frac{1}{\hbar}S_{\text{cl,E}}[\bm{x}]+\int_{-\infty}^\infty\,dt\,\sum_{k=1}^N L_k(t)x_k(t)}\,,\label{pathintegralZ2}
\end{equation}
with
\begin{equation}
\boxed{S_{\text{cl,E}}=\int_{-\infty}^{+\infty} dt \left(\frac{1}{2}\dot{\bm{x}}^2+\sum_{i_1 < \cdots < i_p}\, J_{i_1i_2\cdots i_p} \, {x}_{i_1}(t)\cdots {x}_{i_p}(t)+V({\bm{x}}^2(t))\right)\,.}
\end{equation}
As usual in QFT, the loop expansion of the path integral organizes as a power series in $\hbar$, and the limit $\hbar \to 0$ is identified  with the \textit{classical limit}. The next stage at this step is to construct the averaging over disorder, that we will discuss in the next section. Note finally that we keep $m=1$ and this is not a crude limitation because in the symmetric phase and for $N\to \infty$, this parameter does not renormalizes \cite{lahoche2022functional}.

\subsection{Averaging over disorder and replica}

Note that we are aiming to construct WRG, not for a given sample $J_{i_1\cdots i_p}$ but for the quenched averaging. This is totally justified, in the large $N$ limit, because of the expected self averaging property of the functional self energy \cite{castellani2005spin}:
\begin{equation}
\lim_{N\to \infty} W_{\text{E}}[\bm{J},\bm{L}] \to \overline{W_{\text{E}}[\bm{J},\bm{L}] }\,, \mbox{ with } \, W[\bm{J},\bm{L}]:=\ln \cZ_{\text{E}}[\bm{J},\bm{L}]\,.
\end{equation}
The standard trick to compute the average logarithm is the \textit{replica method}  \cite{mezard1987spin,nishimori2001statistical,castellani2005spin}, coming from the elementary observation that:
\begin{equation}
\overline{W_{\text{E}}[\bm{J},\bm{L}] }=\lim_{n\to 0} \frac{\overline{\cZ_{\text{E}}^n[\bm{J},\bm{L}]}-1}{n}\,.
\end{equation}
The interest is that we can explicitly construct the replicated partition function $\cZ_{\text{E}}^n[\bm{J},L]$:
\begin{equation}
\cZ_{\text{E}}^n[\bm{J},\bm{L}]:=\int \, \prod_{\alpha=1}^n[d x_\alpha]\, e^{-\frac{1}{\hbar}\sum_{\alpha=1}^nS_{\text{cl}}[\bm{x}_\alpha]+\int_{-\infty}^\infty\,dt\,\sum_{k=1,\alpha=1}^{N,n} L_k(t)x_{k,\alpha}(t)}\,,
\end{equation}
where latin indices $i,j,k\cdots$ corresponds again to vector components and greek indices $\alpha,\beta,\delta \cdots$ are for replica. Note that each replica corresponds to the same sample $\bm{J}$, and the average can be easily performed from \eqref{averageJ}. We get:
\begin{equation}
\overline{\cZ_{\text{E}}^n[\bm{J},\bm{L}]}:=\int \, \prod_{\alpha=1}^n[d x_\alpha]\, e^{-\frac{1}{\hbar}\overline{S_{\text{cl}}}[\{\bm{x}\}]+\int_{-\infty}^\infty\,dt\,\sum_{k=1,\alpha=1}^{N,n} L_k(t)x_{k,\alpha}(t)}
\end{equation}
where the \textit{classical averaged action} reads in the large $N$ limit:
\begin{align}\label{classicalaveraged}
\nonumber \overline{S_{\text{cl}}}[\{\bm{x}\}]:&=\int_{-\infty}^{+\infty} dt \bigg(\frac{1}{2}\sum_{\alpha}\dot{\bm{x}}^2_\alpha-\frac{\lambda N}{2}\int_{-\infty}^{+\infty} dt^\prime\sum_{\alpha,\beta}\,  \left(\frac{\bm{x}_\alpha(t)\cdot \bm{x}_\beta(t^\prime)}{N}\right)^p \\
&+\sum_{\alpha}V({\bm{x}}^2_\alpha(t))\bigg)\,,
\end{align}
the symbol $\{\bm{x}\}$ denoting both the replicated coordinates and $\bm{x}_\alpha\cdot \bm{x}_\beta:=\sum_i{x}_{i\alpha}{x}_{i\beta}$ is the standard Euclidean scalar product. Finally, we essentially work in the Fourier space, and we denote by $\omega$ the Fourier variable dual to time $t$.
This formula is formally valid, but the construction to the limit $n\to 0$ is a difficult mathematical problem, opening the possibility of a \textit{replica symmetry breaking} \cite{castellani2005spin}. We  use a less debated formalism and more suitable for functional renormalization group applications, focusing on the \textit{cumulants} of the random observable $W[\bm{J},\bm{L}]$ \cite{Tarjus,Tarjus2}. We then keep $n$ finite, and consider different sources for the replica that break explicitly the replica symmetry before averaging. 
\medskip

We also intend to  investigate in the region of the full theory space where the replica symmetry is broken and study the stability conditions of such an assumption. To this end, we consider the averaged replicated free energy $\mathcal{W}_{\text{E}}^{(n)}[\bm{L}]$ defined as:
\begin{equation}
\mathcal{W}_{\text{E}}^{(n)}[\bm{L}]:= \ln \overline{\cZ_{\text{E}}^n[\bm{J},\mathcal{L}]}\,,
\end{equation}
where now, $\mathcal{L}:=\{\bm{L}_\alpha \}$ is a set of $n$ source fields, one per replica, 
\begin{equation}
\overline{\cZ_{\text{E}}^n[\bm{J},\mathcal{L}]}\equiv \int \, \prod_{\alpha=1}^n[d x_\alpha]\, e^{-\frac{1}{\hbar}\overline{S_{\text{cl}}}[\{\bm{x}\}]+\int_{-\infty}^\infty\,dt\,\sum_{k=1,\alpha=1}^{N,n} L_{k,\alpha}(t)x_{k,\alpha}(t)}.
\end{equation}

\section{Perturbative renormalization theory}\label{sec2}

The WRG \cite{wilson1971renormalization,wilson1983renormalization,Zinn-Justin:1989rgp} is based on the idea that long range physics could be in principle dependent on the microscopic details of the theory. Then, small wavelength phenomena can be integrated out in large scale the partition, to provide effective couplings for long wavelength degrees of freedom in such a way that long range physics is again preserved. Despite its critical phenomena origin, the WRG is indeed a powerful and general tool to investigate large scale regularities of physical systems.
\medskip

This paper  essentially focus on the effective average action (EAA) formalism due to Wetterich and Morris \cite{Berges_2002,MORRIS_1994}. The reason of this choice is because we are aiming to bring the gap with the nonperturbative framework we construct to extend the results of this paper. Accordingly with \cite{Berges_2002}, we add to the classical averaged action $\overline{S}$ a scale depend on mass called \textit{regulator}:
\begin{equation}
\Delta S_k[\{\textbf{x}\}]:=\frac{1}{2}\, \sum_{\alpha=1}^n\sum_{i=1}^N\, x_{i\alpha}(t) R_k(t-t^\prime)x_{i\alpha}(t^\prime)
\end{equation}
Note that for the purpose of this paper, the regulator is expected to be diagonal in the replica space. The role of the regulator is to froze-out large scale degrees of freedom ($\omega \lesssim k$), which acquires a large mass, whereas microscopic degrees of freedom ($\omega \gtrsim k$) essentially unaffected by the regulator are integrated out. More precisely, we require that:

\begin{enumerate}
    \item $\lim_{k\to 0} R_k(t-t^\prime)=0$, meaning that all the degrees of freedom are integrated out as $k\to 0$ (infrared (IR) limit).
    \item $\lim_{k\to \infty} R_k(t-t^\prime)\to\infty$, meaning that no fluctuation is integrated out in the deep ultraviolet (UV) limit $k\to \infty$.
\end{enumerate}
Note that, the procedure is in some sense complementary to the original Wilson point of view because the microscopic cut-off remains fixed as $k$ moves toward long time scales ($k\to 0$). We essentially focus on the Litim regulator for simplicity, but other choices can be also considered. Explicitly, in Fourier space, it reads \cite{litim2000optimisation}:
\begin{equation}
R_k(\omega)=(k^2-\omega^2)\theta(k^2-\omega^2)\,,
\end{equation}
where $\theta$ is the standard Heaviside step function. Finally, because we are ultimately interested by the variation with respect to $k$, effective integrals, due to the shape of the regulator kernel $R_k$ are essentially concentrated in  windows of momenta (frequencies) around $k$, and no divergences are expected in the calculation. For this reason in general, no UV cutoff is required at the end of the loop calculation. 
\medskip

The receipt to construct perturbative $\beta$ functions is the following. First, we construct the loop expansion of vertex functions $\Gamma_k^{(2n)}$, corresponding to the $2n$ order functional derivative of the 1PI generating functional $\Gamma_k$, defined as the slightly modified Legendre transform of the averaged replicated free energy $\mathcal{W}_{\text{E},k}^{(n)}[\mathcal{L}]$:
\begin{equation}
\Gamma_k[\mathcal{M}]+\Delta S_k [\mathcal{M}]= \sum_{\alpha=1}^n \int\, dt \, \bm{L}_\alpha(t) \cdot \bm{M}_\alpha(t)-\mathcal{W}_{\text{E},k}^{(n)}[\mathcal{L}],\label{defGammak}
\end{equation}
where $\mathcal{M}:=\{\mathbf{M}_\alpha \}$ is the set of classical fields (a $N$ dimensional vector per replica), defined as:
\begin{equation}
\frac{\delta }{\delta L_{i\alpha}}\mathcal{W}_{\text{E},k}^{(n)}[\mathcal{L}]=M_{i\alpha}\,,
\end{equation}
and where in $\mathcal{W}_{\text{E},k}^{(n)}[\mathcal{L}]$, we included an index $k$ due to  the modification of the kinetic action by $\Delta S_k$.
\medskip

Second, we take the derivative with respect to $k$, and rescale the couplings from their own canonical dimension (fixed by power counting) and by the wave function renormalization (which  is  equal $1$ in the large $N$ limit). Note that the definition for $\Gamma_k$ given by \eqref{defGammak} ensures that $\Gamma_k$ is a smooth interpolation between the classical averaged action $\overline{S_{\text{cl}}}[\{\bm{x}\}]$ given by \eqref{classicalaveraged} as $k\to \infty$ and the full effective action $\Gamma$, including all the quantum fluctuation and corresponding to the true Legendre transform of the $\mathcal{W}_{\text{E}}^{(n)}[\mathcal{L}]$ (without $k$). 
\medskip

\begin{figure}
\begin{center}
\includegraphics[scale=1.2]{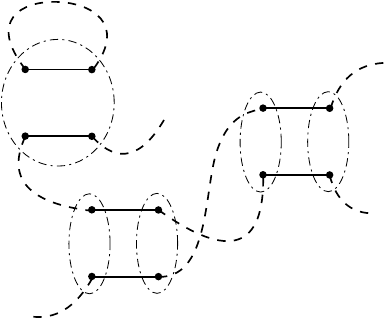}
\end{center}
\caption{A typical Feynman diagram for $p=2$, involving two non local vertices, one local quartic vertex and four external points. }\label{FeynmanDiag}
\end{figure}

Any vertex function $\Gamma_k^{(2n)}$ (with $2n$ external points) then expands in power series in perturbation theory \cite{peskin2018introduction,Zinn-Justin:1989rgp,zinn2007phase}:
\begin{equation}
\Gamma_k^{(2n)}(\omega_1,\cdots \omega_{2n})= \sum_{\mathcal{G}\in\mathbb{G}_{2n}}\,\frac{1}{s(\mathcal{G})} \mathcal{A}_{\mathcal{G}}(\omega_1,\cdots \omega_{2n})\,,
\end{equation}
where $\mathbb{G}_{2n}$ is the set of 1PI Feynman diagrams with $2n$ external points, labelling the Feynman amplitudes $\mathcal{A}_{\mathcal{G}}$, and the additional factor $s(\mathcal{G})$ called symmetry factor, depends on the dimension of the automorphism group of the graph\footnote{If the coupling constant is cleverly normalized, $s(\mathcal{G})$ equals the dimension of the automorphism group.}. A Feynman graph is basically a set of edges and nodes or vertices, which materialize respectively the bare propagator arising from Wick contractions and interactions involved in the bare action. In this paper we adopt the same notation as in our previous work \cite{lahoche2024largetimeeffectivekinetics, lahoche2021functional}, which accommodates the three main features of the effective field theory we consider. Within this convention, Feynman graphs look as \textit{multi-graphs} including edges and \textit{hyper-edges} \cite{bretto2013hypergraph}. Black nodes materialize fields, and there are two different kinds of edges in such a Feynman graph:
\begin{itemize}
    \item Solid edges, materializing latin indices contractions (i.e. scalar product between fields):
\begin{equation}
\vcenter{\hbox{\includegraphics[scale=1.2]{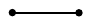}}}\equiv \sum_{i=1}^N \, x_{i\alpha} (t)x_{i\alpha}(t)\,.
\end{equation}
\item Dashed edges, materializing Wick contractions between black nodes.
\end{itemize}
In addition, there are hyper-edges, linking an arbitrary number of nodes, and materialized by a surface with dashed-dotted boundary. All the fields into such a surface have the same replica index, and are taken at the same time. For instance:
\begin{equation}
\vcenter{\hbox{\includegraphics[scale=1.2]{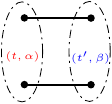}}}\equiv \int {\color{red}dt}\int {\color{blue}dt^\prime} \sum_{{\color{red}\alpha},{\color{blue}\beta}}\, (\textbf{x}_{{\color{red}\alpha}}({\color{red}t})\cdot \textbf{x}_{{\color{blue}\beta}}({\color{blue}t^\prime}))^2\,.\label{equationquartic}
\end{equation}
Such a typical Feynman graph is pictured on Figure \ref{FeynmanDiag} for the $p=2$ model and with $4$ external edges. We call \textit{external nodes} the black points connected with external edges. Furthermore, we define as follows the so called \textit{bubble}:
\begin{definition}
We call bubble any global $O(N)$ invariant set of dashed dotted cells, hooked with solid edges, such that two arbitrary bubbles are linked by at least one connected path made of cells and solid edge. 
\end{definition}\label{bubbledef}

\begin{figure}
\begin{center}
\includegraphics[scale=1.2]{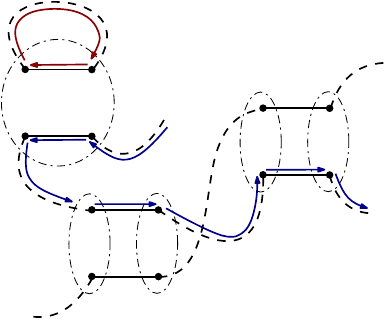}
\end{center}
\caption{Illustration of the concept of face, a closed face (red cycle) and an open face (blue cycle).}\label{figFace}
\end{figure}

A relevant concept which is useful to discuss the large $N$ limit is the notion of face of the Feynman diagrams. We then have the following definition:
\begin{definition}
A face is  a cycle made of an alternative sequence of dashed and solid edges. A face can be closed or open see Figure \ref{figFace} for illustration.
\end{definition}
Note that a closed face shares a global factor $\sum_{i=1}^N=N$. Then, for the $p=2$ theory, if we denote respectively  by $V$ and $F$  the numbers of quartic vertices (local or not) and closed faces of some Feynman graph $\mathcal{G}$, the amplitude scale with $N$ as:
\begin{equation}
\mathcal{A}_{\mathcal{G}} \sim N^{-V+F}\,.
\end{equation}
The leading order graphs are then those which optimize the number of created faces, taking into account all the other constraints imposed for the configurations of the internal dashed edges (not connected to external nodes). The leading order graph structure is more transparent using loop-vertex representation (LVR) \cite{rivasseau2018loop,rivasseau2010feynman}, and for the quartic theory, the leading order graphs look as planar trees in that representation \cite{lahoche2024largetimeeffectivekinetics}. 

\section{Large time one-loop $\beta$-functions}\label{sec3}

In this section we derive one loop $\beta$-functions in the small $p$ regime, typically for $p=2$ and $p=3$. The reason why we choose this regime is: as $p$ increases, the effective behavior of low $p$ vertices involves many loops, and becomes therefore highly quantum regime. In contrast, for $p$ small enough, the quantum to classical limit is expected less sharp, because a view number of loops are required to construct effective effects (each loop share a factor $\hbar$). In the computation, we then use $\hbar$ as a ‘‘small parameter" to construct the loop expansion, despite the fact that we finally set $\hbar =1$ at the end of the computation. 

\subsection{The $p=2$ model with quartic confinement}

\begin{figure}
\begin{center}
\includegraphics[scale=0.8]{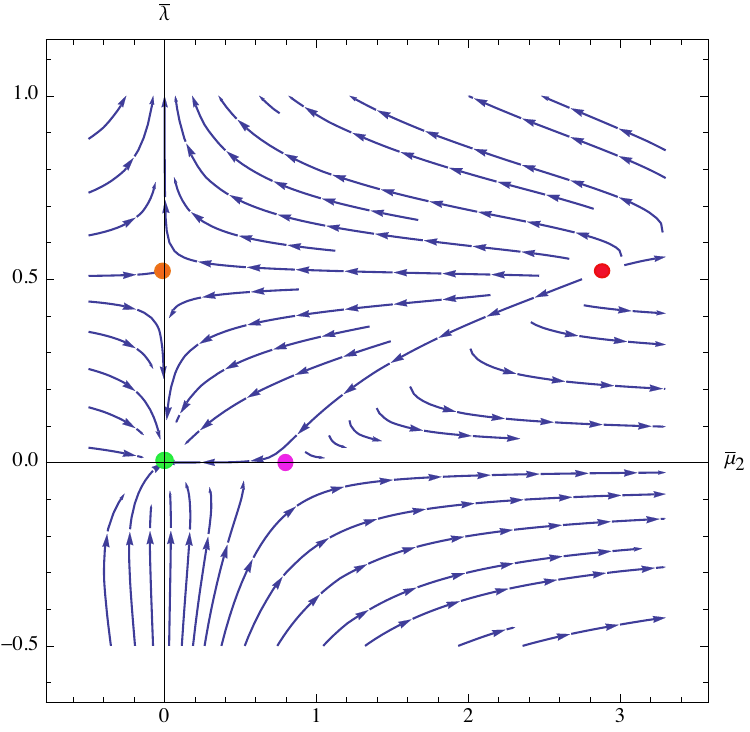}
\end{center}
\caption{Fixed point structure of the renormalization group flow in the deep IR, fixed points FP1, FP2 ans FP3 are respectively materialized by the red, purple and orange points, the green point is the Gaussian fixed point. Coupling $\bar{\mu}_2$ has been rescaled by a factor $2\pi$ for convenience.}\label{figflow1}
\end{figure}

Let us begin by the case $p=2$. The computation of the one-loop $\beta$ function requires two ingredients: The first one is  the one loop \textit{self energy} and the second is the one loop $4$-point vertex function $\Gamma_k^{(4)}$. Let us recall that in perturbation theory, the self energy $\Sigma$ is  defined as the (formal) sum of Feynman amplitudes indexed by 1PI Feynman diagrams. The diagonal part of the bare propagator, moreover, writes in Fourier space as\footnote{Note that bubble definition \ref{bubbledef} assume a sum over replica, that we ignore in the following construction, by considering  only a single replica component.}:
\begin{equation}
C(\omega):=\frac{1}{\omega^2+\mu_1+R_k(\omega)}\,,
\end{equation}
that has no singularity as soon as the bare mass is positive. Graphically, the one loop structure of self energy and $4$-points vertex function is:
\begin{equation}
\Sigma\,=\,\vcenter{\hbox{\includegraphics[scale=0.7]{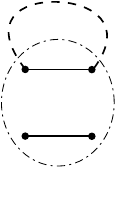}}}\,+\,\vcenter{\hbox{\includegraphics[scale=0.7]{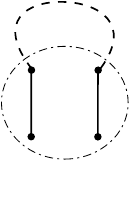}}}\,+\,\vcenter{\hbox{\includegraphics[scale=0.7]{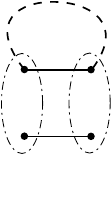}}}\,+\,\vcenter{\hbox{\includegraphics[scale=0.7]{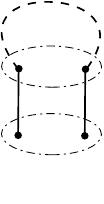}}}\,+\mathcal{O}(\hbar^2)\,,
\end{equation}
\begin{align}
\nonumber\Gamma_k^{(4)}&\,=\, \vcenter{\hbox{\includegraphics[scale=0.55]{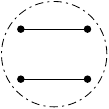}}}\,+\,\vcenter{\hbox{\includegraphics[scale=0.55]{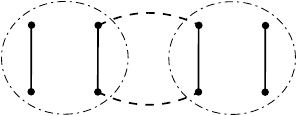}}}\,+\,\vcenter{\hbox{\includegraphics[scale=0.55]{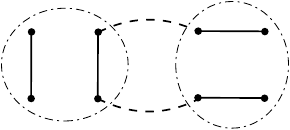}}}\,+\,\vcenter{\hbox{\includegraphics[scale=0.55]{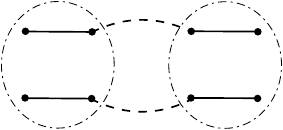}}}\\\nonumber
&\,+\,\vcenter{\hbox{\includegraphics[scale=0.6]{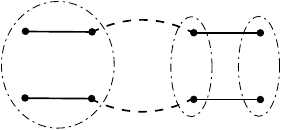}}}
\,+\,\vcenter{\hbox{\includegraphics[scale=0.6]{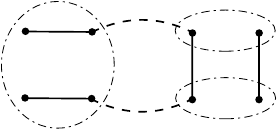}}}\,+\,\vcenter{\hbox{\includegraphics[scale=0.6]{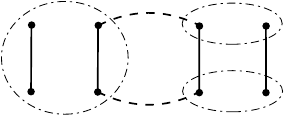}}}\\\nonumber
&\,+\,\vcenter{\hbox{\includegraphics[scale=0.6]{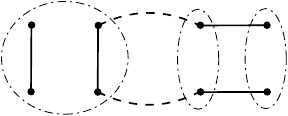}}}\,+\,\vcenter{\hbox{\includegraphics[scale=0.6]{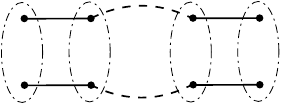}}}\,+\,\vcenter{\hbox{\includegraphics[scale=0.6]{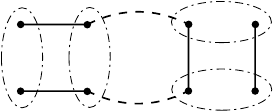}}}\\
&\,+\,\vcenter{\hbox{\includegraphics[scale=0.6]{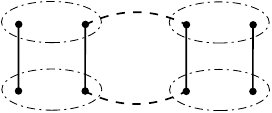}}}\,+\mathcal{O}(\hbar^2)\,.
\end{align}
In the large $N$ limit, only diagrams which maximize the number of faces have to be kept. For instance, one loop diagrams for $\Sigma$ involve respectively one face or zero face:
\begin{equation}
\Sigma\,=\,\vcenter{\hbox{\includegraphics[scale=0.7]{LoopSigma1.pdf}}}\,+\,\vcenter{\hbox{\includegraphics[scale=0.7]{LoopSigma3.pdf}}}\,+\mathcal{O}(\hbar^2,1/N)\,,\label{expansionloopSigma}
\end{equation}
then, only the first one diagram have to be kept in the large $N$ limit. In the same way, we have for the 4 points function:
\begin{equation}
\Gamma_k^{(4)}\,=\, \vcenter{\hbox{\includegraphics[scale=0.55]{V4.pdf}}}\,+\,\vcenter{\hbox{\includegraphics[scale=0.55]{Loop4pts1.pdf}}}\,+\,\vcenter{\hbox{\includegraphics[scale=0.6]{Loop4pts10.pdf}}}\,+\,\vcenter{\hbox{\includegraphics[scale=0.6]{Loop4pts6.pdf}}}\,+\mathcal{O}(\hbar^2,1/N^2)\,.\label{expansionloopGamma4}
\end{equation}
Let us compute each Feynman diagram. For the first term on the right hand side of equation \eqref{expansionloopSigma},  because of the Feynman rules and taking into account symmetry factors we have:
\begin{align}
\nonumber\vcenter{\hbox{\includegraphics[scale=0.7]{LoopSigma1.pdf}}}&\,=\,-4\times \int_{-\infty}^{+\infty} \frac{d\omega}{2\pi} \, \left(\frac{\mu_2}{4N}\right) \left(\sum_{i=1}^N1\right)\, \frac{1}{\omega^2+\mu_1+R_k(\omega)}\\\nonumber
&\,=-\frac{\mu_2}{\pi} \Bigg(\frac{k}{k^2+\mu_1}+\int_k^{+\infty} \,\frac{1}{\omega^2+\mu_1} \Bigg)\\
&\,=-\frac{\mu_2}{\pi} \, \Bigg(\frac{k}{k^2+\mu_1}+\frac{\pi -2 \tan ^{-1}\left(\frac{k}{\sqrt{\mu_1 }}\right)}{2 \sqrt{\mu_1 }} \Bigg)\,.
\end{align}
Note that the above expression is independent of the external frequencies, and then only renormalizes the mass. For the second diagram, we get in the same way, taking into account the non local structure (in time and replica):
\begin{align} \nonumber\vcenter{\hbox{\includegraphics[scale=0.7]{LoopSigma3.pdf}}}&\,=\, \frac{2 \lambda}{\omega^2+\mu_1+R_k(\omega)}=\frac{2 \lambda}{\omega^2+\mu_1}\theta(\omega^2-k^2)+\frac{-2 \lambda}{k^2+\mu_1}\theta(k^2-\omega^2)\\
&\,=\,\frac{2\lambda}{\mu_1} \left(1-\frac{\omega^2}{\mu_1}+\mathcal{O}(\omega^2) \right)\theta(\omega^2-k^2)+\frac{2 \lambda}{k^2+\mu_1}\theta(k^2-\omega^2)\,,
\end{align}
where $\omega$ designates here the external frequency. Because of the non local structure of the interaction, this correction includes not only a finite shift of the mass, but also a non trivial renormalization of the wave function. Indeed, for $\omega$ small enough, the effective propagator reads:
\begin{align}
G_k(\omega)&=\frac{1}{\omega^2+\mu_1+R_k(\omega)-\Sigma(\omega)}\\
&\approx \frac{1}{Z^{(1)}_{\infty}\omega^2+\mu_{1,\infty}^{\text{eff}}+R_k(\omega)}+ \mathcal{O}(\omega^4)\,,
\end{align}
where the one-loop wave function renormalization $Z^{(1)}_{\infty}$ and the one loop effective mass $\mu_{1,\infty}^{\text{eff}}$ are:
\begin{equation}
Z^{(1)}_{\infty}:=1+\frac{\lambda}{\pi\mu_1^2} \,\theta(\omega^2-k^2) 
\end{equation}
\begin{equation}
\mu_{1,\infty}^{\text{eff}}:=\mu_{1}+\frac{\mu_2}{\pi} \, \Bigg(\frac{k}{k^2+\mu_1}+\frac{\pi -2 \tan ^{-1}\left(\frac{k}{\sqrt{\mu_1 }}\right)}{2 \sqrt{\mu_1 }} \Bigg)-\frac{2\lambda}{k^2+\mu_1}\,,\label{effectivemass}
\end{equation}
where the subscript $\infty$ refers to the large $N$ limit. Note that the wave function renormalization takes an unconventional form, having a sharp dependency on the regulator. As $k\neq 0$ however, we can always set $Z^{(1)}_{\infty}=1$, and the anomalous dimension
\begin{equation}
\eta:=\frac{1}{Z^{(1)}_{\infty}} k\, \frac{d}{dk} Z^{(1)}_{\infty}\,,
\end{equation}
The contribution vanishes at one loop. Note that we have discarded the corresponding term in the mass shift. Now, let us proceed with the computation of the 4-point diagrams. Setting the external frequencies to zero (and omitting the global factor $\delta(0)$, which is common to all contributions), we obtain the following:
\begin{align}
\nonumber \vcenter{\hbox{\includegraphics[scale=0.55]{Loop4pts1.pdf}}}&= -\frac{1}{2!}4! \times 2^3 \left(\frac{\mu_2}{4N}\right)^2 \int_{-\infty}^{+\infty} \frac{d\omega}{2\pi}\,\left(\sum_{i=1}^N 1\right) \frac{1}{(\omega^2+\mu_1+R_1(\omega))^2}\\
&=-\frac{3\mu_2^2}{N \pi} \left(\frac{k}{(k^2+\mu_1)^2}+\frac{\cot ^{-1}\left(\frac{k}{\sqrt{\mu_1 }}\right)-\frac{k \sqrt{\mu_1 }}{k^2+\mu_1 }}{2 \mu_1^{3/2}}\right)\,,
\end{align}
\begin{align}
\nonumber \vcenter{\hbox{\includegraphics[scale=0.6]{Loop4pts6.pdf}}}&= 4!\times 2^3\times  \left(\frac{\lambda \mu_2}{8 N^2}\right) \left(\sum_{i=1}^N 1\right) \frac{1}{(k^2+\mu_1)^2}\\
&= 4! \times  \frac{\lambda \mu_2}{N}  \frac{1}{(k^2+\mu_1)^2}\,,
\end{align}
\begin{align}
\nonumber\vcenter{\hbox{\includegraphics[scale=0.6]{Loop4pts10.pdf}}}&=-\frac{4!\times 2^3}{2} \left(\frac{\lambda}{2N}\right)^2 \left(\sum_{i=1}^N 1\right)  \, \frac{\delta(0)}{(k^2+\mu_1)^2}\\
&=-4!\times  \left(\frac{\lambda^2}{N}\right)  \, \frac{\delta(0)}{(k^2+\mu_1)^2}\,.
\end{align}
Then $\Gamma_k^{(4)}$ splits in two contributions, 
\begin{equation}
\Gamma_k^{(4)}=: \Gamma_{k,L}^{(4)}+\Gamma_{k,NL}^{(4)}\,,
\end{equation}
the subscript $L$ and $NL$ meaning respectively ‘‘local'' and ‘‘non-local'' contributions. Explicitly (we choose external indices to be equal):
\begin{equation}
\Gamma_{k,L}^{(4)}:=\frac{6\delta(0)}{N} \Bigg(\mu_2- \frac{\mu_2^2}{2\pi} \left(\frac{k}{(k^2+\mu_1)^2}+\frac{\cot^{-1}\left(\frac{k}{\sqrt{\mu_1 }}\right)-\frac{k \sqrt{\mu_1 }}{k^2+\mu_1 }}{2 \mu_1^{3/2}}\right) + \frac{ 4\lambda \mu_2}{(k^2+\mu_1)^2}  \Bigg)\,,
\end{equation}
\begin{equation}
\Gamma_{k,NL}^{(4)}=\frac{12 \delta^2(0)}{N}\Bigg(\lambda-\frac{2\lambda^2}{(k^2+\mu_1)^2}\Bigg)\,.
\end{equation}
These equations define the \textit{effective couplings}, 
\begin{equation}
\mu_{2,\infty}^{\text{eff}}:=\mu_2- \frac{\mu_2^2}{2\pi} \left(\frac{k}{(k^2+\mu_1)^2}+\frac{\cot^{-1}\left(\frac{k}{\sqrt{\mu_1 }}\right)-\frac{k \sqrt{\mu_1 }}{k^2+\mu_1 }}{2 \mu_1^{3/2}}\right) + \frac{ 4\lambda \mu_2}{(k^2+\mu_1)^2}\,,\label{effectivemu2}
\end{equation}
\begin{equation}
\lambda^{\text{eff}}_\infty:=\lambda-\frac{2\lambda^2}{(k^2+\mu_1)^2}\,.\label{effectivelambda}
\end{equation}
Equations \eqref{effectivemass}, \eqref{effectivemu2}, and \eqref{effectivelambda} define the effective couplings 'at scale $k$, and the one-loop $\beta$-functions can be computed by taking the derivative with respect to $k$. First, we focus on the critical regime, which is approached from above, i.e. $\mu_1 \to 0^+$, and we define the dimensionless couplings as follows:
\begin{equation}
{\lambda}^{\text{eff}}_\infty:=k^4\bar{\lambda}^{\text{eff}}_\infty\,,\quad \mu_{2,\infty}^{\text{eff}}:=k^3\bar{\mu}_{2,\infty}^{\text{eff}}\,, \quad \mu_{1,\infty}^{\text{eff}}:=k^2\bar{\mu}_{1,\infty}^{\text{eff}}\,,
\end{equation}
then, because:
\begin{equation}
\frac{\cot^{-1}\left(\frac{k}{\sqrt{\mu_1 }}\right)-\frac{k \sqrt{\mu_1 }}{k^2+\mu_1 }}{2 \mu_1^{3/2}}= \frac{1}{3k^3}+\mathcal{O}(\mu_1/k^2)\,,
\end{equation}
and:
\begin{equation}
\frac{\pi -2 \tan ^{-1}\left(\frac{k}{\sqrt{\mu_1 }}\right)}{2 \sqrt{\mu_1 }}=\frac{1}{k}+\mathcal{O}(\mu_1/k^2)
\end{equation}
we get, for $k$ small enough (deep IR):
\begin{equation}
\dot{\bar{\mu}}_{1,\infty}^{\text{eff}}:=-2{\bar{\mu}}_{1,\infty}^{\text{eff}}-\frac{2}{\pi} {\bar{\mu}}_{2,\infty}^{\text{eff}} +4{\bar{\lambda}}^{\text{eff}}_\infty\,,
\end{equation}
\begin{equation}
\dot{\bar{\mu}}_{2,\infty}^{\text{eff}}:=-3 {\bar{\mu}}_{2,\infty}^{\text{eff}}+4{\bar{\mu}}_{2,\infty}^{\text{eff}}\left(\frac{{\bar{\mu}}_{2,\infty}^{\text{eff}}}{2\pi}-4 {\bar{\lambda}}^{\text{eff}}_\infty\right)\,,
\end{equation}
\begin{equation}
\dot{\bar{\lambda}}^{\text{eff}}_\infty:=-4{\bar{\lambda}}^{\text{eff}}_\infty+8 ({\bar{\lambda}}^{\text{eff}}_\infty)^2\,,
\end{equation}
where the dot means derivative with respect to $s:=\ln (k)$. In addition to the Gaussian fixed point, these equations admits three non-Gaussian fixed points for the values:
\begin{align}
\text{FP1}:=({\bar{\mu}}_{2,1}^{\text{eff}*},{\bar{\lambda}}^{\text{eff}*}_1)&=\left(\frac{11 \pi}{2},\frac{1}{2}\right)\,,\\
\text{FP2}:=({\bar{\mu}}_{2,3}^{\text{eff}*},{\bar{\lambda}}^{\text{eff}*}_2)&=\left(\frac{3 \pi}{2},0\right)\,,\\
\text{FP3}:=({\bar{\mu}}_{2,3}^{\text{eff}*},{\bar{\lambda}}^{\text{eff}*}_3)&=\left(0,\frac{1}{2}\right)\,,
\end{align}
and the critical exponents\footnote{The opposite values of the stability matrix.} are:
\begin{align}
(\theta_{1},\theta_2)_{\text{FP1}}&=(-11,-4)\,,\label{FP1}\\
(\theta_{1},\theta_2)_{\text{FP2}}&=(4,-3)\,,\label{FP2}\\
(\theta_{1},\theta_2)_{\text{FP3}}&=(11,-4)\label{FP3}\,.
\end{align}
Fixed point FP1 looks as a pure IR attractor, whereas fixed point FP2 and FP3 look as Wilson-Fisher critical fixed points, with one attractive and one repulsive direction. Figure \ref{figflow1} summarize the fixed point structure of the deep IR renormalization group flow (arrows being oriented toward UV scales). Note that the values we find are large with respect to $1$, and should invalidate the perturbative computation. We improve the reliability of the fixed point in the next section using nonperturbative vertex expansion at the leading order of the derivative expansion.

\subsection{The $p=3$ model with sextic confinement}\label{sec5}

Now, let us consider the case where $p=3$, which is equivalent to  the non local sextic interactions, and we add a sextic local potential to the interactions. As in the previous section we will consider the leading orders in the loop expansion, taking care that the disorder effect is of order $\hbar^{-1}$. In the large $N$ limit, the relevant diagrams are the following:
\begin{equation}
\Sigma\,=\,\vcenter{\hbox{\includegraphics[scale=0.7]{LoopSigma1.pdf}}}\,+\,\vcenter{\hbox{\includegraphics[scale=0.7]{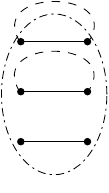}}}\,+\,\vcenter{\hbox{\includegraphics[scale=0.7]{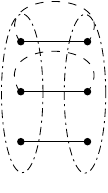}}}\,,
\end{equation}
\begin{align}
\nonumber\Gamma_k^{(4)}\,=\, &\vcenter{\hbox{\includegraphics[scale=0.55]{V4.pdf}}}\,+\,\vcenter{\hbox{\includegraphics[scale=0.55]{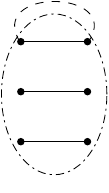}}}\,+\,\vcenter{\hbox{\includegraphics[scale=0.55]{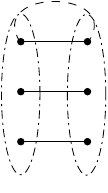}}}\,+\,\vcenter{\hbox{\includegraphics[scale=0.55]{Loop4pts1.pdf}}}\\
&\,+\,\vcenter{\hbox{\includegraphics[scale=0.55]{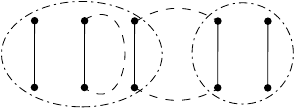}}}\,+\,\vcenter{\hbox{\includegraphics[scale=0.55]{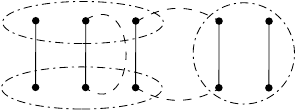}}}\,,
\end{align}
\begin{align}
\Gamma_k^{(6)}\,=\, \vcenter{\hbox{\includegraphics[scale=0.55]{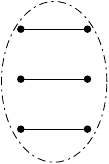}}}\,+\,\vcenter{\hbox{\includegraphics[scale=0.55]{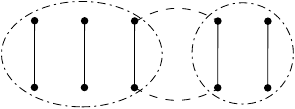}}}\,+\,\vcenter{\hbox{\includegraphics[scale=0.55]{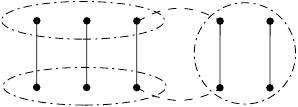}}}\,.
\end{align}
Note that we discarded the one loop term of order $\mu_2^3$. All the loops involved in these equations, for vanishing external momenta, have been computed in the previous subsection, and we can reproduce exactly the same kind of analysis we done in our previous work \cite{lahoche2024largetimeeffectivekinetics}, considering the exact Wetterich-Morris flow equation:
\begin{equation}
\dot{\Gamma}_k=\frac{1}{2}\int_{-\infty}^{+\infty} \frac{d\omega}{2\pi} \sum_{\alpha=1}^n\sum_{i=1}^N\dot{R}_k(\omega) G_{k,ii}(\omega,-\omega,\alpha)\,,\label{Wett}
\end{equation}
where the effective $2$-points function $G_k(\omega,\omega^\prime)$ is defined as:
\begin{equation}
G_{k,ij}^{-1}(\omega,\omega^\prime,\alpha)-R_k(\omega) \delta_{ij}\delta(\omega+\omega^\prime)=\frac{\delta^2 \Gamma_k}{\delta M_{i\alpha}(\omega) \delta M_{j\alpha} (\omega^\prime)}\,.
\end{equation}
Note that the effective propagator in diagonal with respect to the replica indices, and local in time. This assumption make sense in the vicinity of the Gaussian fixed point, in the perturbative region, but could become wrong in some regions of the phase space \cite{lahoche2024largetimeeffectivekinetics}. Taking the second derivative with respect to the classical field $M_{i\alpha}(\omega)$, and assuming to be is the symmetric phase (odd vertex function vanish on shell):
\begin{equation}
\dot{\Gamma}_{k,i_1i_2}^{(2)}(\omega_1,\omega_2)=-\int_{-\infty}^{+\infty} \frac{d\omega}{4\pi} \sum_{i,j,j^\prime=1}^N\, \dot{R}_k(\omega) \tilde{G}_{k,ij}(\omega,\alpha)\tilde{G}_{k,ij^\prime}(\omega,\alpha) \Gamma_{k,jj^\prime i_1 i_2}^{(4)}(\omega,-\omega,\omega_1,\omega_2)\,,
\end{equation}
where:
\begin{equation}
G_{k,ij}(\omega,\omega^\prime,\alpha)=\tilde{G}_{k,ij}(\omega,\alpha) \delta(\omega+\omega^\prime)\,,
\end{equation}
and where we we omitted the replica indices for the $4$-points function for simplicity. Assuming to be close enough to the Gaussian fixed point, we may use of the local approximation for $\Gamma_k^{(4)}$, and we define the effective coupling $u_4$ as:
\begin{equation}
\Gamma_{k,i_1i_2i_3i_4}^{(4)}(\omega_1,\omega_2,\omega_3,\omega_4)=: \frac{2u_4}{N} \delta\left(\sum_{i=1}^4\omega_i\right) W_{i_1i_2i_3i_4}\,,\label{defcouplingu4}
\end{equation}
where:
\begin{equation}
W_{i_1i_2i_3i_4}:=\delta_{i_1i_2}\delta_{i_3i_4}+\delta_{i_1i_3}\delta_{i_2i_4}+\delta_{i_1i_4}\delta_{i_2i_3}\,.
\end{equation}
Furthermore, close enough to the Gaussian fixed point and in the critical regime:
\begin{equation}
\tilde{G}_{k,ij}(\omega,\alpha) \approx \frac{\delta_{ij}}{\omega^2+R_k(\omega)}\,+\, \mathcal{O}(u_2)\,.
\end{equation}
Moreover, 
\begin{equation}
{\Gamma}_{k,i_1i_2}^{(2)}(\omega_1,\omega_2)\approx\delta(\omega_1+\omega_2)\delta_{ij}\left(u_2(k)+\omega^2_1\right)
\end{equation}
and the flow equation becomes:
\begin{equation}
\dot{u}_2=-\frac{k^2 u_4}{\pi} \frac{1}{k^4} \int_{-\infty}^{\infty} d\omega\, \theta(k^2-\omega^2)=-\frac{2 u_4}{\pi}\, \frac{1}{k}\,.
\end{equation}
Defined dimensionless couplings as previously, $\bar{u}_{2}:=k^{-2} u_2$ and $\bar{u}_{4}:=k^{-3} u_4$, we get:
\begin{equation}
\dot{\bar{u}}_2=-2 \bar{u}_2-\frac{2 \bar{u}_4}{\pi}\,.
\end{equation}
We will proceed in the same way for the other couplings, and we define the local and non-local sextic couplings by:
\begin{equation}
\Gamma_{k,i_1i_2i_3i_4i_5i_6}^{((6),\text{L})}(\omega_1,\omega_2,\omega_3,\omega_4,\omega_5,\omega_6)=: \frac{u_6}{6N^2} \delta\left(\sum_{i=1}^6\omega_i\right) W_{i_1i_2i_3i_4i_5i_6}^{(\text{L})}\,,\label{defu6}
\end{equation}
\begin{equation}
\Gamma_{k,i_1i_2i_3i_4i_5i_6}^{((6),\text{NL})}(\omega_1,\omega_2,\omega_3,\omega_4,\omega_5,\omega_6)=: \frac{\tilde{u}_6}{6N^2} W_{i_1i_2i_3i_4i_5i_6}^{(\text{NL})}(\omega_1,\omega_2,\omega_3,\omega_4,\omega_5,\omega_6)\,,
\end{equation}
where:
\begin{equation}
W_{i_1i_2i_3i_4i_5i_6}^{(\text{L})}=\sum_{\pi}\, \delta_{i_{\pi(1)}i_{\pi(2)}}\delta_{i_{\pi(3)}i_{\pi(4)}}\delta_{i_{\pi(5)}i_{\pi(6)}}\,,
\end{equation}
$\pi$ is the set of the permutation on $[6]^*:=\{1,2,\cdots,6\}$
and
\begin{align}
\nonumber W_{i_1i_2i_3i_4i_5i_6}^{(\text{NL})}&(\omega_1,\omega_2,\omega_3,\omega_4,\omega_5,\omega_6)=\sum_{\pi}\, \delta_{i_{\pi(1)}i_{\pi(2)}}\delta_{i_{\pi(3)}i_{\pi(4)}}\delta_{i_{\pi(5)}i_{\pi(6)}}\\
&\times \delta \left(\omega_{\pi(1)}+\omega_{\pi(3)}+\omega_{\pi(5)}\right)\delta \left(\omega_{\pi(2)}+\omega_{\pi(4)}+\omega_{\pi(6)}\right)\,.
\end{align}
By taking fourth and sextic functional derivations of the exact RG equation \eqref{Wett}, and identifying local and non-local interactions on both sides of the equations, we deduce the corresponding coupling RG flow equations. It is suitable to introduce a graphical notation, and the corresponding equations reads\footnote{Once again, bubbles involve in principle a sum over replica index, which exists also on the left hand side. We assume these additional factors are cancelled in the diagrammatic construction of the flow equation, as we assumed it for perturbation theory.}:
\begin{align}
\dot{u}_4\,=\,-\,\vcenter{\hbox{\includegraphics[scale=0.6]{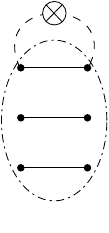}}}\,-\,\vcenter{\hbox{\includegraphics[scale=0.6]{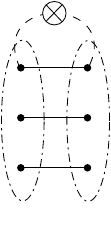}}}\,+\,\vcenter{\hbox{\includegraphics[scale=0.6]{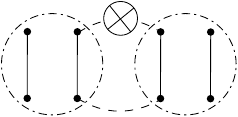}}}\,,\label{sextictrunc1}
\end{align}
\begin{align}
\dot{u}_6\,=\,\vcenter{\hbox{\includegraphics[scale=0.6]{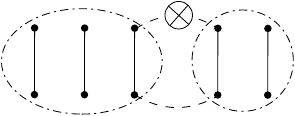}}}\,+\,\vcenter{\hbox{\includegraphics[scale=0.6]{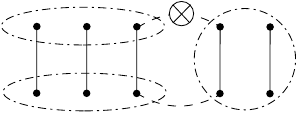}}}\,-\,\vcenter{\hbox{\includegraphics[scale=0.6]{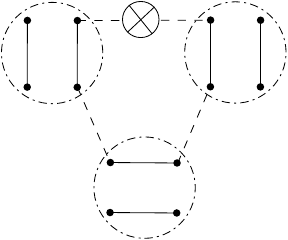}}}\,,\label{sextictrunc2}
\end{align}
\begin{align}
\dot{\tilde{u}}_6\,=\,0\,,
\end{align}
where bubbles represent effective vertices rather than perturbative vertices, dashed edges represent effective propagator, and crossed circles are $\dot{R}_k$ insertions. The computation of the loop integrals leads the equations\footnote{Factors $\pi$ differ from the previous calculation because of our conventions for the definition of effective couplings.}:


\begin{align}
 \dot{\bar{u}}_4&=-3 {\bar{u}}_4 - \frac{4 \bar{u}_6}{ \pi}\, -\frac{2\bar{\tilde{u}}_6}{\pi}  + \frac{16\bar{u}_4^2}{\pi} \,,\\
\dot{\bar{u}}_6&=-4 {\bar{u}}_6 + 48 \frac{\bar{u}_4\bar{u}_6}{\pi} + 24 \frac{\bar{u}_4\bar{\tilde{u}}_6}{\pi} \,,\\
\dot{\bar{\tilde{u}}}_6&=-5 {\bar{\tilde{u}}}_6 \,.
\end{align}
In contrast with what we obtained for $p=2$, the resulting flow equations have no global fixed point because of the flow equation for ${\bar{\tilde{u}}}_6$. The origin of this difference can be traced from the fact that quartic interactions are the only ones whose flow depends on their square. However, it may exist fixed trajectories, such that $\dot{\bar{u}}_4=\dot{\bar{u}}_6=0$. We find three such a trajectory. The first solution has a positive value for the sextic coupling, and up to order $1$ in ${\bar{\tilde{u}}}_6$ it reads:
\begin{align}
\bar{u}_{4*}&=\frac{3 \pi }{16}-\frac{8 \overline{\tilde{u}}_6}{15 \pi }+\mathcal{O}(\overline{\tilde{u}}_6^2)\,,\\
 \bar{u}_{6*}&=-\frac{9 \overline{\tilde{u}}_6}{10} +\mathcal{O}(\overline{\tilde{u}}_6^2)\,,
\end{align}
with critical exponents:
\begin{align}
\theta_{1,*}&=-3+\frac{832 \overline{\tilde{u}}_6}{15 \pi ^2}+\mathcal{O}(\overline{\tilde{u}}_6^2)\,,\\
\theta_{2,*}&=-5-\frac{64 \overline{\tilde{u}}_6}{5 \pi ^2}+\mathcal{O}(\overline{\tilde{u}}_6^2)\,.
\end{align}

The two directions are irrelevant, and asymptotically, the fixed point behaves as an attractor. Note that ${\bar{\tilde{u}}}_6$ must be negative, consistent with the hypothesis that this interaction stems from the integration of the disorder tensor. The two other solutions are characterized by the fact that the sextic coupling is negative, which raises concerns about the stability of the path integral. We will return to and further discuss this issue in the next section (subsection \ref{subsectionp3}, remark \ref{remarkconv}). For the moment, we consider these solutions formally, with the assumption that higher-order relevant interactions, generated at later stages of the perturbative expansion, could enhance the convergence of the integral. The two solutions are as follows (with $+$ and $-$ labels distinguishing them):
\begin{align}
\bar{u}_{4+}&=-\frac{2 \overline{\tilde{u}}_6}{3 \pi }+\mathcal{O}(\overline{\tilde{u}}_6^2)\,,\\
\bar{u}_{6+}&=0+\mathcal{O}(\overline{\tilde{u}}_6^2)\,,\\
\bar{u}_{4-}&=\frac{\pi }{12}+\frac{6 \overline{\tilde{u}}_6}{5 \pi }+\mathcal{O}(\overline{\tilde{u}}_6^2)\,,\\
\bar{u}_{6-}&=-\frac{5 \pi ^2}{144}-\frac{3 \overline{\tilde{u}}_6}{5} +\mathcal{O}(\overline{\tilde{u}}_6^2)\,,
\end{align}
with critical exponents:
\begin{align}
\theta_{1+}&=4-\frac{64 \overline{\tilde{u}}_6}{\pi ^2}+\mathcal{O}(\overline{\tilde{u}}_6^2)\,,\\
\theta_{2+}&=3+\frac{352 \overline{\tilde{u}}_6}{3 \pi ^2}+\mathcal{O}(\overline{\tilde{u}}_6^2)\,,\\
\theta_{1-}&=2.75-4.42\overline{\tilde{u}}_6+\mathcal{O}(\overline{\tilde{u}}_6^2)\,,\\
\theta_{2-}&=-2.42-5.30\overline{\tilde{u}}_6+\mathcal{O}(\overline{\tilde{u}}_6^2)\,.
\end{align}
The first fixed point has two relevant directions and appears to be a UV attractor. The second fixed point resembles a Wilson-Fisher fixed point, with one relevant and one irrelevant direction. However, at this stage, one may question the reliability of this second fixed point, given the large values of the couplings, which seem sufficient to invalidate perturbation theory. The same concern applies to the quartic coupling at the first fixed point. Therefore, only the second fixed point, identified as a UV attractor, appears to be a reliable prediction from perturbation theory—a conclusion that we will confirm using nonperturbative methods in the next section.

\begin{figure}
\begin{center}
\includegraphics[scale=0.5]{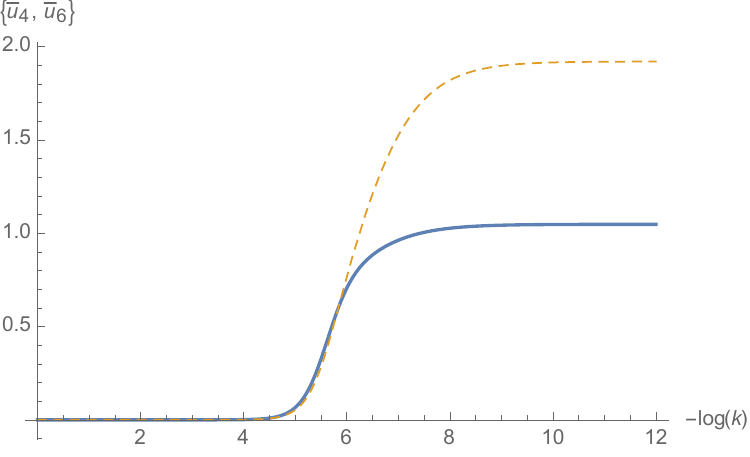}\qquad \includegraphics[scale=0.5]{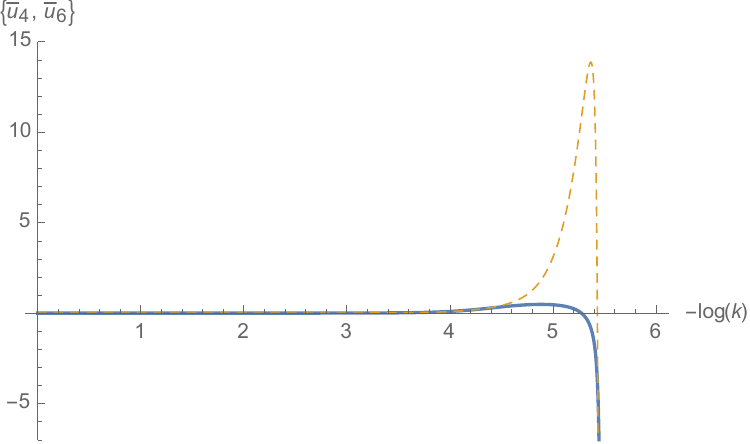}
\end{center}
\caption{Behvaior of the RG flow near the Gaussian fixed point for vanishing disorder (on left) and for a small disorder (on right). Blue curve is for $\bar{u}_4$ and yellow curve for $\bar{u}_6$.}\label{figpert}
\end{figure}

To conclude this preliminary discussion, let us comment on the results shown in Figure \ref{figpert}, which illustrates the flow behavior for initial conditions close to the Gaussian fixed point. We set $\bar{u}_4(0) = 10^{-10}$, $\bar{u}_6(0) = 10^{-7}$, and for the figure on the right, $\bar{\tilde{u}}_6(0) = 10^{-7}$. As previously observed in classical spin glass dynamics \cite{lahoche2022functional,lahoche2023functional} and in the case of $2+p$ quantum spin glass \cite{lahoche2024largetimeeffectivekinetics}, the presence of disorder seems to generate finite-scale divergences, which is also noted in the literature for similar problems. Clearly, discussing these divergences in a perturbative framework is challenging, and we will close this section with that observation.

\section{Derivative expansion in the symmetric phase}\label{sec4}

In this section, we construct nonperturbative approximate solutions for the exact flow equation \eqref{Wett} within the symmetric phase. Recall that the symmetric phase refers to the region of phase space where the expansion around the vacuum (with zero field) is stable. The vertex expansion is the simplest approximation suitable for constructing nonperturbative solutions in this phase. It involves expanding the effective average action $\Gamma_k$ as a power series in the field, consistent with the symmetries of the theory. For our purposes, each monomial is $O(N)$-invariant, but it can be local, bilocal, trilocal, and so on. We will focus on the bilocal expansion, which for $p=3$ is the only contribution in the large $N$ limit in the symmetric phase, as the propagator is local, i.e., proportional to $\delta(\omega + \omega')$ in Fourier space.

The vertex expansion generally assumes a derivative expansion, where at leading order, the kinetic part of the classical action is given by \cite{Delamotte_2012,canet2003nonperturbative}:
\begin{equation}
\Gamma_{k,\text{kin}}=\sum_{\alpha} \int d\omega M_\alpha(-\omega) (Z(k)\omega^2+u_2(k)) M_\alpha(\omega)+\mathcal{O}(\omega^3)\,,
\end{equation}
where the fields share the same replica index. It is well known that for such an $O(N)$ theory, the field strength renormalization does not flow for large $N$ in the symmetric phase, as we will explicitly show below—see also \cite{lahoche2022functional,lahoche2024largetimeeffectivekinetics}. Notice that $Z(k)$ acts like the classical mass $m$ (see equation \eqref{classicalequation}).

The vertex expansion has a significant limitation in this context: power counting makes interactions of arbitrarily high valence relevant, leading to a breakdown in the approximation. Therefore, we expect the ordinary vertex expansion to be inadequate, as it overlooks relevant effects. To address this issue, we consider an improved version of the vertex expansion known as the Effective Vertex Expansion (EVE) \cite{lahoche2023functional,lahoche2018nonperturbative}. This approach leverages exact relations between observables that arise in the symmetric phase from perturbation theory to close the hierarchical flow equations. We will explore this method in the final subsection of this part.

\subsection{The $p=2$ model}\label{vertexexpp2}

\begin{figure}
\begin{center}
\includegraphics[scale=0.8]{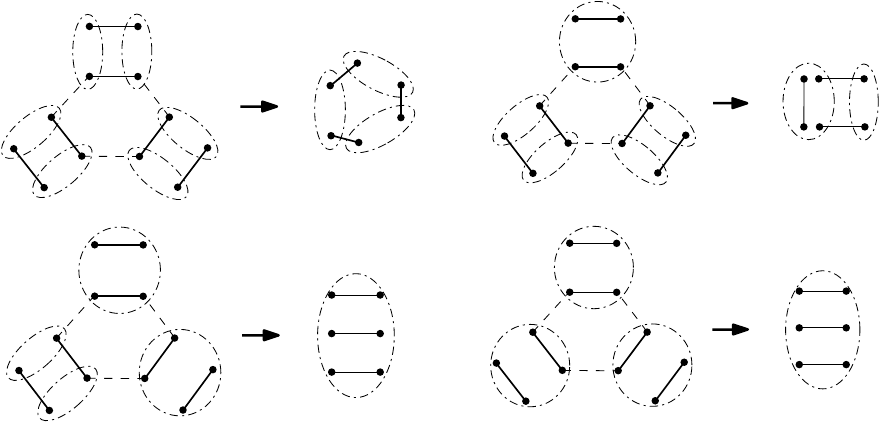}
\end{center}
\caption{Relevant contributions for sextic interactions at the one loop order. Bubbles at the end of the arrows are the corresponding effective interactions.}\label{figlargeN}
\end{figure}

Let us start by the $p=2$ model, meaning that the initial classical action involves only non-local quartic interactions of type \eqref{equationquartic}. Analytic insight can be found in our previous work \cite{lahoche2024largetimeeffectivekinetics} Usually vertex expansion assumes to project systematically the flow on the sub-space of the full phase space spanned by a finite set of local interactions. In our case, the expansion involves local but also bi-local, tri-local and so on \cite{lahoche2022functional,Tarjus,Tarjus2}, what we call a bubble expansion. Explicitly:
\begin{align}
\nonumber\Gamma_{k,\text{int}}&=\underbrace{\left(\vcenter{\hbox{\includegraphics[scale=0.55]{V4.pdf}}}+\vcenter{\hbox{\includegraphics[scale=0.55]{V6.pdf}}}+\vcenter{\hbox{\includegraphics[scale=0.55]{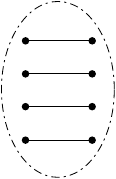}}}+\cdots\right)}_{\text{Local}}\\\nonumber
&+\underbrace{\left(\vcenter{\hbox{\includegraphics[scale=0.55]{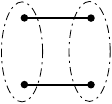}}}+\vcenter{\hbox{\includegraphics[scale=0.55]{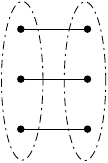}}}+\vcenter{\hbox{\includegraphics[scale=0.55]{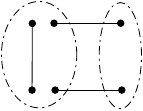}}}+\cdots\right)}_{\text{Bi-local}}+\underbrace{\left(\vcenter{\hbox{\includegraphics[scale=0.55]{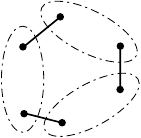}}}+\cdots\right)}_{\text{Tri-local}}\,+\,\cdots \,,
\end{align}
where the subscript ‘‘int" means that this expansion concern monomial of degree higher than 2. The large $N$ limit reduces the type of bubbles that we have to consider. Indeed, because of the symmetric phase assumption, the vertex function we have to consider are those which can be constructed perturbatively, and the large $N$ limit reduces drastically the kind of bubble that we have to consider. We will return precisely on this point in the subsection \ref{EVE}, but Figure \ref{figlargeN} illustrates this point at the one-loop order for sextic interactions. In particular, we show that the second bi-local interaction in the second line of the previous expansion is not generated by the leading order perturbation theory.
\medskip

We will consider a sextic truncation, meaning that we stop the expansion for vertices having a valence smaller than $6$ with the sharp constraint:
\begin{equation}
\Gamma^{(2n)}_k=0\,, \qquad n>3\,.
\end{equation}

The minimal truncation then writes as:
\begin{equation}
\boxed{\Gamma_{k,\text{int}}=\underbrace{\vcenter{\hbox{\includegraphics[scale=0.55]{V4.pdf}}}}_{u_4}\,+\,\underbrace{\vcenter{\hbox{\includegraphics[scale=0.55]{V6.pdf}}}}_{u_6}\,+\,\underbrace{\vcenter{\hbox{\includegraphics[scale=0.55]{V22.pdf}}}}_{\tilde{u}_{4}}\,+\,\underbrace{\vcenter{\hbox{\includegraphics[scale=0.55]{V42.pdf}}}}_{\tilde{u}_{6,1}}\,+\,\underbrace{\vcenter{\hbox{\includegraphics[scale=0.55]{V222.pdf}}}}_{\tilde{u}_{6,2}} \,,}
\end{equation}
where we indicated the coupling constant below each interaction, accordingly with the definition of them provided in the subsection \ref{sec5}. In particular bubbles involving $n$ solid lines have a factor $N^{-n+1}$. The relevant contributions for the couplings flow equations are, graphically\footnote{To simplify the notations, we omitted the crossed circle materializing the regulator. The reader has to keep in mind that, as soon as such a diagram is involved in a nonperturbative equation one of the dashed edges has to involves a regulator insertion.}

\begin{align}
\dot{u}_2\,&=\,-\,\vcenter{\hbox{\includegraphics[scale=0.6]{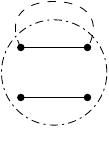}}}\,-\,\vcenter{\hbox{\includegraphics[scale=0.6]{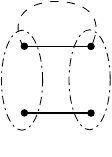}}}\,,\\
\dot{u}_4\,&=\, -\, \vcenter{\hbox{\includegraphics[scale=0.6]{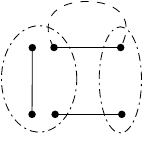}}}\,-\,\vcenter{\hbox{\includegraphics[scale=0.6]{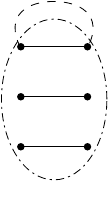}}}\,+\,\vcenter{\hbox{\includegraphics[scale=0.6]{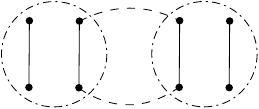}}}\,+\, \vcenter{\hbox{\includegraphics[scale=0.6]{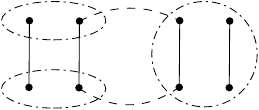}}}\,,\\
\dot{\tilde{u}}_{4}\,&=\, -\, \vcenter{\hbox{\includegraphics[scale=0.6]{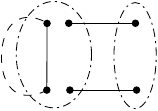}}}\,-\,\vcenter{\hbox{\includegraphics[scale=0.6]{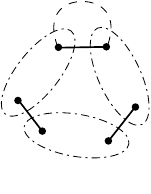}}}\,+\,\vcenter{\hbox{\includegraphics[scale=0.6]{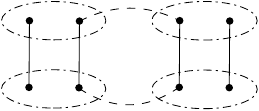}}}\,,
\end{align}
and:

\begin{align}
\dot{u}_6\,&=\,\vcenter{\hbox{\includegraphics[scale=0.6]{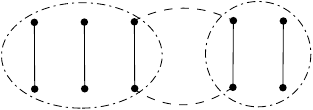}}}\,+\,\vcenter{\hbox{\includegraphics[scale=0.6]{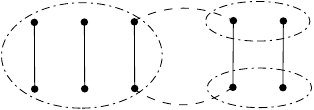}}}\,+\,\vcenter{\hbox{\includegraphics[scale=0.6]{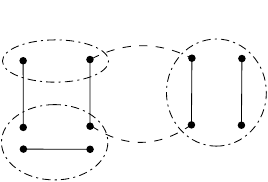}}}\\
&\,+\,\vcenter{\hbox{\includegraphics[scale=0.6]{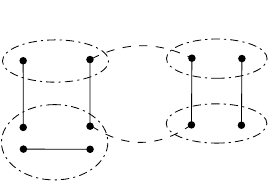}}}\,-\,\vcenter{\hbox{\includegraphics[scale=0.7]{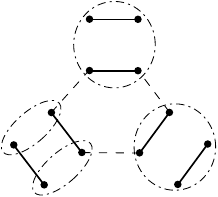}}}\,-\, \vcenter{\hbox{\includegraphics[scale=0.7]{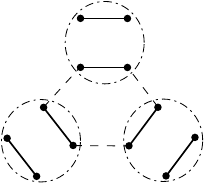}}}\,,\\
\dot{\tilde{u}}_{6,1}\,&=\, \vcenter{\hbox{\includegraphics[scale=0.6]{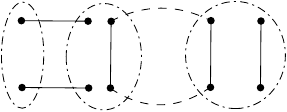}}}\,+\, \vcenter{\hbox{\includegraphics[scale=0.6]{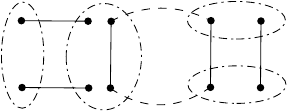}}}\,+\, \vcenter{\hbox{\includegraphics[scale=0.6]{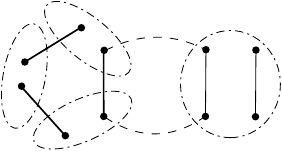}}}\\
&\,-\,\vcenter{\hbox{\includegraphics[scale=0.7]{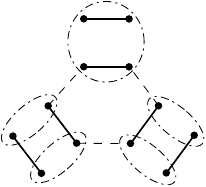}}} \,,\\
\dot{\tilde{u}}_{6,2}\,&=\, \vcenter{\hbox{\includegraphics[scale=0.6]{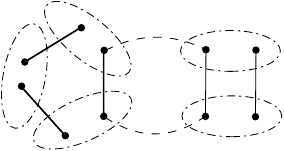}}}\,-\,\vcenter{\hbox{\includegraphics[scale=0.6]{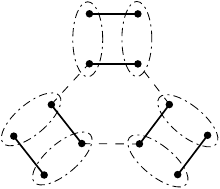}}}\,.
\end{align}

Computing each diagram, we get, using dimensionless couplings:
\begin{align}
\dot{\bar{u}}_2&=-2 \bar{u}_2-\frac{2 \bar{u}_4}{\pi} \frac{1}{(1+\bar{u}_1)^2} - \frac{\bar{\tilde{u}}_4}{\pi} \frac{1}{(1+\bar{u}_1)^2}\,,\label{betau2p2}\\
\dot{\bar{u}}_4&=-3 \bar{u}_4-\frac{4}{3\pi} \frac{\bar{\tilde{u}}_{6,1}}{(1+\bar{u}_2)^2}-\frac{4}{\pi} \frac{\bar{{u}}_{6}}{(1+\bar{u}_2)^2}+ \frac{16}{\pi}\frac{\bar{u}_4^2}{(1+\bar{u}_2)^3}+ \frac{16}{\pi}  \frac{\bar{u}_4\bar{\tilde{u}}_4}{(1+\bar{u}_2)^3}\label{betau4p2}\\
\dot{\bar{\tilde{u}}}_4&=-4 \bar{\tilde{u}}_4-\frac{4}{3\pi} \frac{\bar{\tilde{u}}_{6,1}}{(1+\bar{u}_2)^2}-\frac{2}{\pi} \frac{\bar{\tilde{u}}_{6,2}}{(1+\bar{u}_2)^2}+\frac{8}{\pi}\frac{\bar{\tilde{u}}_4^2}{(1+\bar{u}_2)^3}\,,\label{betau4tp2}\\
\nonumber \dot{\bar{u}}_6&=-4 \bar{u}_6 +\frac{6}{\pi} \frac{\bar{u}_6\bar{\tilde{u}}_{4} }{(1+\bar{u}_2)^3}+\frac{12}{\pi} \frac{\bar{{u}}_{6} \bar{u}_4}{(1+\bar{u}_2)^3}+\frac{4}{\pi} \frac{\bar{\tilde{u}}_{6,1} \bar{u}_4}{(1+\bar{u}_2)^3}+\frac{4}{\pi} \frac{\bar{\tilde{u}}_{6,1} \bar{\tilde{u}}_4}{(1+\bar{u}_2)^3}\\
& \quad - \frac{48}{\pi} \frac{\bar{u}_4^3}{(1+\bar{u}_2)^4}- \frac{72}{\pi} \frac{\bar{u}_4^2\bar{\tilde{u}}_4}{(1+\bar{u}_2)^4}\,,\\
\dot{\bar{\tilde{u}}}_{6,1}&=-5 \bar{\tilde{u}}_{6,1}+\frac{4}{\pi} \frac{\bar{\tilde{u}}_{6,1} \bar{u}_4}{(1+\bar{u}_2)^3}+\frac{2}{\pi} \frac{\bar{\tilde{u}}_{6,1} \bar{\tilde{u}}_4}{(1+\bar{u}_2)^3}+\frac{6}{\pi} \frac{\bar{\tilde{u}}_{6,2} \bar{u}_4}{(1+\bar{u}_2)^3}- \frac{72}{\pi} \frac{\bar{u}_4\bar{\tilde{u}}_4^2}{(1+\bar{u}_2)^4}\,,\\
\dot{\bar{\tilde{u}}}_{6,2}&=-6 \bar{\tilde{u}}_{6,2}+\frac{6}{\pi} \frac{\bar{\tilde{u}}_{6,2} \bar{\tilde{u}}_4}{(1+\bar{u}_2)^3}- \frac{24}{\pi} \frac{\bar{\tilde{u}}_4^3}{(1+\bar{u}_2)^4}\,.
\end{align}

Let us start  by investigating the fixed point solutions of these equations. We get three formal fixed points, 
\begin{align}
\text{NPFP1}&=(\bar{u}_2= -0.20, \bar{u}_4=0, \bar{\tilde{u}}_4\approx 0.80)\,,\\
\text{NPFP2}&=(\bar{u}_2= -0.16, \bar{u}_4=0.35, \bar{\tilde{u}}_4=0)\,,\\
\text{NPFP3}&=(\bar{u}_2=0.07, \bar{u}_4 \approx-1.19, \bar{\tilde{u}}_4\approx 1.91)\,.\,.
\end{align}
with critical exponents:
\begin{align}
\Theta_1&=(-5, -5, 2.)\,\\
\Theta_2&=(4, -3.68, 1.93)\,\\
\Theta_3&=(5.33, -3.89, 1.81)\,.
\end{align}
Including sextic interactions into the truncation, we get once again three fixed points;
\begin{align}
\text{NPFP1}^\prime&=(\bar{u}_2\approx -0.14, \bar{u}_4 =0, \bar{\tilde{u}}_4 \approx 0.66, \bar{u}_6=0, \bar{\tilde{u}}_{6,1}=0, \bar{\tilde{u}}_{6,2}\approx -1.01)\,,\\
\text{NPFP2}^\prime&=(\bar{u}_2\approx -0.11, \bar{u}_4 \approx 0.27, \bar{\tilde{u}}_4 =0, \bar{u}_6\approx -0.18, \bar{\tilde{u}}_{6,1}=0, \bar{\tilde{u}}_{6,2} = 0)\,,\\\nonumber
\text{NPFP3}^\prime&=(\bar{u}_2\approx 3.77, \bar{u}_4 \approx -400.5, \bar{\tilde{u}}_4 \approx 261,\\
& \qquad \bar{u}_6\approx -34052.43, \bar{\tilde{u}}_{6,1}\approx 307093.64, \bar{\tilde{u}}_{6,2} \approx -184993.0)\,,
\end{align}
and critical exponents are:
\begin{align}
\Theta_1^\prime&=(6.39,-4.58,4.30 -1.52 i,4.30 +1.52i,-4.39,2)\,\\
\Theta_2^\prime&=(6, 4.52, 4.42, 4., -3.12, 1.95)\,\\
\Theta_3^\prime&=(37.8,4.90 -4.34 i,4.90 +4.34 i,5.36,-4.82,0.18)\,.
\end{align}
The first and second fixed points, $\text{NPFP1}^\prime$ and $\text{NPFP2}^\prime$, are reminiscent of the fixed points $\text{NPFP1}$ and $\text{NPFP2}$ obtained from the quartic truncation. These, in turn, recall the fixed points $\text{FP1}$ and $\text{FP2}$ (see \eqref{FP1} and \eqref{FP2}) derived using perturbation theory, especially $\text{FP2}$. However, the reliability of the third fixed point is rather questionable, as its characteristics vary significantly across different truncations, unlike the first two fixed points, which remain consistent even up to orders 8 and 10. At this point, we conclude that there are two interacting fixed points:

\begin{claim} We claim that there exists a one-dimensional stable manifold, spanned by the single relevant direction of $\text{NPFP2}^\prime$, and a two-dimensional stable manifold, spanned by the two relevant directions of $\text{NPFP1}^\prime$. These fixed points appear to be multi-critical, with the number of relevant directions increasing with the order of truncation. 
\end{claim}

\subsection{The $p=3$ model}\label{subsectionp3}
Now, let us turn to the case $p=3$. In this scenario, the non-local interaction is of sixth order in the microscopic action. In the symmetric phase, and based on similar reasoning as before, non-local quartic interactions cannot be generated in the large $N$ limit. Therefore, the vertex expansion for the interaction part of the effective average action, graphically, is:
\begin{equation}
\Gamma_{k,\text{int}}=\underbrace{\left(\vcenter{\hbox{\includegraphics[scale=0.55]{V4.pdf}}}+\vcenter{\hbox{\includegraphics[scale=0.55]{V6.pdf}}}+\vcenter{\hbox{\includegraphics[scale=0.55]{V8.pdf}}}+\cdots\right)}_{\text{Local}}+\underbrace{\left(\vcenter{\hbox{\includegraphics[scale=0.55]{V33.pdf}}}+\vcenter{\hbox{\includegraphics[scale=0.55]{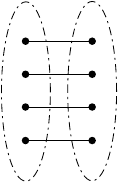}}}+\cdots\right)}_{\text{Bi-local}}\,.
\end{equation}
Note that, in contrast to the case of $p=2$, the multi-local expansion terminates at the bi-local sector in the large $N$ limit, with higher-order multi-local contributions being subleading. The minimal truncation involves only sextic non-local interactions. By restricting ourselves to a sextic truncation for the local part of the effective action, we obtain exactly the flow equations previously encountered in equations \eqref{sextictrunc1} and \eqref{sextictrunc2}, which lead to:
\begin{align}
\dot{\bar{u}}_2&=-2 \bar{u}_2-\frac{2 \bar{u}_4}{\pi} \frac{1}{(1+\bar{u}_1)^2}\,,\\
 \dot{\bar{u}}_4&=-3 {\bar{u}}_4 - 4\frac{ \bar{u}_6}{ \pi} \frac{1}{(1+\bar{u}_1)^2}\, -2\frac{\bar{\tilde{u}}_6}{\pi} \frac{1}{(1+\bar{u}_1)^2} + \frac{16\bar{u}_4^2}{\pi} \frac{1}{(1+\bar{u}_1)^3}\,,\\
 \dot{\bar{u}}_6&=-4 {\bar{u}}_6 + \frac{12}{(1+\bar{u}_1)^3} \frac{\bar{u}_4\bar{u}_6}{\pi} + \frac{6}{(1+\bar{u}_1)^3} \frac{\bar{u}_4\bar{\tilde{u}}_6}{\pi} - \frac{48}{(1+\bar{u}_1)^4} \frac{\bar{u}_4^3}{\pi} \\
 \dot{\bar{\tilde{u}}}_6&=-5 {\bar{\tilde{u}}}_6 \,.
\end{align} 

\begin{remark}\label{remarkconv}
Note that convergence of the integral is not guaranteed in the sextic sector. However, it is important to keep in mind that this approximation neglects relevant effects that could ultimately ensure convergence. Introducing octic couplings, for instance—assuming they are positive—should, in principle, guarantee convergence. However, this approximation again neglects higher-order contributions, which are more relevant than the octic terms, leading to an unsolvable infinite hierarchy of interactions.
Moreover, it should be noted that in the real-time formalism—where we apply a Wick rotation (see \cite{lahoche2024largetimeeffectivekinetics})—increasing the order of the local potential does not improve the convergence of the path integral due to the global factor ii in front of the classical action. This makes the definition of the path integral highly formal, even though the flow equations remain the same as those written here. Ultimately, only the Effective Vertex Expansion (EVE), which we will discuss in the next section, addresses this problem more satisfactorily.
\end{remark}

\begin{figure}
\begin{center}
\includegraphics[scale=0.5]{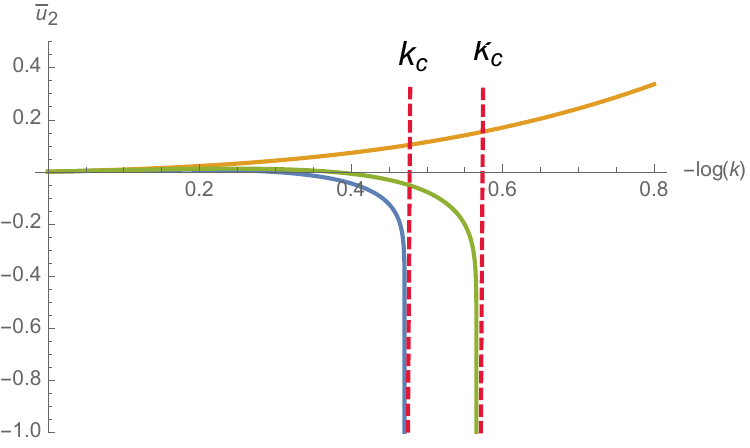}\qquad \includegraphics[scale=0.5]{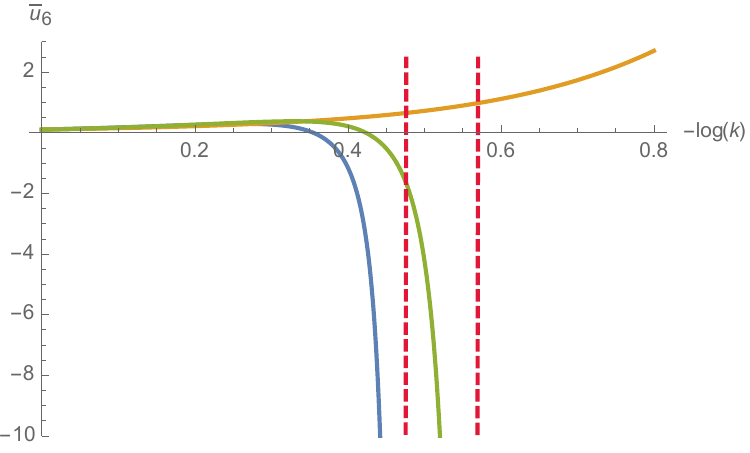}
\end{center}
\caption{On the left: Behavior of the mass coupling $\bar{u}_2$ for $\bar{\tilde{u}}_6(0)=1$ (blue curve, $t_c:=\ln k_c\approx 0.47$), $\bar{\tilde{u}}_6(0)=0.8$ (green curve, $t_c:=\ln k_c\approx 0.57$) and $\bar{\tilde{u}}_6(0)=0.001$ (yellow curve). On the right, behavior of $\dot{\bar{u}}_6$ for $\bar{\tilde{u}}_6(0)=1$. }\label{figdiv2}
\end{figure}

As in the perturbation theory in the previous section, we investigate the fixed trajectories, such that $\dot{\bar{u}}_2=\dot{\bar{u}}_4= \dot{\bar{u}}_6=0$. There is only one fixed point which survives in the deep IR i.e. for arbitrary large values of ${\bar{\tilde{u}}}_6$, but unfortunately this solution lies below the frontier $\bar{u}_2=-1$, arising because of the truncation in the symmetric phase. Hence, no reliable solutions are found in the deep IR, and we return to this question using EVE in the next section. With this non-perturbative formalism we can also return to this phenomenon of finite-scale divergence that we have observed perturbatively. We show in figure \ref{figdiv2} the behavior of typical trajectories, for initial conditions: $\bar{u}_4(0)=\bar{u}_6(0)=0.1$ and $\bar{u}_2(0)=0$. 
\medskip

The results confirm what we observed in perturbation theory and are consistent with our conclusions from previous work \cite{lahoche2024largetimeeffectivekinetics}. The presence of sufficiently large disorder leads to a finite-scale singularity, preventing the flow from reaching long time scales. In other words, beyond a certain characteristic time scale, the assumptions used to construct the parameterization of the phase space (through the choice of the ansatz for $\Gamma_k$) break down. These singularities may indicate an instability in the theory space, associated with couplings that lie beyond the reach of perturbation theory.

This phenomenon is related to critical phenomena, particularly ferromagnetism, where flow singularities (such as at $\bar{u}_2=-1$ with the Litim regulator) can signal the emergence of symmetry breaking. In this context, it is linked to non-zero magnetization, which perturbation theory (based on the Gaussian measure) cannot predict. In our previous work, we found indirect evidence suggesting that these operators, which are "overlooked" by perturbation theory, might correspond to the emergence of strong correlations between replicas. We will explore this further in the following subsection.
\medskip

Numerically, we find for these initial conditions that the critical value is $\bar{\tilde{u}}_{6,c}(0)\approx 0.269$, below which the singularity at $k_c$ disappears. Note that the exact dependency on the initial couplings is difficult to determine. It is not our goal in this paper to reconstruct the phase space; instead, we focus on constructing non-perturbative solutions, with plans to address this problem in the future. This step will indeed be crucial, as it will allow for direct comparison with other analytical results to evaluate the robustness of our approximations.

\medskip

Finally, let us conclude this subsection with two important remarks. The first concerns the nature of the singularities. In our previous work \cite{lahoche2024largetimeeffectivekinetics}, we associated these singularities with the appearance of finite correlations between the replicas, which are of order $1/N$ according to perturbation theory. We maintain this interpretation here, and we will establish it more precisely in subsection \ref{2PI}. In the context of this article, the presence of singularities also has implications for time-translation symmetry. Invariance under time translation imposes Ward identities, which cannot be integrated along the flow (and therefore the symmetry cannot be preserved) in the presence of singularities \cite{lahoche2022functional}.

\medskip

\begin{figure}
\begin{center}
\includegraphics[scale=0.8]{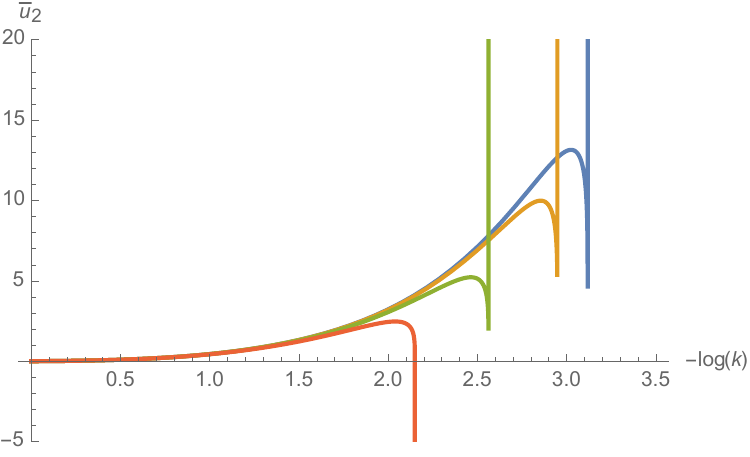}
\end{center}
\caption{Behavior of the mass coupling $\bar{u}_2$ for $\bar{\tilde{u}}_6(0)=0.269$ (blue curve), $\bar{\tilde{u}}_6(0)=0.27$ (yellow curve), $\bar{\tilde{u}}_6(0)=0.275$ (green curve) and $\bar{\tilde{u}}_6(0)=0.29$ (red curve).}\label{figdiv3}
\end{figure}

The second remark concerns the origin of these singularities. In Figure \ref{figdiv3}, we see that the singularity near the critical value $\bar{\tilde{u}}_{6,c}(0)$ does not occur because the flow reaches the singularity at $\bar{u}_2=-1$, which is associated with the ‘‘ferromagnetic’’ transition. Instead, it arises in the positive regime. However, for higher values of $\bar{\tilde{u}}_6$, the flow appears to be directed towards this singularity (see Figure \ref{figdiv2}). This phenomenon reveals something about the global phase space and suggests the existence of regions where ‘‘glassy’’ and ‘‘ferromagnetic’’ phases are present. Moreover, near the critical value, the singularity at $k_c$ is preceded by a sharp singularity of the ‘‘cusp’’ type, as we can see for the blue, yellow, and green curves in Figure \ref{figdiv3}.

\subsection{Leading order effective vertex expansion}\label{EVE}
Working in the symmetric phase, the EVE method exploits the exact relations between observables (Schwinger-Dyson equations) derived from the large $N$ assumption. This method was initially applied in the context of tensorial field theory—see \cite{lahoche2018nonperturbative,lahoche2020pedagogical} and references therein—where the exact relations between sextic and quartic local observables in the symmetric phase allow for the closure of the flow equation hierarchy. This method is particularly relevant in our case, as high-valence interactions are arbitrarily significant, as highlighted in our previous work \cite{lahoche2023functional} on out-of-equilibrium classical spin glasses. We will begin by discussing the quartic sector ($p=2$) and then extend to the sextic case ($p=3$) using the methods described in \cite{lahoche2019ward}.

\medskip

\begin{figure}
\begin{center}
\includegraphics[scale=0.8]{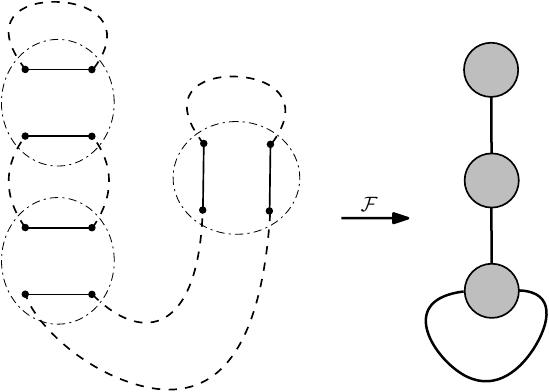}
\end{center}
\caption{Illustration of the LVR construction. Gray discs materialize loops (loop vertices) and fat edges materialize vertices in the original representation. }\label{FigLVR}
\end{figure}

\paragraph{Quartic theory.} Let us start with the case $p=2$, assuming that the confinement potential is quartic. Due to the scaling of the quartic interaction, the Feynman amplitude $\mathcal{A}_{\mathcal{G}}$ for the Feynman graph $\mathcal{G}$ scales as $\mathcal{A}_{\mathcal{G}} \sim N^{1-(V-F+1)}=
N^{1-\omega}$, and the leading order graphs are those that maximize the number of faces $F$, while keeping the number of vertices $V$ and the number of external edges fixed. The structure of the leading order graphs becomes clearer using the so-called \textit{loop vertex representation} (LVR) \cite{rivasseau2018loop,rivasseau2010feynman,lahoche2022functional}, which establishes a correspondence between vertices and edges on one hand, and loops and vertices on the other. The precise construction is illustrated in Figure \ref{FigLVR} for a quartic theory involving only local $O(N)$ vertices. Explicitly, for a purely local $O(N)$ model, the loop vertex representation maps as $\mathcal{F}:\mathcal{G}\to \mathcal{G}^\prime$, such that $\mathcal{F}(V)=\mathcal{E}$ (the number of edges in $\mathcal{G}^\prime$) and $\mathcal{F}(F)=\mathcal{V}$ (the number of vertices in $\mathcal{G}^\prime$). Then, for a vacuum graph:

\begin{equation} \omega=\mathcal{E}-\mathcal{V}+1 \geq 0,,\label{powercount} \end{equation}

The last inequality follows from the well-known fact that the left-hand side quantity is equivalent to the number of spanning trees in graph theory. To distinguish between the vertices in the LVR representation and the vertices in the original representation, we typically refer to the former as \textit{loop vertices}. Including the non-local vertices that arise due to averaging over disorder does not change the power counting in \label{powercount}, but $V$ now becomes a sum $V=V_{\text{L}}+V_{\text{NL}}$, where $V_{\text{L}}$ and $V_{\text{NL}}$ represent the local and non-local vertices, respectively.
\medskip

\begin{figure}
\begin{center}
\includegraphics[scale=0.8]{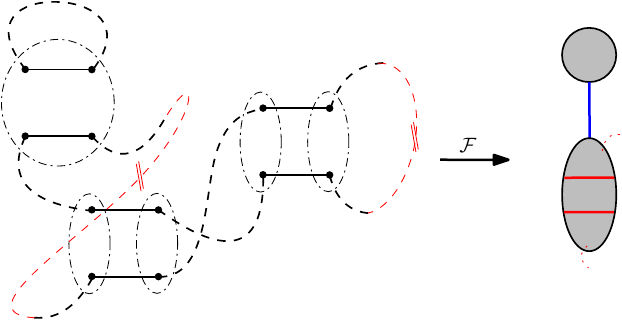}
\end{center}
\caption{Illustration of the map $\mathcal{F}$ for a non-vacuum Feynman graph with local and non-local vertices. Red dotted edges are the positions we perform the cut.} \label{FigLVR2}
\end{figure}

The mapping $\mathcal{F}$ assigns vertices to two different types of edges, depending on whether the vertex is local or non-local. Let $v$ be a vertex in $\mathcal{G}$; if $v$ is local, we represent $\mathcal{F}(v)$ as a blue edge, and as a red edge if $v$ is non-local. In Figure \ref{FigLVR}, we considered a vacuum graph, and since any diagram can be derived from a vacuum diagram by "cutting" a number of lines, we represent non-empty diagrams by a vacuum diagram where some lines are marked with a "cilium" to indicate they are open. Figure \ref{FigLVR2} provides an example. Due to the power counting in \eqref{powercount}, it follows that the leading order graphs are trees in the LVR \cite{lahoche2022functional,benedetti20182pi,lahoche2024largetimeeffectivekinetics}, which is a well-known property of $O(N)$ vector models \cite{Zinn-Justin:1989rgp}.
\medskip

\begin{figure}
\begin{center}
\includegraphics[scale=0.8]{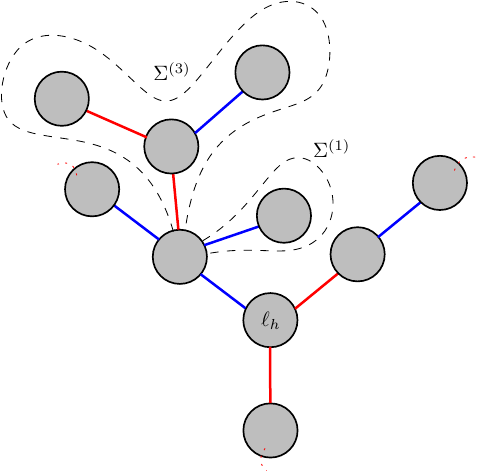}
\end{center}
\caption{An tree with $3$ opened leaves corresponding to a $1PI$ diagram with $6$ external edges. $\Sigma^{(n)}$ denote order $n$ contributions to the self energy $\Sigma$.} \label{FigLVR3}
\end{figure}

\begin{figure}
\begin{center}
$\underset{\underset{\includegraphics[scale=0.7]{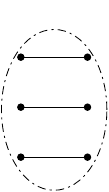}}{\downarrow}}{\includegraphics[scale=0.8]{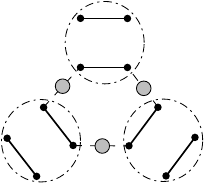}}$\quad $\underset{\underset{\includegraphics[scale=0.7]{V6V2.pdf}}{\downarrow}}{\includegraphics[scale=0.8]{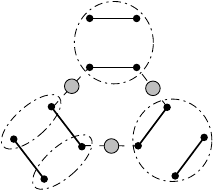}}$\quad $\underset{\underset{\includegraphics[scale=0.8]{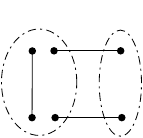}}{\downarrow}}{\includegraphics[scale=0.8]{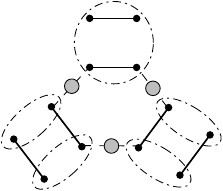}}$\quad $\underset{\underset{\includegraphics[scale=0.7]{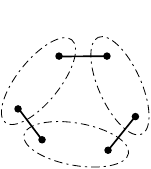}}{\downarrow}}{\includegraphics[scale=0.8]{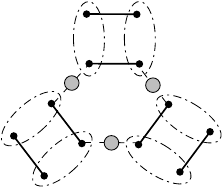}}$
\end{center}
\caption{The four relevant configurations for $6$-points functions. We indicated the corresponding effective vertices under each diagram, and we denoted the effective propagators $G_k$ with a gray disc, to distinguish them from the bare propagator (dashed lines) in this context.}\label{figconf6pts}
\end{figure}

Any non-vacuum $1PI$ Feynman diagram can be derived from leading-order vacuum graphs by opening certain internal edges (in the original representation). Additionally, these open lines must be placed on one of the leaves of the LVR tree to ensure that the resulting graph remains $1PI$. Figure \ref{FigLVR3} shows an example of a sextic $1PI$ diagram. The diagram includes specific contributions, which we have highlighted with dotted black circles. These contributions represent the perturbative expansions of the self-energy $\Sigma$ involved in the Dyson series:
\begin{equation}
G_k=C_k+C_k\Sigma C_k+ C_k\Sigma C_k \Sigma C_k+\cdots = \frac{1}{C^{-1}_k+\Sigma}\,.
\end{equation}
Here, $C_k$ is the bare propagator (including the regulator), $G_k$ is the effective propagator, and $\Sigma$ is the self-energy. These contributions can be re-summed by replacing the bare propagators with effective propagators in the loop vertices, which we refer to as effective loop vertices. The perturbative expansion can then be reorganized as a sum over the number of effective loop vertices, with the corresponding diagrams all sharing the same structure: three 'arms' consisting of a succession of effective loop vertices, all connected to the same central effective loop vertex $v_0$. Once these arms are summed, they correspond to the 4-point effective functions. It is immediately apparent that these functions can only be local or bi-local at leading order in $N$ (see the discussion in \cite{lahoche2021no}). Therefore, in practice, there are only four possible configurations for the 6-point functions, as illustrated in Figure \ref{figconf6pts}.
\medskip

Considering that the propagator is local, we find only three types of effective vertices (labeled under each diagram in the figure). These are precisely the same ones we previously considered in the truncations of subsection \ref{vertexexpp2}, and the only ones we can construct at large $N$. We denote them as $\Gamma_{k,\text{L}}^{(6)}$, $\Gamma_{k,\text{NL}}^{(6,1)}$, and $\Gamma_{k,\text{NL}}^{(6,2)}$, following the notation for the couplings used in subsection \ref{vertexexpp2}. Their explicit expressions (for zero external momenta, with field indices omitted to simplify the notation) are:
\begin{align}
\nonumber \Gamma_{k,\text{L}}^{(6)}&=\delta(0) \bigg[ K_1 \, \int_{-\infty}^{+\infty} \frac{d\omega}{2\pi}\,\Tr\, \tilde{G}_{k}^3(\omega,\alpha) \left(\gamma_{k,\text{L}}^{(4)}(\omega)\right)^3 \,,\\
&\quad + \frac{K_1^\prime}{2\pi} \,\Tr\, \tilde{G}_{k}^3(0,\alpha) \left(\gamma_{k,\text{L}}^{(4)}(0)\right)^2\gamma_{k,\text{NL}}^{(4)}(0)\bigg]\,,\\
\Gamma_{k,\text{NL}}^{(6,1)}&=\frac{K_2}{2\pi}\,\delta^2(0) \,\Tr\, \tilde{G}_{k}^3(0,\alpha) \left(\gamma_{k,\text{NL}}^{(4)}(0)\right)^2 \gamma_{k,\text{L}}^{(4)}(0)\,,\\
\Gamma_{k,\text{NL}}^{(6,2)}&= \frac{K_3}{2\pi}\,\delta^3(0)\,\Tr\, \tilde{G}_{k}^3(0,\alpha) \left(\gamma_{k,\text{NL}}^{(4)}(0)\right)^3 \,\,,
\end{align}
were $K_1$, $K^\prime_1$, $K_2$, $K_3$ are numerical constants and $\gamma_{k,\text{L}}^{(4)}(\omega)$, $\Tr$ is the trace over field indices, and $\gamma_{k,\text{NL}}^{(4)}(\omega)$ are even functions defined by (see the notations of subsection \ref{sec5})
\begin{align}
\Gamma_{k,i_1i_2i_3i_4\text{L}}^{(4)}(\omega,\omega^\prime,0,0)&=\delta(\omega+\omega^\prime) \gamma_{k,\text{L}}^{(4)}(\omega) W_{i_1i_2i_3i_4}\\
\nonumber \Gamma_{k,i_1i_2i_3i_4\text{NL}}^{(4)}(\omega,\omega^\prime,0,0)&=\delta(\omega+\omega^\prime)\delta(0)  \gamma_{k,\text{NL}}^{(4)}(\omega)\delta_{i_1i_3}\delta_{i_2i_4}\\
&+\delta(\omega)\delta(\omega^\prime) \gamma_{k,\text{NL}}^{(4)}(0)(\delta_{i_1i_2}\delta_{i_3i_4}+\delta_{i_1i_4}\delta_{i_2i_3})\,.
\end{align}
Because of the definition \eqref{defcouplingu4} for the coupling $u_4$, we must have:
\begin{equation}
\gamma_{k,\text{L}}^{(4)}(0)=\frac{2u_4}{N}\,.
\end{equation}
In the same way, defining the coupling $\tilde{u}_4$ as:
\begin{equation}
\Gamma_{k,i_1i_2i_3i_4\text{NL}}^{(4)}(0,0,0,0)=\frac{2 \tilde{u}_4}{N} \delta^2(0) W_{i_1i_2i_3i_4}\,,
\end{equation}
such that, again:
\begin{equation}
\gamma_{k,\text{NL}}^{(4)}(0)=\frac{2\tilde{u}_4}{N}\,.
\end{equation}
Then, the computation of the two contributions $\Gamma_{k,\text{NL}}^{(6,1)}$ and $\Gamma_{k,\text{NL}}^{(6,2)}$ is easy because $G_k(0,\alpha)$ is given by the renormalization condition:
\begin{equation}
G_k(0,\alpha)=\frac{1}{k^2+u_2}\,.
\end{equation}
The computation of $\Gamma_{k,\text{L}}^{(6)}$ is more challenging. In \cite{lahoche2018nonperturbative}, the authors demonstrated that it can be consistent to approximate these integrals by replacing $G_k$ with its leading-order approximation in the derivative expansion, with the neglected terms being irrelevant. However, in this case, power counting suggests that these effects could be relevant, though still sub-dominant. Additionally, the $\omega$ dependence of the effective vertices is harder to disregard for similar reasons: while the derivative couplings are less relevant, their dimension remains positive. Here, we will adopt the same drastic approximation that we employed in \cite{lahoche2023functional}, specifically:
\begin{enumerate}
\item We replace the effective propagator $G_k$ by its approximation given by the derivative expansion to leading order.
\item We ignore the frequency dependence of the effective vertex, replacing $\gamma_{k,\text{L}}(\omega)$ with $\gamma_{k,\text{L}}(0)$.
\end{enumerate}
This approximation captures the most relevant effects, however it should be complemented by a specific study of corrections to higher orders of the derivative expansion (also taking into account the frequency dependence of the vertices), we reserve this study for further work. Note that quartic theory is perfect for assessing the robustness of these approximations, since the path integral can be computed exactly in the limit $N\to \infty$ -- see \cite{lahoche2024largetimeeffectivekinetics}. Explicitly:
\begin{align}
\Gamma_{k,\text{L}}^{(6)}&=\frac{8u_4^2(k)}{N^2}\frac{\delta(0)}{2\pi} \bigg[ K_1 \,u_4 \int_{-\infty}^{+\infty}\,\frac{d\omega}{(\omega^2+u_2+R_k(\omega^2))^3}  +{K_1^\prime} \, \frac{\tilde{u}_4}{(k^2+u_2)^3}\bigg]\,,\\
\Gamma_{k,\text{NL}}^{(6,1)}&=\frac{8 \tilde{u}_4^2 u_4}{N^2}\frac{K_2}{2\pi}\, \frac{\delta^2(0)}{(k^2+u_2)^3}\,,\\
\Gamma_{k,\text{NL}}^{(6,2)}&=\frac{8 \tilde{u}_4^3}{N^2} \frac{K_3}{2\pi}\, \frac{\delta^3(0)}{(k^2+u_2)^3} \,\,.
\end{align}
Because of the definition of coupling $u_6$ given by equations \eqref{defu6}, and defining couplings $\tilde{u}_{6,1}$ and $\tilde{u}_{6,2}$ as (assuming we set to zero external frequencies and all external indices to be equals, namely $i_1=i_2=\cdots=i_6=1$):
\begin{align}
\Gamma_{k,\text{NL}}^{(6,1)}&=\frac{5!\tilde{u}_{6,1}}{N^2}\delta^2(0) \\
\Gamma_{k,\text{NL}}^{(6,2)}&=\frac{5!\tilde{u}_{6,1}}{N^2}\delta^3(0)\,.
\end{align}
The numerical constants can be computed using perturbation theory, from the one-loop graphs, and we get finally:
\begin{align}
\bar{u}_6&=\frac{12\bar{u}_4^2}{\pi}\bigg[ \bar{u}_4 \left(-\frac{3+5 \bar{u}_2}{\bar{u}_2^2 \left(1+\bar{u}_2\right){}^2}+\frac{8}{\left(\bar{u}_2+1\right){}^3}+\frac{3 \tan ^{-1}\left(\sqrt{\bar{u}_2}\right)}{\bar{u}_2^{5/2}}\right)+\frac{12 \bar{\tilde{u}}_4}{(1+\bar{u}_2)^3}\bigg]\,,\\
\bar{\tilde{u}}_{6,1}&=\frac{144\bar{\tilde{u}}_4^2 \bar{u}_4}{\pi} \frac{1}{(1+\bar{u}_2)^3}\,,\\
\bar{\tilde{u}}_{6,2}&=\frac{48 \bar{\tilde{u}}_4^3}{\pi} \, \frac{1}{(1+\bar{u}_2)^3}\,.
\end{align}



\begin{figure}
\begin{center}
\includegraphics[scale=0.5]{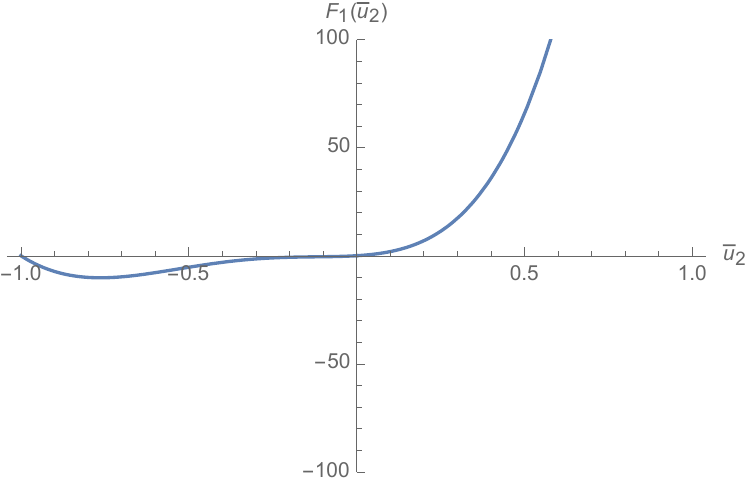}\qquad \includegraphics[scale=0.5]{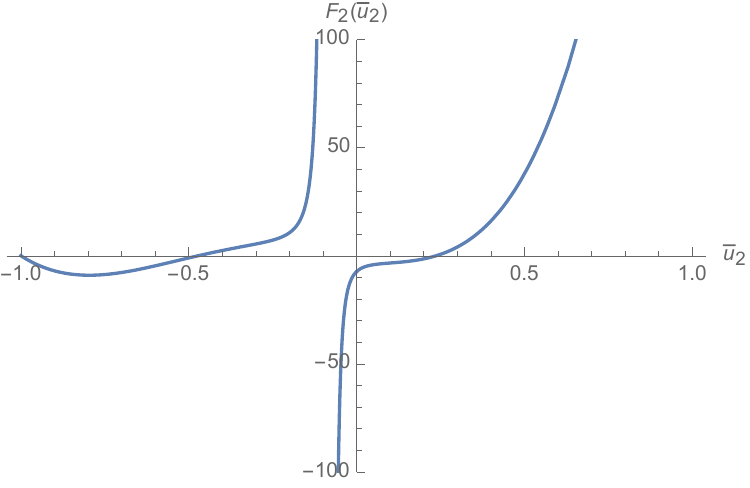}
\end{center}
\caption{Plots of the two functions $F_1$ and $F_2$.}\label{figF1F2} 
\end{figure}

These expressions can thus be reported in the flow equations \eqref{betau2p2}-\eqref{betau4tp2} for $u_2$, $u_4$ and $\tilde{u}$, and then depend only on these three couplings. These three equations can be solved numerically. The solution of the equation $\dot{\bar{u}}_2=0$ gives us $\bar{u}_4(\bar{u}_2,\bar{\tilde{u}}_4)$. Inserting this relation into the third equation \eqref{betau4tp2}, and solving it $\dot{\tilde{u}}_4=0$, we get two solutions:
\begin{equation}
{\tilde{u}}_4=0\,,\qquad {\tilde{u}}_4=\frac{\pi  \left(1+\bar{u}_2\right){}^3}{48 \bar{u}_2+2}\,.
\end{equation}
Plugging these solution into the remaining flow equation for $u_4$, we obtain two functions depending only on $\bar{u}_2$, $F_1(\bar{u}_2)$ and $F_2(\bar{u}_2)$, as the figure \ref{figF1F2} shows. The first solution $F_1$ has only one vanishing point for $\bar{u}_2=0$, leading to $\bar{u}_4=\bar{\tilde{u}}_4=0$, which is nothing but the Gaussian fixed point solution. The second function has two solutions for the equation $F_2(\bar{u}_2)=0$, 
\begin{align}
\text{sol}_1&=(\bar{u}_2 \approx -0.47, \bar{u}_4\approx 0.44, \bar{\tilde{u}}_4\approx -0.05)\,,\\
 \text{sol}_2&=(\bar{u}_2 \approx 0.24, \bar{u}_4\approx -1.53, \bar{\tilde{u}}_4\approx 0.77)\,.
\end{align}
The second solution has the wrong sign for the disorder coupling $\bar{\tilde{u}}_4$ which has to be negative; only the first fixed point is physically relevant. Computing the critical exponents, we get:

\begin{equation}
\Theta_{\text{EVE}}=(\theta_1=20.86,\theta_2=-8.08,\theta_3=1.92)\,.
\end{equation}

The fixed point  has two relevant and one irrelevant direction, spanning a stable subspace of dimension $1$, which we identify with the fixed point $\text{NPFP2}^\prime$ discovered from vertex expansion.

\noindent
\paragraph{Sextic theory.} For $p=3$, the bare action include quartic and sextic local interactions, and sextic bi-local interactions. The power counting is slightly modified, and the amplitude of some vacuum Feynman graph $\mathcal{G}$ scale as $\mathcal{A}_{\mathcal{G}} \sim N^{1-\omega(\mathcal{G})}$, with:
\begin{equation}
\omega(\mathcal{G})=1+V_4(\mathcal{G})+2V_6(\mathcal{G})-F(\mathcal{G})\,,
\end{equation}
where $V_{2n}$ is the number of vertices with valence $2n$. The presence of sextic vertices complicates the construction of graphs, which is always simpler in quartic theory where the leading order graphs are planar trees. However, we can formally reduce the sextic theory to a quartic theory by decomposing the quartic bubbles into the sum of two sextic bubbles. The sum of two local bubbles is defined as follows:

\begin{definition}
Let $C$ the local (bare) propagator, and $(v_1,,v_2)$ two local bubbles. Let $n_1 \in v_1$ and $n_2\in v_2$ two black nodes in $v_1$ and $v_2$ and let $\ell:=(n_1,n_2)$ some dashed edge connecting $n_1$ and $n_2$. The sum $v=v_1\sharp_{n_1n_2} v_2$ of the two bubbles is the local bubble defined as the contraction of the edge $\ell$, such that:
\begin{enumerate}
\item We delete the edge $\ell$ and the nodes $n_1,n_2$.
\item We connect the solid lines together.
\item We merge the dashed-dotted bubbles.
\end{enumerate}
\end{definition}

\begin{figure}
\begin{center}
\includegraphics[scale=1]{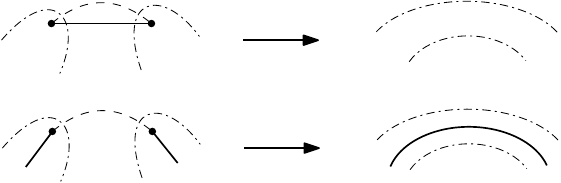}
\end{center}
\caption{The contraction of some edge between two local bubbles}\label{figsum1}
\end{figure}

\begin{figure}
\begin{center}
\includegraphics[scale=1]{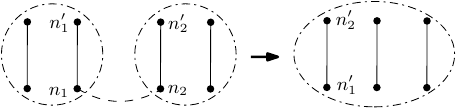}
\end{center}
\caption{The sum of two quartic bubbles}\label{figsum2}
\end{figure}

\begin{figure}
\begin{center}
\includegraphics[scale=1]{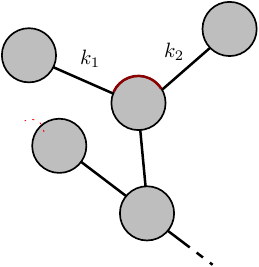}
\end{center}
\caption{Example of corner (the red arc between $k_1$ and $k_2$). It should be noticed that such an arc corresponds to a dashed edge in the original representation. We furthermore indicated the contracted corners with two bars. Note that the edges are black because the graph involves a single kind of interaction.}\label{figcorner}
\end{figure}

Figures \ref{figsum1} and \ref{figsum2} illustrate us the sum of some local bubbles and show in particular that any sextic bubble can be obtained as the sum of two quartic bubbles. The benefit of this observation is that we can construct the leading order diagrams for a local theory involving sextic interactions from the ones of a quartic theory. As we recalled in the previous paragraph, the leading order vacuum diagrams for a quartic theory are unrooted planar trees. Such a Feynman graph has two kinds of loop vertices; the leafs are the simpler ones hooked to a single edge, but there are also effective vertices hooked to more than one edge. We call them \textit{cornered loop vertices}, and we call corner the dotted edge $\widehat{k_1k_2}$ between two edges $k_1$ and $k_2$ (see Figure \ref{figcorner}). Let $\mathcal{G}$ be some leading order Feynman graph for the quartic theory, involving $V(\mathcal{G})$ vertices, and let $\mathcal{G}^\prime$ its LVR. We then obtain a leading order Feynman graph with $V-n$ vertices (for some integer $n < V$) by contracting $n$ corners $\mathcal{C}_n:=(c_1,\cdots,c_n)$, and we have the following statement:

\begin{proposition}
The contraction of a corner on the graph $\mathcal{G}$ does not change the power counting $\omega(\mathcal{G})$. 
\end{proposition}

\noindent
\textit{Proof.} The statement is obvious because the contracted corner cannot be on a leaf. Then, the contraction does not change the number of faces $F$. Furthermore, the number of quartic vertices decreases by $2$, and the number of sextic vertices increases by one. The total variation of the power counting is then $\delta \omega=\delta V_4+2 \delta V_6-\delta F= -2+2-0=0$. 
\begin{flushright}
$\square$
\end{flushright}

We denote by $\mathcal{G}/\mathcal{C}_n$ the graph obtained by contracting the corners in $\mathcal{C}_n$. It is important to note that the corners in the set $\mathcal{C}_n$ cannot be arranged arbitrarily. For instance, two corners $c_i, c_j \in \mathcal{C}_n$ cannot both be attached to the same quartic vertex, as this configuration would generate an octic vertex.

\begin{figure}
\begin{center}
\includegraphics[scale=0.8]{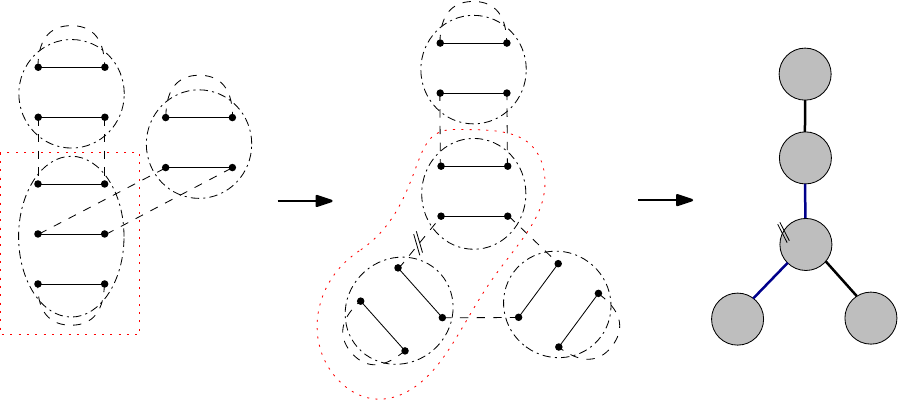}
\end{center}
\caption{Example of the construction of leading order vacuum diagrams for sextic theory.}\label{construction}
\end{figure}

Let $c_i \in \mathcal{C}_n$, and denote by $v_i$ and $w_i$ the two quartic vertices to which $c_i$ is connected. At tree level $\mathcal{G}^\prime$, $c_i$ is the corner between the two lines corresponding to vertices $v_i$ and $w_i$. We will mark the corner $c_i$ with two bars, as shown in Figure \ref{figcorner}, and color the two lines corresponding to $v_i$ and $w_i$ in blue. Figure \ref{construction} summarizes the construction using an explicit example.
Finally, non-vacuum 1PI diagrams can be obtained from vacuum diagrams by removing the dashed lines from the leaves.
\medskip

This construction applies to local bubbles, but the challenge is to generalize it for diagrams involving non-local vertices. The method we propose here differs slightly from the one presented in our previous work \cite{lahoche2024largetimeeffectivekinetics}. It is important to note that the power counting in $N$ does not depend on the number of local components in the bubbles. In other words, we can substitute a non-local vertex for any local vertex in a diagram without changing the power of $N$, as long as we preserve the position of the solid lines (i.e., maintain the number of faces).

\begin{figure}
\begin{center}
\includegraphics[scale=0.8]{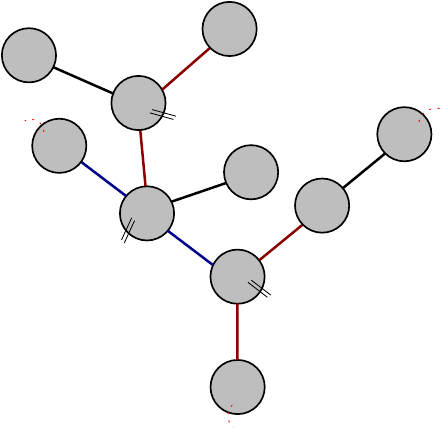}
\end{center}
\caption{A typical tree including red and blue edges, as well as contracted corners and cilia. Such a tree is for a 1PI Feynman diagram with six external points.}\label{figLVRbluered}
\end{figure}

On the corresponding tree, we will indicate the quartic vertices in blue. Once the corner between them is contracted, they will yield the local sextic interaction, which we replace with a non-local vertex—see Figure \ref{figLVRbluered}.
\medskip

\begin{figure}
\begin{center}
$\underset{a}{\includegraphics[scale=0.7]{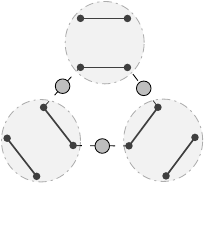}}\qquad \underset{b}{\includegraphics[scale=0.7]{Effphi6sex2.pdf}}\qquad \underset{c}{\includegraphics[scale=0.7]{Effphi6sex22.pdf}}\qquad \underset{d}{\includegraphics[scale=0.7]{Effphi6sex23.pdf}}$\\
$\underset{e}{\includegraphics[scale=0.7]{Effphi6sex2NL.pdf}}\qquad \underset{f}{\includegraphics[scale=0.7]{Effphi6sex22NL.pdf}}\qquad \underset{g}{\includegraphics[scale=0.7]{Effphi6sex23NL.pdf}}$
\end{center}
\caption{The four allowed configurations for the sextic 1PI diagrams. As in the previous section we materialized the effective propagators with a grey circle to avoid confusion; we materialized the effective quartic local vertices in gray for the same reason to distinguish them from the bare vertices on the figure.}\label{figsecticconfs}
\end{figure}

Now, let us move on to the characterization of sextic 1PI diagrams. In the previous section, we characterized sextic diagrams in the context of quartic theory. After formally resumming the propagators along the corners, all these graphs consist of three arms, each carrying a cilium at one end and connected at their other end to a loop vertex ($v_0$ in Figure \ref{FigLVR3}).
For diagrams involving sextic interactions, we need to distinguish between two cases:
\begin{enumerate}
\item If $v_0 \cap \mathcal{C}_n = 0$, no corner is contracted on $v_0$, and the corresponding configuration is pictured on the Figure \ref{figsecticconfs}a. 
\item If  $v_0 \cap \mathcal{C}_n \neq 0$, one of the corners of $v_0$ has to be contracted. The contraction leads to a sextic vertex, and the three allowed configurations are pictured on Figure \ref{figsecticconfs}b,c,d for the local case, and on Figure \ref{figsecticconfs}e,f,g when the central bare vertex is non-local.
\end{enumerate}

Then, graphically, the full $6$-points local and non-local functions $\Gamma_{k,\text{L}}^{(6)}$ and $\Gamma_{k,\text{NL}}^{(6)}$ are (remember that effective propagators remain locals and that contraction of non local components with a local one gives a local bubble):
\begin{align}
\nonumber \Gamma_{k,\text{L}}^{(6)}&=\vcenter{\hbox{\includegraphics[scale=0.6]{V6.pdf}}}+\vcenter{\hbox{\includegraphics[scale=0.7]{Effphi6sex1}}}+\vcenter{\hbox{\includegraphics[scale=0.7]{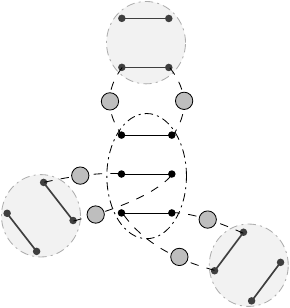}}}+\vcenter{\hbox{\includegraphics[scale=0.7]{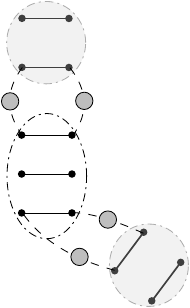}}}+\vcenter{\hbox{\includegraphics[scale=0.7]{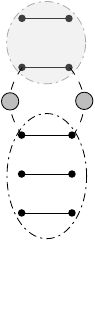}}}\\
&+\vcenter{\hbox{\includegraphics[scale=0.7]{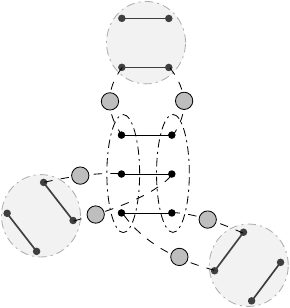}}}+\vcenter{\hbox{\includegraphics[scale=0.7]{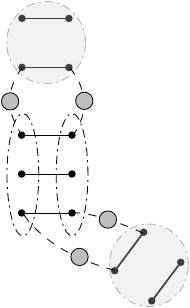}}}+\vcenter{\hbox{\includegraphics[scale=0.7]{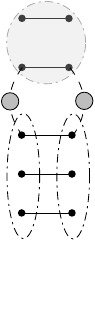}}}\,,
\end{align}
and
\begin{equation}
\Gamma_{k,\text{NL}}^{(6)}=\vcenter{\hbox{\includegraphics[scale=0.7]{V33.pdf}}}\,.
\end{equation}
In particular, we confirm that the non-local vertex does not undergo renormalization. As before, we can compute each diagram using the leading order of the derivative expansion and the local approximation for vertices. This allows us to establish a relationship between the bare local and non-local couplings, $u_6(0)$ and $\tilde{u}_6(0)$, and the effective local coupling $u_6(k)$:
\begin{equation}
\boxed{u_6(0)=k^4\frac{\bar{u}_6(k)-24 L_3(\bar{u}_2) \bar{u}_4^3+\bar{\tilde{u}}_6(0) \left(12 \bar{u}_4^3 \tilde{L}_4(\bar{u}_2)-8 \bar{u}_4^2  \tilde{L}_3(\bar{u}_2)+3 \bar{u}_4  \tilde{L}_2(\bar{u}_2)\right)}{1-12 \bar{u}_4^3 L_2^3(\bar{u}_2)+8 \bar{u}_4^2  L_2^2(\bar{u}_2)-3 \bar{u}_4  L_2(\bar{u}_2) }\,,}\label{u60}
\end{equation}
where, using Litim regulator\footnote{As before, integrals are computed using local potential approximation.}:
\begin{equation}
  \tilde{L}_2:=\frac{1}{2\pi} \frac{1}{(1+\bar{u}_2)^2} \,,
\end{equation}
\begin{equation}
 \tilde{L}_3:=\frac{1}{2\pi} \frac{\frac{\sqrt{\bar{u}_2} \left(\bar{u}_2 \left(\bar{u}_2 \left(15 \bar{u}_2-73\right)-55\right)-15\right)}{\left(\bar{u}_2+1\right){}^4}+15 \tan ^{-1}\left(\sqrt{\bar{u}_2}\right)}{24 \bar{u}_2^{7/2}}\,,
\end{equation}
\begin{equation}
  \tilde{L}_4:=\frac{1}{(2\pi)^2} \int \prod_{i=1}^3d\omega_i\, G^2_k(\omega_1) G^2_k(\omega_2) G^2_k(\omega_3) \delta(\omega_1+\omega_2+\omega_3) \,,
\end{equation}
\begin{equation}
L_2(\bar{u}_2):=\frac{1}{2\pi} \left(-\frac{1-\bar{u}_2}{\bar{u}_2 \left(1+\bar{u}_2\right){}^2}+\frac{\tan ^{-1}\left(\sqrt{\bar{u}_2}\right)}{\bar{u}_2^{3/2}}\right)\,,
\end{equation}
\begin{equation}
L_3(\bar{u}_2):=\frac{1}{2\pi}\frac{-\frac{3+5 \bar{u}_2}{\bar{u}_2^2 \left(1+\bar{u}_2\right){}^2}+\frac{8}{\left(\bar{u}_2+1\right){}^3}+\frac{3 \tan ^{-1}\left(\sqrt{\bar{u}_2}\right)}{\bar{u}_2^{5/2}}}{4}\,.
\end{equation}
Note that the integral $ \tilde{L}_4$ have to be computed numerically. The flow equation for $\bar{u}_6$ can be derived from the observation that $u_6(0)$ does not depends on $k$. Taking the derivative of \eqref{u60} with respect to $k$, we get:
\begin{align}
\boxed{\dot{\bar{u}}_6=(u_6+a) \left(\frac{\dot{\bar{u}}_2\partial_{\bar{u}_2} b +\dot{\bar{u}}_4\partial_{\bar{u}_4} b}{b} -4\right)-(\dot{\bar{u}}_2\partial_{\bar{u}_2} a +\dot{\bar{u}}_4\partial_{\bar{u}_4} a)+5a\,,}
\end{align}
where $a$ and $b$ are two functions depending on $\bar{u}_2$ and $\bar{u}_4$, defined as:
\begin{align}
a(\bar{u}_2,\bar{u}_4)&:=-24 L_3(\bar{u}_2) \bar{u}_4^3+\bar{\tilde{u}}_6(0) \left(12 \bar{u}_4^3 \tilde{L}_4(\bar{u}_2)-8 \bar{u}_4^2  \tilde{L}_3(\bar{u}_2)+3 \bar{u}_4  \tilde{L}_2(\bar{u}_2)\right)\,,\\
b(\bar{u}_2,\bar{u}_4)&:=1-12 \bar{u}_4^3 L_2^3(\bar{u}_2)+8 \bar{u}_4^2  L_2^2(\bar{u}_2)-3 \bar{u}_4  L_2(\bar{u}_2)\,.
\end{align}

\begin{figure}
\begin{center}
\includegraphics[scale=0.7]{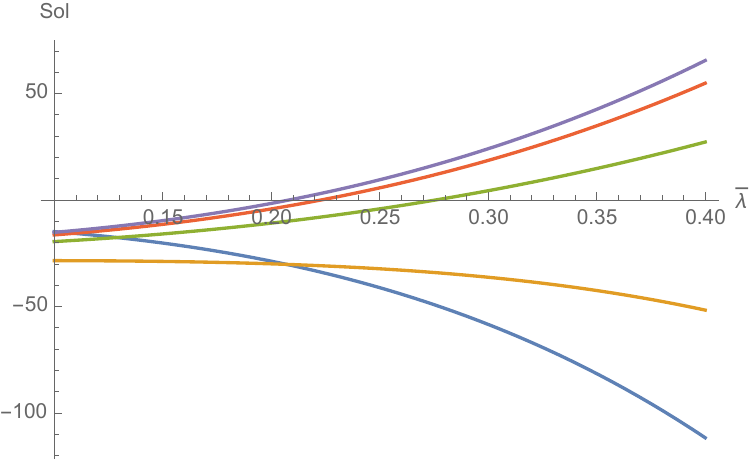}
\end{center}
\caption{Behavior of the function $\text{sol}$ for $\bar{\tilde{u}}_6=-1$ (blue curve), $\bar{\tilde{u}}_6=-10$ (yellow curve), $\bar{\tilde{u}}_6=-30$ (green curve), $\bar{\tilde{u}}_6=-100$ (red curve) and $\bar{\tilde{u}}_6=-1000$ (purple curve). Note that we rescaled the two last functions to keep them on the same figure.}\label{figsol}
\end{figure}

As before, let us begin by investigating the fixed trajectory solutions, focusing on the deep IR regime. Since the non-local coupling does not renormalize, this is equivalent to examining the limit $\bar{\tilde{u}}_6 \to \infty$. Following the same procedure to find fixed-point solutions, we construct a function $\text{sol}(\bar{u}_2)$ from the two conditions $\dot{\bar{u}}_2 = \dot{\bar{u}}_4 = 0$, as shown in Figure \ref{figsol}. The figure indicates that a fixed-point solution emerges when $\bar{\tilde{u}}_6$ becomes sufficiently large (typically $\bar{\tilde{u}}_6 \sim -30$). Moreover, this solution quickly converges to a finite value for $\bar{u}_2$, given numerically by:
\begin{equation}
\bar{u}_{2,\infty}\approx 0.21\,,
\end{equation}
where the subscript $\infty$ indicates the limit $\bar{\tilde{u}}_6 \to \infty$ (numerically, $\bar{\tilde{u}}_6 = 10^5$; the figure shows that the convergence is sufficiently fast). However, this fixed-point solution has a poor critical exponent:
\begin{equation}
\Theta_{EVE,6}=(\theta_1=-\infty,\theta_2\approx 1.41)\,,
\end{equation}
with eigenvectors converging toward $(0,1)$ and $(-0.32, 0.95)$, corresponding to the effective mass and quartic coupling directions, respectively. While the quartic direction is highly stable, the reliability of this result is limited, given the approximations we have made and the important effects we have neglected, such as higher-order derivative couplings, as previously discussed.
Nevertheless, we will assume that the physical conclusions are valid, leaving a deeper investigation of the missing relevant effects for future work.
\medskip

\begin{remark}
No fixed point is found in the negative mass region. Moreover, no asymptotic fixed point with positive mass is found assuming the initial classical action is quartic in the local sector, but surprisingly a novel fixed point is found in the negative region,
\begin{equation}
\bar{u}_{2,\infty}\approx -0.74\,,
\end{equation}
for critical exponents:
\begin{equation}
\Theta_{EVE,60}=(\theta_1=\infty,\theta_2\approx 1.42)\,,
\end{equation}
and eigen directions $(0,1)$ and $(0.66,0.75)$. Interestingly, the quartic coupling axis becomes a very unstable direction is that case. Note however that this fixed point is poorly reliable than the positive one, higher order interactions having a negative sign. In contrast the positive fixed point we find in the sextic theory seems to be stable, the sign of higher order effective local couplings being always positive (we tested it until valence 10). 
\end{remark}

\begin{figure}
\begin{center}
\includegraphics[scale=0.5]{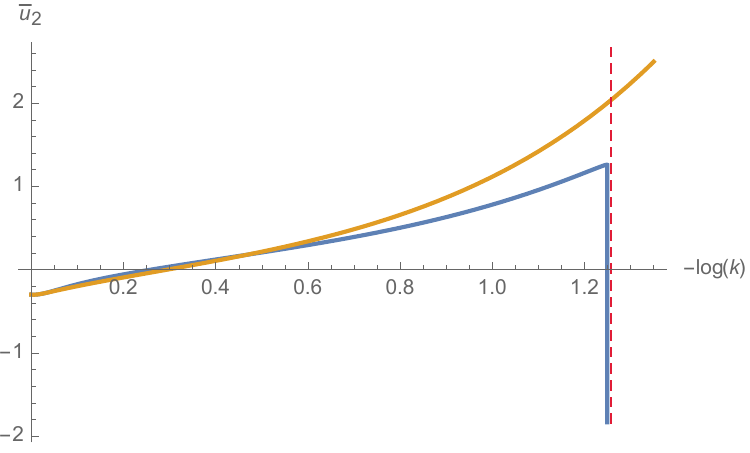}\qquad \includegraphics[scale=0.5]{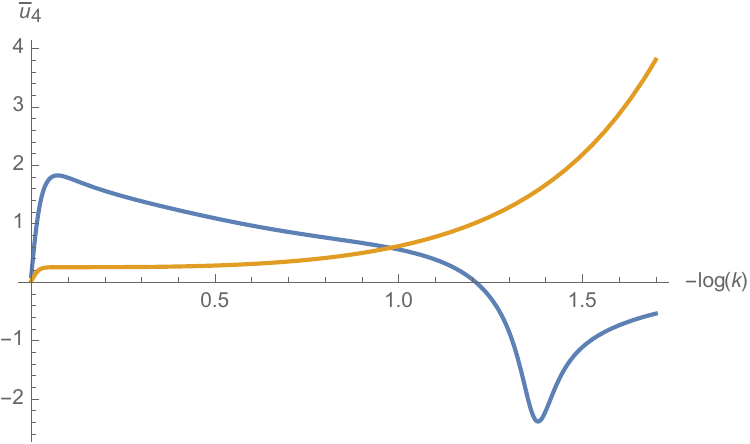}
\end{center}
\caption{On the left: behavior of the effective mass for $\bar{\tilde{u}}_6(0)=-0.8$ (blue curve) and for $\bar{\tilde{u}}_6=0$ (yellow curve). On the right: Behavior of the quartic coupling for $\bar{\tilde{u}}_6(0)=-0.7$ (blue curve) and for $\bar{\tilde{u}}_6=0$ (yellow curve).}\label{figplottrajectories1}
\end{figure}

Next, let us examine the behavior of some RG trajectories and their dependence on the disorder. Figure \ref{figplottrajectories1} illustrates the behavior of several trajectories with and without disorder for the following initial conditions: $\bar{u}_2(0) = -0.3$, $\bar{u}_4(0) = 0.1$, and $\bar{u}_6(0) = 1$. From the initial conditions we investigated, we found a critical value of $\bar{\tilde{u}}_{6,c}(0) \sim -0.7$, below which the RG trajectory exhibits a finite-scale singularity at approximately $t_c \approx 1.25$.

For sufficiently small values of $\bar{\tilde{u}}_6(0)$, the trajectories behave almost as if there were no disorder. However, as we approach the critical value, the trajectory becomes noticeably different after the critical scale $t_c$, displaying angular points that signal the onset of a singularity, as shown on the right side of Figure \ref{figplottrajectories1}.

Finally, although it may appear that the singularity drives the flow toward the region of negative masses (as seen on the left side of Figure \ref{figplottrajectories1}), numerical evidence suggests that for values of $\bar{\tilde{u}}_6(0)$ close to the critical value, the vertical asymptote is actually oriented toward positive masses.
\medskip

\begin{figure}
\begin{center}
\includegraphics[scale=0.5]{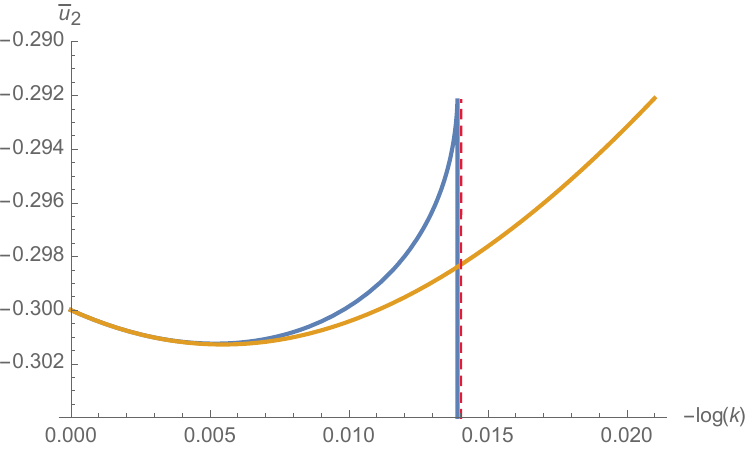}\qquad \includegraphics[scale=0.5]{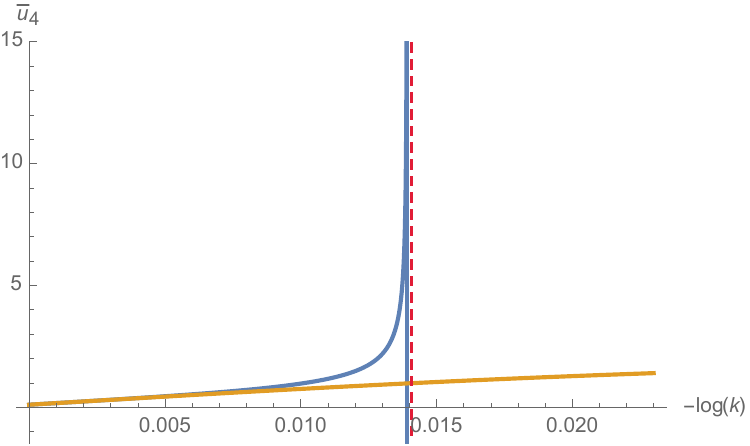}
\end{center}
\caption{In yellow: Behavior of the RG flow for the quartic theory (${u}_6(0)=0$) with the same initial conditions as before. In blue: behavior of the RG flow for $\bar{\tilde{u}}_6(0)=-0.37$. }\label{figplottrajectories1Q}
\end{figure}

Investigating the behavior of the quartic theory (without local sextic coupling in the initial action), for the same initial conditions, we recover essentially the same behavior, for a critical $\bar{\tilde{u}}_{6,c}(0)\approx -0.36$, see Figure \ref{figplottrajectories1Q}. Note that as before, near the transitions, the trajectories exhibit angular points, announcing the divergence.
\medskip

\begin{figure}
\begin{center}
\includegraphics[scale=0.5]{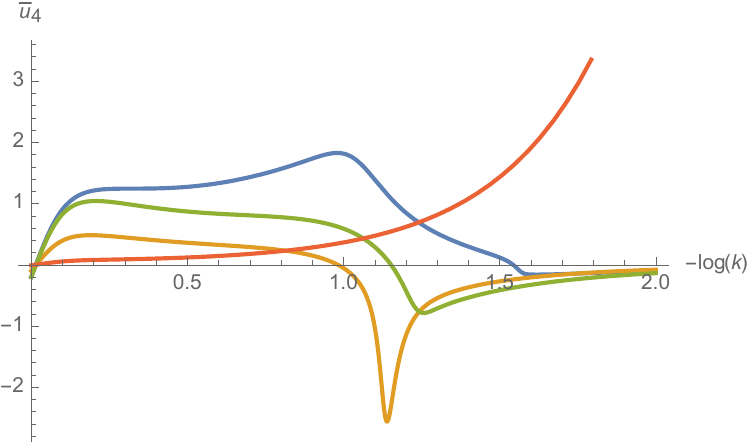}\qquad \includegraphics[scale=0.5]{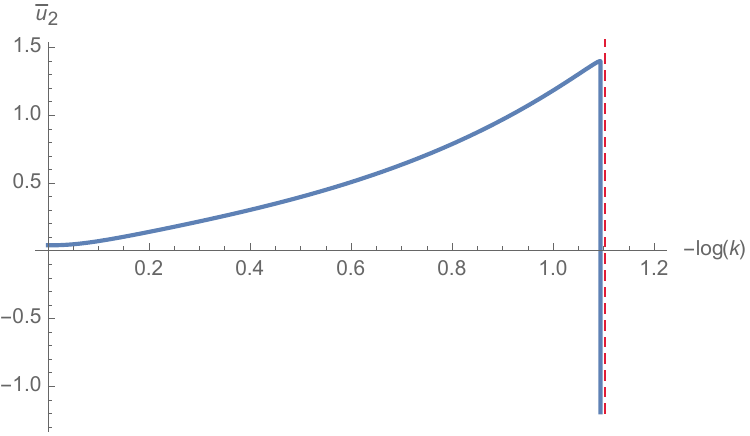}
\end{center}
\caption{On the left: Behavior of the quartic  coupling for $\bar{\tilde{u}}_6(0)=0$ (red curve), $\bar{\tilde{u}}_6(0)=-1.8$ (blue curve), $\bar{\tilde{u}}_6(0)=-2.8$ (green curve) and $\bar{\tilde{u}}_6(0)=-3.3$ (yellow curve). On the right: Behavior of the effective mass for $\bar{\tilde{u}}_6(0)=-3.4$.}\label{figpositive1}
\end{figure}

The previous conclusions appear to hold for any set of initial conditions. Consider, for example, the positive quadrant of the $(\bar{u}_2, \bar{u}_4)$ plane, with the initial conditions $\bar{u}_2(0) = 0.01$, $\bar{u}_4(0) = 0.1$, and $\bar{u}_6(0) = 4$. The results, summarized in Figure \ref{figpositive1}, essentially mirror those obtained for the negative mass region, though the critical value of $\bar{\tilde{u}}_6(0)$ is larger: $\bar{\tilde{u}}_6(0) \approx -3.4$. Similar behavior is observed in the region where $\bar{u}_4 < 0$, with a critical value for $\bar{\tilde{u}}_6(0)$ of the same order.
These results suggest the existence of a region around the Gaussian point, from which the flow always becomes singular at a finite time, once the disorder reaches a critical value that depends on the initial conditions. Notably, in the continuous limit, this critical value tends to zero, implying that the singular behavior of the flow appears even with the presence of infinitesimal disorder.
To summarize:

\begin{claim}
Our numerical investigations demonstrate that the presence of disorder significantly affects the flow, leading to the emergence of singular behavior at a finite time. This occurs when the disorder reaches a critical value, though the exact dependence of this critical value on the initial conditions has yet to be determined.
\end{claim}

\subsection{The 2PI formalism, a first look}\label{2PI}

As mentioned above and highlighted in our previous work \cite{lahoche2024largetimeeffectivekinetics}, and following the interpretation in \cite{gredat2014finite}, we view these finite-scale divergences as evidence that interactions, which are forbidden in perturbation theory (i.e., in the symmetric phase), are dynamically generated by the flow. In other words, these divergences indicate that the flow reaches an unstable region of phase space where these interactions, incompatible with perturbation theory, become relevant.

Moreover, we anticipate that once the RG flow is fully accounted for, these divergences can be suppressed. In this section, we aim to evaluate the stability of the phase space in relation to interactions whose flow is suppressed (at large $N$) by perturbation theory. To do this, we will consider the following truncation of the kinetic term:
\begin{equation}
\Gamma_{k,\text{kin}}=\frac{1}{2} \int dt \,\sum_{\alpha} \left(\dot{\bm{M}}^2_\alpha(t)+u_2(k)\bm{M}^2_\alpha(t)+\sum_{\beta} q_{\alpha\beta}\, \bm{M}_\alpha(t) \cdot \bm{M}_\beta(t) \right)\,,\label{ansatz2PI}
\end{equation}
and the inverse $2$-point function reads, in the Fourier space:
\begin{equation}
G^{-1}_{k,\alpha\beta}(\omega)= (\omega^2+u_2(k)+R_k(\omega)) \delta_{\alpha\beta}+q_{\alpha\beta}(k)\,.
\end{equation}
for simplicity, we assume $q_{\alpha\beta}(k)=-q(k)$ (the same value for all replica indices). Reversing the matrix then become a simple issue, and we get:
\begin{equation}
G_{k,\alpha\beta}(\omega)= \underbrace{\vcenter{\hbox{\includegraphics[scale=1]{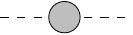}}}}_{\frac{\delta_{\alpha\beta}}{\omega^2+u_2(k)+R_k(\omega)}}\,+\,\underbrace{\vcenter{\hbox{\includegraphics[scale=1]{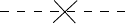}}}}_{\frac{q}{(\omega^2+u_2(k)+R_k(\omega))(\omega^2+u_2(k)-n q+R_k(\omega))}}\,.
\end{equation}
The operator $q_{\alpha\beta}$ plays the role of an order parameter, and the 2PI formalism is generally more suitable to investigate this case \cite{benedetti20182pi,dupuis2014nonperturbative}, see also the short presentation given in \cite{lahoche2024largetimeeffectivekinetics}. However, it appears that in the large limit $N$, and in the limit where the nonlocal source associated with the two-point function vanishes, the self energy formally satisfies the same equation as that deduced from perturbation theory (see Appendix \ref{App1}), replacing the two-point function by the ansatz \eqref{ansatz2PI}. 
\begin{align}
u_2&= \, \vcenter{\hbox{\includegraphics[scale=0.8]{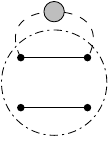}}}\,+\,  \vcenter{\hbox{\includegraphics[scale=0.8]{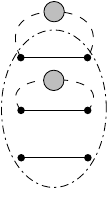}}}\,+\, \vcenter{\hbox{\includegraphics[scale=0.8]{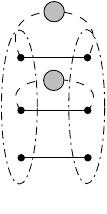}}}\,+\,\mathcal{O}(q)\\
q_{\alpha\beta}&=-q=\, \vcenter{\hbox{\includegraphics[scale=0.8]{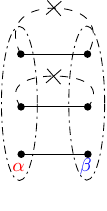}}}=:q^2F(u_2,q)\,.
\end{align}
For a detailed derivation of this result in a closely related context, the reader can refer to \cite{lahoche2023functional,lahoche2024largetimeeffectivekinetics} by the same authors. Note that the approximations here should be understood in the sense of the Ginzburg-Landau approximation: we treat $q$ as a small parameter, so that an expansion around $q = 0$ is always justified. As we will see below, this approximation is actually inadequate since the transition is of first order. However, it will suffice for the purposes of this article. We plan to revisit this formalism with more rigorous approximations in a forthcoming article. 
\medskip

\begin{figure}
\begin{center}
\includegraphics[scale=0.5]{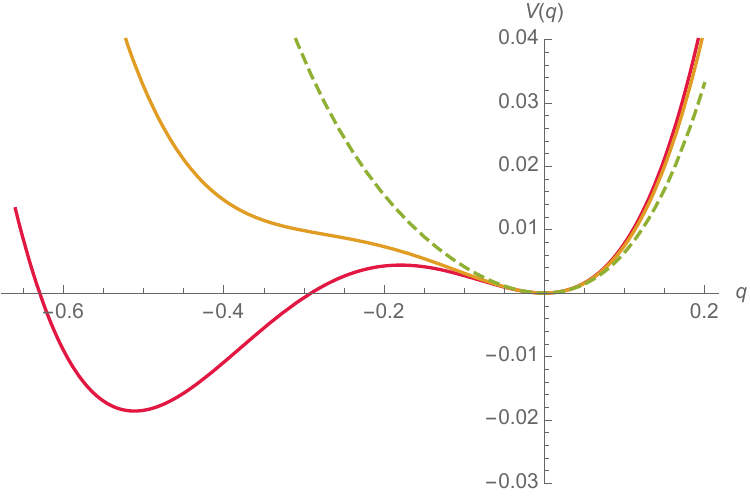}\qquad \includegraphics[scale=0.5]{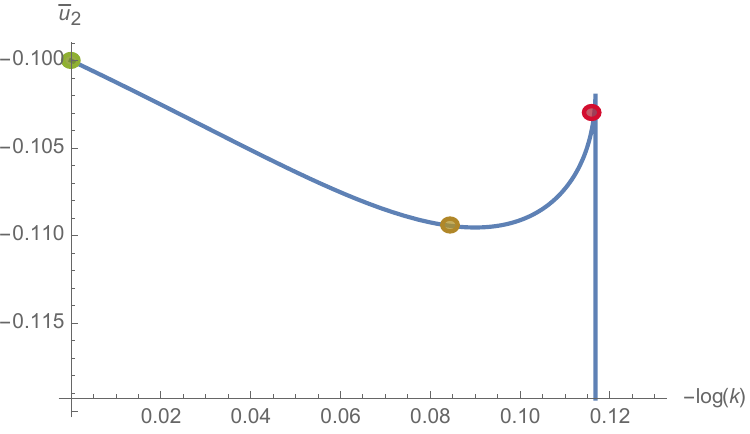}
\end{center}
\caption{The evolution of the shape of the potential $V(q)$  (on the left) along some RG trajectory with non-vanishing disorder (on the right), for $t=0$, $t=0.085$ and $t=0.111$. The critical time scale is for $t_c=0.113$.}\label{fig2PI}
\end{figure}

We then assume that $F(q)$ can be reliably approached by the first terms of its expansion around zero:
\begin{equation}
F(u_2,q)\approx F(u_2,0)+q F'(u_2,0)+\mathcal{O}(q^2)\,. 
\end{equation}
But we have to take into account that the flow of $u_2$ expands also in power of $q$, namely:
\begin{equation}
u_2(k)=u_2^{(0)}(k)+q u_2^{(1)}(k) + \mathcal{O}(q^2)\,,
\end{equation}
then,
\begin{equation}
-q=q^2\left(F(u_2^{(0)},0)+q \left(F'(u_2^{(0)},0)+u_2^{(1)}(k)\frac{\partial }{\partial u_2} F(u_2^{(0)},0)\right)\right)+\mathcal{O}(q^4)\,.\label{equationQ}
\end{equation}
Let us compute $u_2^{(1)}(k)$. For simplicity, we will use the same graphical notation as before, and we will introduce the following notation for q insertions in the expansion of the crossed propagator:
\begin{equation}
\vcenter{\hbox{\includegraphics[scale=1]{NlocPot.pdf}}}=\, \underbrace{\vcenter{\hbox{\includegraphics[scale=1]{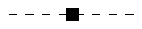}}}}_{\mathcal{O}(q)}\,+\, \underbrace{\vcenter{\hbox{\includegraphics[scale=1]{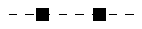}}}}_{\mathcal{O}(q^2)}\,+\, \mathcal{O}(q^3)\,,
\end{equation}
where each square insertion corresponds to a factor $q$. Then:
\begin{align} q u_2^{(1)}=\,\vcenter{\hbox{\includegraphics[scale=0.8]{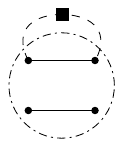}}}\,+\,2\,\vcenter{\hbox{\includegraphics[scale=0.8]{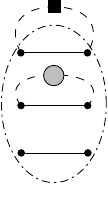}}}\,+\,2\,\vcenter{\hbox{\includegraphics[scale=0.8]{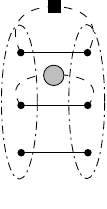}}}\,,
\end{align}
and we have also:
\begin{equation}
u_2^{(1)}=\, -\frac{d}{du_2^{(0)}}\, \left(\, \vcenter{\hbox{\includegraphics[scale=0.7]{V41eff1.pdf}}}\,+\,  \vcenter{\hbox{\includegraphics[scale=0.7]{V61eff1.pdf}}}\,+\, \vcenter{\hbox{\includegraphics[scale=0.7]{V62eff1.pdf}}}\,\right)=-\frac{d}{du_2^{(0)}}\,  u_2^{(0)} =-1\,.
\end{equation}
Computing $F(u_2,0)$ and $F'(u_2,0)$, we get:
\begin{align}
F(u_2,0)&= -6 \tilde{u}_6(0)\int \frac{d\omega}{2\pi}\, G_{k,0}^4(\omega^2) \,,\\
F'(u_2,0)&=-12 n \tilde{u}_6(0) \int \frac{d\omega}{2\pi}\, G_{k,0}^5(\omega^2) \,,
\end{align}
where $G_{k,0}$ is the local propagator, for $q=0$. Finally the equation \eqref{equationQ} fixing the value for $q$ can be rewritten as an equilibrium condition:
\begin{equation}
V'(q)=0\,,
\end{equation}
where the function $V(q)$ (a simplified version of the \textit{Ward-Luttinger functional}) is, at the leading order:
\begin{equation}
V(q)=\frac{q^2}{2}+\frac{q^3}{3}F(u_2^{(0)},0)+\frac{q^4}{4} \left(F'(u_2^{(0)},0)-\frac{\partial }{\partial u_2} F(u_2^{(0)},0)\right)\,,
\end{equation}
or, 
\begin{equation}
\boxed{V(q)=\frac{q^2}{2}+\frac{q^3}{3}F(u_2^{(0)},0)+\frac{q^4}{4}\frac{n+1}{n} F'(u_2^{(0)},0)\,.}
\end{equation}
In this equation, the flow for coupling constants $F(u_2^{(0)},0)$ and $F^\prime(u_2^{(0)},0)$ can be computed from the 1PI flow for $u_2$ in the symmetric phase. The result is shown on Figure \ref{fig2PI} for some trajectory computed using EVE with non vanishing disorder, and we have the explicit formula:
\begin{equation}
F(u_2,0)\equiv \frac{f(\bar{u}_2)}{k^7}=:\frac{\sqrt{\bar{u}_2} \left(\bar{u}_2 \left(\bar{u}_2 \left(15 \bar{u}_2-73\right)-55\right)-15\right)+15 \left(1+\bar{u}_2\right){}^4 \tan ^{-1}\left(\sqrt{\bar{u}_2}\right)}{48 \pi  k^7 \bar{u}_2^{7/2} \left(1+\bar{u}_2\right){}^4}\,.
\end{equation}
In Figure \ref{fig2PI}, we explicitly show the evolution of the potential $V(q)$ along the singular trajectory depicted on the left of Figure \ref{figplottrajectories1}, for the initial condition $\bar{\tilde{u}}_6(0)=-1$. As suggested above, we observe that the presence of the singularity is accompanied upstream by the emergence of a second stable vacuum at $q \neq 0$, which becomes progressively deeper as we approach the singularity. The vacuum at $q=0$ also remains stable up to the singularity, suggesting a first-order transition that cannot be completely described within this formalism. To summarize:
\begin{claim}
We have evidence of a relationship between the singular flow behavior described above and a potential first-order phase transition for the $q$ parameter that couples the replicas.
\end{claim}

\begin{remark}
Note that correlations between replicas are not the only ones we have 'ignored' in this treatment. Derivative interactions for local potentials could also influence the results and may be as relevant as some effects considered in our study. We will completely disregard them in this work and reserve their investigation for future research, focusing, for instance, on the advanced method proposed by Blaizot, Méndez, and Wschebor (BMW) in \cite{blaizot2006new}.
\end{remark}

\pagebreak
\section{Multi-local expansion around local potential solution for $p=3$}\label{sectionbeyond}

The finite (time) scale singularities we observed for the RG flow constructed from the truncation inspired from large $N$ perturbation theory are reminiscent to the so-called Larkin scale singularities \cite{Tarjus,Tarjus2,dupuis2020bose}, and are usually interpreted as the signal that some interaction forbidden by perturbation theory dominates the flow up to this scale. This assumption seems to confirmed by the heuristic computation done in section \ref{2PI}, and our investigation for the ‘‘2+p" model \cite{lahoche2024largetimeeffectivekinetics,lahoche2024Ward,achitouv2024timetranslationinvariancesymmetrybreaking}. This section aims to provide an additional argument in the sense of this hypothesis, by considering the influence of these forgotten interactions on the flow. Our approximations will still suffer, at this stage, from the shortcomings already mentioned above, in particular concerning the highly unsatisfactory character of a truncation of local and non-local interactions from a certain rank.
Note that this shortcoming is probably less important in our previous works cited above, where the truncations consider the whole relevant sector.
More advanced methods mentioned above will allow to further refine the validity of our conclusions in future investigations.

\subsection{Principles of the approximation}

This section follows the construction in \cite{lahoche2024Ward}. We will consider in the truncation interactions which are not allowed by the large $N$ perturbation theory, but expected to correspond to metastable states. A first limitation of our approach, motivated by the heuristic observations of section \ref{2PI} as well as by the field theory of spin glasses \cite{de2006random} is that we will assume that these metastable states only concern the $2$-point function. Thus, we will assume that it is the instabilities of the $2$ point function which induce the growth of other “deviant” interactions with respect to the theory space reachable by perturbation theory.

We will essentially consider one kind of deviant interaction, based on the kinetic effective action considered as before, equation \eqref{ansatz2PI}, for $q_{\alpha\beta}=-q$ but taking $u_2=0$. Another relevant piece for $\Gamma_k$ is the local part, defining the local potential. In the first part of this work, we assumed the validity of an expansion around the vanishing classical field (symmetric phase), agreeing with the restriction to the perturbative phase space. Here, we consider an expansion around the minimum of the effective potential, which becomes a running coupling constant among the others. This approach is quite standard in the local potential approximation formalism and has been also considered in \cite{Tarjus} for instance. Finally, the last piece of our approximation will be a projection of the flow equation on a uniform solution; for $n=2$ (two replicas), we set:
\begin{equation}
\bm M_1(t)\to \bm M_{01}:=\sqrt{2 \rho_1 N} \begin{pmatrix} 1\\
\bm 0_\bot
\end{pmatrix}\,,\quad \bm M_2(t)\to \bm M_{02}=\sqrt{2 \rho_2 N} \begin{pmatrix} \theta\\
\bm e_\bot
\end{pmatrix}\,,\label{homogeneous}
\end{equation}
where $\theta$ depends on $k$, $\bm 0_\bot$ denotes the null vector in $\mathbb{R}^{N-1}$, $\bm 0_\bot^2=0$, and $\bm{e}_\bot\in \mathbb{R}^{N-1}$ such that:
\begin{equation}
\bm{e}_\bot\cdot \bm{e}_\bot= 1-\theta^2\,,\quad \theta \in [-1,1]\,.
\end{equation}
We will denote as $\bm e_1:=(1,\bm{0}_\bot)^T$ and $\bm e_1:=(\theta,\bm{e}_\bot)^T$ the two vectors along each replica. 

In these definitions, $\rho_I$ is \textit{time independent}, and different for the two replicas. Ultimately in the flow equations we will set $\rho_1=\rho_2=\kappa$, where $\kappa(k)$ is the running minimum of the effective potential:

\begin{equation}
\frac{U_k(\bm M^2)}{N}=\frac{u_4}{2} \left(\frac{\bm M^2}{2N}-\kappa \right)^2+\frac{u_6}{3} \left(\frac{\bm M^2}{2N}-\kappa \right)^3+\cdots\,,
\end{equation}

which is the same for all replicas. Let us recall at this stage that the replica symmetry is nerveless explicitly broken by construction, the source fields being different. The non-local part of the effective action will depend on the nature of the metastable state we choose to consider. If we focus on the truncation \eqref{ansatz2PI}, we then consider:
\begin{align}
\nonumber \Gamma_{k,\text{int}}&= \int d t \bigg(\sum_{\alpha}\underbrace{U_k[\bm{M}_\alpha^2(t)]}_{\text{Local}}+\frac{1}{2}\sum_{\alpha,\beta} \underbrace{V_k(\Vert \bm{M}_\alpha (t)\Vert,\Vert \bm{M}_\beta(t) \Vert,u_{\alpha \beta}(t,t))}_{\text{Coupling replica}}\\
&+\frac{1}{2} \int d t^\prime\sum_{\alpha,\beta} \underbrace{W_k(\Vert \bm{M}_\alpha (t)\Vert,\Vert \bm{M}_\beta(t^\prime) \Vert,u_{\alpha \beta}(t,t^\prime))}_{\text{Non-local}}\bigg)\,,\label{truncationS1}
\end{align}
where:
\begin{equation}
u_{\alpha \beta}(t,t^\prime):=\frac{\bm{M}_{\alpha}(t)}{\Vert \bm{M}_\alpha(t) \Vert}\cdot \frac{\bm{M}_{\beta}(t^\prime)}{\Vert \bm{M}_\beta (t^\prime)\Vert},
\end{equation}
with in particular $u_{\alpha \beta}(t,t^\prime)=\theta$ as we project along the homogeneous solution \eqref{homogeneous}. 

We will simplify the construction of the approximation by considering only the interactions which emerge linearly with $q$. Focusing only on the second cumulant for correlations between replica and/or time, we have explicitly:

\begin{equation}
\int d t \sum_{\alpha,\beta}\, V_k(\Vert \bm{M}_\alpha (t)\Vert,\Vert \bm{M}_\beta(t) \Vert,u_{\alpha \beta}(t,t))= v(k)\int d t \sum_{\alpha,\beta}\, (\bm{M}_\alpha(t) \cdot \bm{M}_\beta(t))^2\equiv \vcenter{\hbox{\includegraphics[scale=0.8]{Vk41Loc}}}\,,
\end{equation}
and (coupling constants are implied):
\begin{equation}
\frac{1}{2} \int d t d t^\prime\sum_{\alpha,\beta} W_k(\Vert \bm{M}_\alpha (t)\Vert,\Vert \bm{M}_\beta(t^\prime) \Vert,u_{\alpha \beta}(t,t^\prime))=\vcenter{\hbox{\includegraphics[scale=0.8]{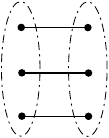}}}\,.
\end{equation}
Once we project potentials $V_k$ and $W_k$ along the uniform solutions, we get (a relation which defines the coupling constants):
\begin{equation}
V_k(\Vert \bm{M}_{0 \alpha}(t)\Vert,\Vert \bm{M}_{0 \beta}(t) \Vert,u_{\alpha \beta})=2N v(k) \Big(\rho_1^2+\rho_2^2+2 \theta^2 \rho_1 \rho_2 \Big)\,,
\end{equation}
and
\begin{equation}
W_k(\Vert \bm{M}_{0 \alpha}(t)\Vert,\Vert \bm{M}_{0 \beta}(t) \Vert,u_{\alpha \beta})=2N w(k) \Big(\rho_1^{3}+\rho_2^{3}+2 \theta^3 \sqrt{\rho_1^3 \rho_2^3}\,\Big)\,.
\end{equation}

\subsection{Flow equations for the first truncation}

Deriving twice with respect to $M_{\alpha i}(t)$ and $M_{\beta j}(t^\prime)$, and projecting along $M_{0\alpha}$, we get:
\begin{align}
\Gamma_{k}^{(2)}(t,t^\prime)+R_k(\vert t-t^\prime \vert )\delta_{ij} \delta(t-t^\prime)&=\bm{A} \delta(t-t^\prime)+\bm{B}
\end{align}
where entries of the matrix $\bm{A}$ is\footnote{We will use a somewhat cavilier notation here, noting the matrix and its entries in the same way.}:
\begin{align}
 \bm{A}&:=\delta_{\alpha \beta}\delta_{ij} \Big(-\frac{d^2}{d t^2} + \frac{\partial U_k}{\partial \rho_\alpha}+R_k\Big)+\Big(-q+4v\sqrt{\rho_\alpha \rho_\beta}\, \bm{e}_\alpha\cdot \bm{e}_\beta  \Big)\delta_{ij}+ \bm{D}\,,
\end{align}
and:
\begin{equation}
\bm{B}:=12 w_{6,1} \rho_\alpha \rho_\beta\, (\bm{e}_\alpha\cdot \bm{e}_\beta)^2 \delta_{ij} \,,
\end{equation}
where moreover:
\begin{align}
\nonumber \bm{D}=&+2 \rho_\alpha \frac{\partial^2 U_k}{\partial \rho_\alpha^2}\,\delta_{\alpha \beta}(\bm{e}_\alpha)_i(\bm{e}_\alpha)_j+4 v\left(\sum_{\beta^\prime} \rho_{\beta^\prime}\, (\bm{e}_{\beta^\prime})_i (\bm{e}_{\beta^\prime})_j \right) \delta_{\alpha\beta}\\
&+12 \beta w \sqrt{\rho_\alpha} \sum_{\beta^\prime} \rho_{\beta^\prime}^{3/2}
(\bm{e}_{\beta^\prime} \cdot \bm{e}_{\alpha}) (\bm{e}_{\beta^\prime})_i (\bm{e}_{\beta^\prime})_j \delta_{\alpha\beta}\,. 
\end{align}
At this stage, it should be noticed that $\bm D$ has Frobenius norm of order one,
\begin{equation}
\Vert \bm D \Vert := \sqrt{\Tr (\bm D \bm D^T)}=\mathcal{O}(1)\,,
\end{equation}
and in contrast $\bm{\mathcal{A}}:=\bm A-\bm D$ has norm of order $N$. Then, in the large $N$ limit, we can ignore the contribution of $\bm D$ in the flow equation. 

Recalling that we define Fourier transform of some gentle function $f(t)$ as:
\begin{equation}
f(t):= \frac{1}{\beta}\sum_{\omega} \tilde{f}(\omega) e^{-i \omega t}\,,\qquad  \tilde{f}(\omega):= \int_{-\beta/2}^{\beta/2} d t \, f(t) e^{i \omega t}\,,
\end{equation}
where frequencies are assumed to be quantified $\omega=2 \pi n/\beta$, $n\in \mathbb{N}$; the propagator $G_k(t,t^\prime)$ is defined by the condition:
\begin{equation}
\bm{\mathcal{A}} G(t,t^\prime)+\bm{B} \int d t^{\prime\prime} G(t^{\prime\prime},t^\prime)= \delta(t-t^\prime)\,,
\end{equation}
which can be rewritten in the Fourier mode as:
\begin{equation}
\bm{\mathcal{\tilde{A}}}(\omega^2) \,G(\omega,\omega^\prime)+\bm{B} \beta \, G(0,\omega^\prime)\delta_{\omega,0}=\beta \, \delta_{\omega,-\omega^\prime}\,.
\end{equation}
where $\bm{\mathcal{\tilde{A}}}$ is the Fourier transform of $\bm{\mathcal{A}}$, and for $\omega^2<k^2$, we have:
\begin{align}
\bm{\mathcal{\tilde{A}}}(\omega^2<k^2)&:=\delta_{\alpha \beta}\delta_{ij} \underbrace{\Big(k^2 + \frac{\partial U_k}{\partial \rho_\alpha}\Big)}_{:=C}+\Big(-q+4v\sqrt{\rho_\alpha \rho_\beta}\, \bm{e}_\alpha\cdot \bm{e}_\beta  \Big)\delta_{ij} \,,
\end{align}
or in the matrix form:
\begin{equation}
\bm{\mathcal{\tilde{A}}}(\omega^2<k^2)=\delta_{ij} \begin{pmatrix}
C+(-q+4v \rho_1\,) & -q+4 \theta v\sqrt{\rho_1\rho_2} \\-q+
4 \theta v\sqrt{\rho_1\rho_2} & C+(-q+4v \rho_2\,)
\end{pmatrix}\,.
\end{equation}
After some straightforward algebraic manipulations, we get:
\begin{equation}
\boxed{G_k(\omega,\omega^\prime)= \beta \left(\bm{\mathcal{\tilde{A}}}^{-1}(\omega^2) \delta_{\omega,-\omega^\prime}-(\bm{\mathcal{\tilde{A}}}(0)+\beta \bm{B})^{-1} \bm{B} \beta \bm{\mathcal{\tilde{A}}}^{-1}(0) \delta_{\omega,0} \delta_{\omega^\prime,0}\right)\,,} 
\end{equation}
and, before projecting along the symmetric solution $\rho_1=\rho_2=\kappa$,
\begin{align}
\nonumber \bm{\mathcal{\tilde{A}}}^{-1}(\omega^2<k^2)\big\vert_{\rho_i=\kappa}=\frac{\delta_{ij}}{(k^2+4v \kappa (1-\theta))(k^2-2 q+4v \kappa (1+\theta)} \\
\qquad \times \begin{pmatrix}
k^2+(-q+4v \kappa\,) & q-4 \theta v \kappa \\q-4 \theta v\kappa & k^2+(-q+4v \kappa\,)
\end{pmatrix}\,.
\end{align}
The bilocal contributions in the right hand side of the flow equation can be identified according to their power of $\beta$, i.e. they should be proportional to $\beta^2$. However, because $\beta$ is assumed to be a large parameter, the expansion of the previous equation is a little subtle. Indeed, if we take the limit $\beta \to \infty$ naively, we conclude that the left-hand side no longer depends on $B$, and hence on the non-local coupling. However, a moment of reflection shows that $\bm{\mathcal{\tilde{A}}}(0)=k^2 \mathcal{O}(1)$; moreover $\bm{B}$ has a dimension $3$ regarding the power counting, $\bm{B}=k^3 \bar{\bm{B}}$. Then, 
\begin{equation}
\bm{\mathcal{\tilde{A}}}(0)+\beta \bm{B}= \bm{\mathcal{\tilde{A}}}(0)\left(1+ \frac{\beta k}{\mathcal{O}(1)} \bar{\bm{B}} \right)\,.
\end{equation}
The condition to expand the denominator in power of $\bm B$ is then $\bar{\bm B} \beta k \ll 1$, which can be translated as a condition for $\kappa$ (see \cite{lahoche2024Ward} for more details):
\begin{equation}
\boxed{\bar{\kappa}^2 \ll \mathcal{O}(1) \, \frac{k^{-1}}{\beta \bar{w}}\,.}
\end{equation}
From this condition, we then define two effective propagators, respectively for the local and the non-local sector:

\begin{equation}
G_k^{\text{L}}(\omega,\omega^\prime)= \beta \left(\bm{\mathcal{\tilde{A}}}^{-1}(\omega^2) \delta_{\omega,-\omega^\prime}-\bm{\mathcal{\tilde{A}}}^{-1}(0) \bm{B} \beta \bm{\mathcal{\tilde{A}}}^{-1}(0) \delta_{\omega,0} \delta_{\omega^\prime,0}\right)\,,
\end{equation}
and
\begin{equation}
G_k^{\text{N.L}}(\omega,\omega^\prime)= \beta^3\bm{\mathcal{\tilde{A}}}^{-1}(0) \bm{B} \bm{\mathcal{\tilde{A}}}^{-1}(0)\bm{B}  \bm{\mathcal{\tilde{A}}}^{-1}(0) \delta_{\omega,0} \delta_{\omega^\prime,0}\,.
\end{equation}
Then, denoting as $\Gamma_k^{\text{L}}$ and $\Gamma_k^{(\text{N.L})}$ respectively the local and non-local (bi-local) contributions to $\Gamma_k$ defined by the truncation \eqref{truncationS1}, we have from the exact flow equation \eqref{Wett} :

\begin{equation}
\dot{\Gamma}_k^{\text{L}}=\frac{1}{2}\int_{-\infty}^{+\infty} \frac{d\omega}{2\pi}  \Tr \dot{R}_k(\omega) G_{k}^{\text{L}}(\omega,-\omega)\,,\label{Wett2}
\end{equation}

\begin{equation}
\dot{\Gamma}_k^{\text{N.L}}=\frac{1}{2}\int_{-\infty}^{+\infty} \frac{d\omega}{2\pi} \Tr\dot{R}_k(\omega) G_{k,ii}^{\text{N.L}}(\omega,-\omega)\,,\label{Wett3}
\end{equation}
where trace here are over field component indices and replica. The derivation of the flow equations is quite standard from this step, and we sketch here only the general strategy for the main steps (see also \cite{lahoche2024Ward} for more details). Consider  For the non-local part of the equation, because the left hand side reads:
\begin{align}
\dot{\Gamma}_k^{\text{N.L}}=N \beta^2 \left(\dot{w}(k) \Big(\rho_1^{3}+\rho_2^{3}+2 \theta^3 \sqrt{\rho_1^3 \rho_2^3}\,\Big)+w(k) \Big(\rho_1^{3}+\rho_2^{3}+6\theta^2 \dot{\theta} \sqrt{\rho_1^3 \rho_2^3}\,\Big)\right)\,,
\end{align}
Then, setting $\rho_1=\rho_2=\kappa$, we obtain a first relation between $\dot{w}$ and $\dot{\theta}$, explicitly:
\begin{align}
&\nonumber \dot{w}(k)(1+\theta^3)+ 3 w(k) \theta^2 \dot{\theta}=\frac{k^5}{(1+4 (1-\theta) \bar{\bar{\kappa} }  \bar{v} )^3 (-2 \bar{q} +4 (\theta +1) \bar{\kappa}   \bar{v} +1)^3}\\\nonumber
&\times \frac{1}{\pi}\Big( 288 \bar{\kappa}  w^2 \big(\theta ^4-4 \left(\theta ^2-1\right)^2 \bar{q} ^3+6 \left(\theta ^2-1\right)^2 \bar{q} ^2 (4 (\theta +1) \bar{\kappa}   \bar{v} +1)\\\nonumber 
&-3 \left(\theta ^2-1\right)^2 \bar{q}  (4 (\theta +1) \bar{\kappa}   \bar{v} +1)^2+64 (\theta -1)^2 (\theta  (\theta  (\theta  (3 \theta +4)+6)+2)+1) \bar{\kappa}  ^3 \bar{v} ^3\\
&+48 \left(\theta ^6+\theta ^4-4 \theta ^3+\theta ^2+1\right) \bar{\kappa}  ^2 \bar{v} ^2+12 \left((\theta -2) \theta ^3+1\right) \bar{\kappa}   \bar{v} +1\big) \Big)\,.
\end{align}
Another independent relation between $\dot{w}$ and $\dot{\theta}$ can be estimated by taking derivatives with respect to $\sqrt{\rho_1}$ and $\sqrt{\rho_2}$ before setting $\rho_1=\rho_2=\kappa$, we have:
\begin{align}
\nonumber &  \theta^3 \dot{w}+ 3 w \theta^2  \dot{\theta}= -\frac{k^5}{\pi  \left(4 (\theta -1) \bar{\kappa } \bar{v}-1\right)^5 \left(2 \bar{q}-4 (\theta +1) \bar{\kappa } \bar{v}-1\right)^5}\\\nonumber
& \times 2304 \bar{w}^2 \bar{\kappa }^3 \Big(-12288 (\theta -1)^3 (\theta +1)^2 \Big(\theta  \Big(\theta  \Big(\theta ^2+\theta +12\Big)+8\Big)+8\Big) \bar{v}^7 \bar{\kappa }^7\\\nonumber
&+1024 (\theta -1)^2 \bar{v}^5 \Big(2 (\theta +1) \Big(\theta  (\theta  (\theta  (\theta  (10 \theta -13)-7)+64)+72)\\\nonumber
&+(\theta -1) (\theta  (\theta  (\theta  (\theta  (3 \theta +5)+33)+124)+111)+72) \bar{q}+72\Big) \bar{v}\\\nonumber
&+3 \Big(\Big(\theta ^2+1\Big)^2 \bar{q} (\theta -1)^3+\theta  (\theta  (\theta  (\theta  (\theta  (5 \theta +8)+21)+14)+13)+2)+1\Big) \bar{u}_4\Big) \bar{\kappa }^6\\\nonumber
&-256 (\theta -1) \bar{v}^4 \Big(\Big(-2 (\theta -1)^2 (\theta  (\theta  (\theta  (\theta  (3 (\theta -7) \theta +10)-152)-335)-371)-162) \bar{q}^2\\\nonumber
&+2 (\theta -1) (\theta  (\theta  (\theta  (\theta  (\theta  (3 \theta  (\theta +2)+106)-70)+132)+418)+605)+360) \bar{q}\\\nonumber
&+\theta  (\theta  (\theta  (\theta  (\theta  (3 \theta +119)-203)-155)+20)+360)+360\Big) \bar{v}\\\nonumber
&+15 (\theta -1) \Big(\theta  \Big(\theta  \Big(\theta  (\theta +1) \Big(\theta ^2+\theta +8\Big)+9\Big)+2\Big)+1\Big) \Big(2 \bar{q}-1\Big) \bar{u}_4\Big) \bar{\kappa }^5\\\nonumber
&+128 \bar{v}^3 \Big(15 \Big(2 \bar{q}-1\Big) \Big(\bar{q} (\theta +1)^5-\theta  (\theta  (\theta  (3 \theta +4)+6)+2)-1\Big) \bar{u}_4 (\theta -1)^2\\\nonumber
&+2 \Big((\theta -1)^3 \Big(\theta  \Big(\theta  \Big(5 \theta ^3+3 \theta +89\Big)+160\Big)+95\Big) \bar{q}^3\\\nonumber
&-(\theta -1)^2 (\theta  (\theta  (3 \theta  (\theta  (5 (\theta -6) \theta -7)-30)-197)-480)-335) \bar{q}^2\\\nonumber
&+(\theta -1) (\theta  (\theta  (\theta  (\theta  (3 \theta  (5 (\theta -3) \theta +58)-92)-50)-135)+265)+360) \bar{q}\\\nonumber
&-\theta ^2 \Big(\theta  \Big(\theta  \Big(\theta  \Big(5 \theta ^3+63 \theta -83\Big)+27\Big)-15\Big)+105\Big)+120\Big) \bar{v}\Big) \bar{\kappa }^4\\\nonumber
&-16 \bar{v}^2 \Big(30 \Big(4 \bar{q}^3 (\theta +1)^4-6 \bar{q}^2 (\theta +1)^4-\theta  (\theta  (\theta  (\theta +2)+4)+2)\\\nonumber
&+4 (\theta  (\theta +1) (\theta  (\theta +2)+3)+1) \bar{q}-1\Big) \bar{u}_4 (\theta -1)^2\\\nonumber
&+\Big(4 (\theta -1)^3 \Big(\theta  \Big(5 (\theta -1) \theta ^2+\theta +53\Big)+50\Big) \bar{q}^4
\end{align}

\begin{align}
&\nonumber +4 (\theta -1)^2 (\theta  (\theta  (\theta  (5 \theta  (3 \theta +16)+72)+98)+313)+310) \bar{q}^3\\\nonumber
&-4 (\theta  (\theta  (\theta  (\theta  (3 \theta  (5 \theta  (2 \theta -13)+19)+13)+390)-495)-315)+515) \bar{q}^2\\\nonumber
&+4 (\theta  (\theta  (\theta  (2 \theta  (\theta  (5 (\theta -22) \theta +57)-33)+307)-340)-105)+360) \bar{q}\\\nonumber
&+5 \theta ^2 (\theta  (\theta  (\theta  (44 \theta -27)+24)-61)+56)-360\Big) \bar{v}\Big) \bar{\kappa }^3\\\nonumber
&+4 \bar{v} \Big(2 \Big(4 (\theta -1)^3 \Big(\theta  (\theta -1)^2+4\Big) \bar{q}^5+4 (\theta -1)^2 (\theta  (\theta  (5 \theta  (2 \theta +5)+13)+47)+61) \bar{q}^4\\\nonumber
&+2 (\theta -1) (\theta  (\theta  (\theta  (5 \theta  (7 \theta +40)+219)-143)-50)+315) \bar{q}^3\\\nonumber
&-2 (\theta  (\theta  (\theta  (\theta  (5 \theta  (25 \theta +34)+73)-524)+137)+290)-345) \bar{q}^2\\\nonumber
&+5 (\theta  (\theta  (\theta  (\theta  (17 \theta  (2 \theta +1)+35)-121)+36)+29)-72) \bar{q}\\\nonumber
&-\theta ^2 \Big(\theta  \Big(34 \theta ^3+50 \theta -121\Big)+38\Big)+72\Big) \bar{v}\\\nonumber
&+15 (\theta -1) \Big(8 (\theta -1) (\theta +1)^3 \bar{q}^4-16 (\theta -1) (\theta +1)^3 \bar{q}^3+12 (\theta -1) (\theta +1)^3 \bar{q}^2\\\nonumber
&+(\theta  (\theta  (2-\theta  (3 \theta +8))+8)+5) \bar{q}+\theta  ((\theta -1) \theta -1)-1\Big) \bar{u}_4\Big) \bar{\kappa }^2+3 \bar{u}_4 \bar{\kappa }\\\nonumber
&+\Big(-10 \Big(\bar{q}-1\Big) \bar{q} \Big(2 \bar{q}-1\Big) \Big(2 \bar{q} \Big(\bar{q}+7\Big)-7\Big) \bar{v} \theta ^5\\\nonumber
&-\Big(2 \bar{q}-1\Big) \Big(3 \Big(4 \Big(\bar{q}-1\Big) \bar{q} \Big(2 \Big(\bar{q}-1\Big) \bar{q}+1\Big)+1\Big) \bar{u}_4\\\nonumber
&-4 \Big(4 \bar{q} \Big(2 \bar{q} \Big(\Big(\bar{q}-7\Big) \bar{q}+9\Big)-11\Big)+11\Big) \bar{v}\Big) \theta ^4\\\nonumber
&-3 \Big(2 \bar{q}-1\Big) \Big(4 \bar{q} \Big(\bar{q} \Big(2 \Big(\bar{q}-12\Big) \bar{q}+33\Big)-21\Big)+21\Big) \bar{v} \theta ^3\\\nonumber
&+8 \Big(\bar{q}-1\Big) \bar{q} \Big(2 \bar{q}-1\Big) \Big(13 \Big(2 \bar{q}-1\Big) \bar{v}+\Big(6 \Big(\bar{q}-1\Big) \bar{q}+3\Big) \bar{u}_4\Big) \theta ^2\\\nonumber
&+2 \Big(\bar{q}-1\Big) \bar{q} \Big(2 \bar{q}-1\Big) \Big(22 \bar{q}^2-86 \bar{q}+43\Big) \bar{v} \theta -8 \Big(2 \bar{q}-1\Big) \Big(4 \Big(\bar{q}-1\Big) \bar{q} \Big(\Big(\bar{q}-1\Big) \bar{q}+3\Big)+3\Big) \bar{v}\\\nonumber 
&-6 \bar{q} \Big(2 \bar{q} \Big(2 \bar{q} \Big(\bar{q} \Big(2 \bar{q}-5\Big)+5\Big)-5\Big)+3\Big) \bar{u}_4\Big) \bar{\kappa }\\
&+\Big(2 \theta ^2 \Big(\bar{q}-1\Big)-\bar{q}\Big) \Big(1-2 \bar{q}\Big)^2 \Big(\theta ^2 \Big(4 \bar{q}^2-2 \bar{q}+1\Big)-4 \Big(\bar{q}-1\Big) \bar{q}\Big)\Big)\,.
\end{align}

Moreover, interestingly, setting $\bar{q}\to 0$ and $ \bar{v}\to 0$, we get:
\begin{equation}
\dot{\theta}=\frac{\bar{w} \left(576 \bar{\kappa } \bar{w} \left(24 \bar{\kappa } \bar{u}_4-7\right)-5 \pi \right)}{\pi }\,.
\end{equation}

Other relations between couplings can be derived in the same way. For the local part (equation \eqref{Wett2}), we get the functional relation:
\begin{align}
\nonumber & k \frac{d}{d k} \bigg[\sum_{\alpha=1}^2 \bigg( \frac{{u}_4}{2} \big(\rho_1-\kappa\big)^2 + \frac{{u}_6}{3} \big(\rho_1-\kappa\big)^3 \bigg)+ v(k)\Big(\rho_1^2+\rho_2^2+2 \theta^2 \rho_1 \rho_2 \Big) \\\nonumber 
&- q\Big(\rho_1+\rho_2+2 \theta \sqrt{\rho_1 \rho_2} \Big)\bigg] = \\\nonumber 
&k \bigg(\frac{2 \left(-\bar{q}+\bar{u}_6 \left(\bar{\rho }_1-\bar{\kappa }\right){}^2+\bar{u}_4 \left(\bar{\rho }_1-\bar{\kappa }\right)+2 \left(\bar{\rho }_1+\bar{\rho }_2\right) \bar{v}+1\right)}{A(\bar{\rho}_1,\bar{\rho}_2)}\\\nonumber
& - \frac{12 \bar{w}}{B(\bar{\rho}_1,\bar{\rho}_2)} \bigg(  16 \theta ^3 \bar{\rho }_1^{3/2} \bar{\rho }_2^{3/2} \bar{v} \Big(\bar{q}+\bar{\kappa } \Big(\bar{u}_4-\bar{\kappa } \bar{u}_6\Big)-2 \bar{\rho }_2 \bar{v}-1\Big)\\\nonumber
&+2 \bar{\rho }_1 \bar{\rho }_2 \Big(-2 \theta ^2 \bar{q} \Big(\bar{q}+\bar{\kappa } \Big(\bar{u}_4-\bar{\kappa } \bar{u}_6\Big)-1\Big)+\bar{\rho }_2 \Big(\bar{q} \Big(2 \bar{\kappa } \bar{u}_6-\bar{u}_4+4 \theta ^2 \bar{v}\Big)\\\nonumber
&-\Big(\bar{u}_4-2 \bar{\kappa } \bar{u}_6\Big) \Big(\bar{\kappa } \Big(\bar{u}_4-\bar{\kappa } \bar{u}_6\Big)-1\Big)\Big)+4 \bar{\rho }_2^2 \bar{v} \Big(-2 \bar{\kappa } \bar{u}_6+\bar{u}_4+2 \theta ^2 \bar{v}\Big)\Big)\\\nonumber
&-8 \theta  \bar{\rho }_1^{5/2} \sqrt{\bar{\rho }_2} \bar{v} \Big(\bar{q}+2 \theta ^2 \bar{\rho }_2 \Big(-2 \bar{\kappa } \bar{u}_6+\bar{u}_4+2 \bar{v}\Big)\Big)\\\nonumber
&+2 \bar{\rho }_1^3 \Big(2 \theta ^2 \bar{\rho }_2 \Big(\bar{q} \bar{u}_6+4 \bar{v}^2\Big)-\Big(\bar{q}+\bar{\kappa } \Big(\bar{u}_4-\bar{\kappa } \bar{u}_6\Big)-1\Big) \Big(-2 \bar{\kappa } \bar{u}_6+\bar{u}_4+4 \bar{v}\Big)\\\nonumber
&+\bar{\rho }_2^2 \bar{u}_6 \Big(\bar{u}_4-2 \bar{\kappa } \bar{u}_6\Big)\Big)+\bar{\rho }_1^2 \Big(2 \bar{q} \Big(\bar{\kappa } \bar{u}_4-\bar{\kappa }^2 \bar{u}_6-\bar{\rho }_2^2 \bar{u}_6+2 \theta ^2 \bar{\rho }_2 \Big(-2 \bar{\kappa } \bar{u}_6+\bar{u}_4+2 \bar{v}\Big)-1\Big)\\\nonumber
&+2 \bar{q}^2+\bar{\rho }_2^2 \Big(6 \bar{\kappa }^2 \bar{u}_6^2+2 \bar{u}_6 \Big(1-3 \bar{\kappa } \bar{u}_4\Big)+\bar{u}_4^2\Big)+\Big(\bar{\kappa }^2 \bar{u}_6-\bar{\kappa } \bar{u}_4+1\Big){}^2\\\nonumber
&+8 \bar{\rho }_2^3 \bar{u}_6 \bar{v}\Big)+\bar{\rho }_2^2 \Big(-2 \bar{q} \Big(\bar{\kappa }^2 \bar{u}_6-\bar{\kappa } \bar{u}_4+4 \bar{\rho }_2 \bar{v}+1\Big)+2 \bar{q}^2+\Big(\bar{\kappa }^2 \bar{u}_6-\bar{\kappa } \bar{u}_4+4 \bar{\rho }_2 \bar{v}+1\Big){}^2\Big)\\\nonumber
&+\bar{\rho }_1^4 \Big(\bar{u}_6 \Big(-2 \bar{q}+\bar{\rho }_2^2 \bar{u}_6+2\Big)+6 \bar{\kappa }^2 \bar{u}_6^2-2 \bar{\kappa } \bar{u}_6 \Big(3 \bar{u}_4+8 \bar{v}\Big)+\Big(\bar{u}_4+4 \bar{v}\Big){}^2\Big)\\
&-8 \theta  \bar{q} \sqrt{\bar{\rho }_1} \bar{\rho }_2^{5/2} \bar{v}+\bar{\rho }_1^6 \bar{u}_6^2-16 \theta ^3 \bar{\rho }_1^{7/2} \bar{\rho }_2^{3/2} \bar{u}_6 \bar{v}+2 \bar{\rho }_1^5 \bar{u}_6 \Big(-2 \bar{\kappa } \bar{u}_6+\bar{u}_4+4 \bar{v}\Big)\bigg)\bigg)\,,
\end{align}

where:
\begin{align}
\nonumber & A(\bar{\rho}_1,\bar{\rho}_2):=\pi  \Big(\Big(-\bar{q}+\bar{u}_6 \Big(\bar{\rho }_1-\bar{\kappa }\Big){}^2+\bar{u}_4 \Big(\bar{\rho }_1-\bar{\kappa }\Big)+4 \bar{\rho }_1 \bar{v}+1\Big)\\
&\times  \Big(-\bar{q}+\Big(\bar{\rho }_1-\bar{\kappa }\Big) \Big(\bar{u}_6 \Big(\bar{\rho }_1-\bar{\kappa }\Big)+\bar{u}_4\Big)+4 \bar{\rho }_2 \bar{v}+1\Big)-\Big(\bar{q}-4 \theta  \sqrt{\bar{\rho }_1} \sqrt{\bar{\rho }_2} \bar{v}\Big){}^2\Big)\,,
\end{align}

and 
\begin{align}
\nonumber & B(\bar{\rho}_1,\bar{\rho}_2):=4 \pi  \Big(-2 \bar{q} \Big(\Big(\bar{\kappa }-\bar{\rho }_1\Big) \Big(\bar{u}_6 \Big(\bar{\kappa }-\bar{\rho }_1\Big)-\bar{u}_4\Big)-4 \theta  \sqrt{\bar{\rho }_1} \sqrt{\bar{\rho }_2} \bar{v}+2 \bar{\rho }_1 \bar{v}+2 \bar{\rho }_2 \bar{v}+1\Big)\\\nonumber
&+4 \bar{\rho }_2 \bar{v} \Big(\Big(\bar{\kappa }-\bar{\rho }_1\Big) \Big(\bar{u}_6 \Big(\bar{\kappa }-\bar{\rho }_1\Big)-\bar{u}_4\Big)-4 \Big(\theta ^2-1\Big) \bar{\rho }_1 \bar{v}+1\Big)\\
&+\Big(\Big(\bar{\kappa }-\bar{\rho }_1\Big) \Big(\bar{u}_6 \Big(\bar{\kappa }-\bar{\rho }_1\Big)-\bar{u}_4\Big)+1\Big) \Big(\Big(\bar{\kappa }-\bar{\rho }_1\Big) \Big(\bar{u}_6 \Big(\bar{\kappa }-\bar{\rho }_1\Big)-\bar{u}_4\Big)+4 \bar{\rho }_1 \bar{v}+1\Big)\Big){}^2
\end{align}

All the remaining coupling constant flows can be derived from this equation as follow. First, setting $\rho_1=\rho_2=\rho$ in both sides, we obtain three independent relations taking first, second and third derivatives with respect to $\rho$, before setting $\rho=\kappa$. Finally, another relation can be obtained by taking derivative with respect to $\sqrt{\rho}_1$ one time and derivative with respect to $\sqrt{\rho}_2$ one time, before setting $\rho=\kappa$. Finally, deriving once time more with respect to $\sqrt{\rho}_1$ before setting $\rho=\kappa$ provides the last relation we need to extract all the flow equations. We summarized our conclusions on Figures \ref{Figsum1}, \ref{Figsum2} and \ref{Figsum3} below, where we defined $\theta=:\cos(\eta)$, for initial conditions:

\begin{equation}
S:=\{\bar{\kappa}(0)=0.1,\bar{u}_4(0)=0.01,\bar{u}_6(0)=0.008\}\,.
\end{equation}
\medskip

For aligned replica configurations, we recover essentially the behavior we observed in the rest of this paper. The disorder enforces finite scale singularities, making contact with the notion of Larkin time. As explained many time in this paper, we expect these singularities arises because interactions forbidden by perturbation theory are associated with metastable states, which dominate the flow near the singularity. This assumption was investigated in our previous work \cite{lahoche2024Ward}, and we showed explicitly that taken into account such an interaction like $q$ or $v$ suppresses the singularities for some trajectories. We recover essentially this result here, as shown explicitly on Figure \ref{Figsum1}. Figures moreover shows the behavior of $q$ and $v$, reached a finite value reminiscent of a ‘‘stable equilibrium" for long times. There is however another parameter, the cosine between replica $\theta$, which plays a significant role. As $\theta=0$ (aligned replica solutions), the overlap $q$ between replica becomes very huge, typically, for $\bar{w}(0)=-0.1$, and $\eta(0)=0.001$, we get  $\bar{q}(0)\simeq -10^{4}$ as $\bar{v}(0)\simeq 1$. Note that the crucial role of the coupling $v$, already pointed out in \cite{lahoche2024Ward} seems to indicate that the true order parameter is not only the $2$-point function in that case. In contrast, for $\eta(0)=0.9$ for instance, convergence is improved, finite scale singularities arise also for small disorder $\bar{w}(0)\simeq -10^{-5}$, but can be removed for a large enough $-\bar{q}(0) \geq 0.0975$ (keeping $\bar{v}(0)=0$ in that case). Interestingly, the effective angle $\eta(k)$ reach a finite value, close to $\pi/2$ (but not equal to it). This behavior, observed for different initial values seems to indicate the existence of a fixed direction for the RG flow, favoring orthogonal configurations for long times. This, physically agree with the intuition that, as the overlap is small, replica are essentially independent, and the average cosine is almost zero. The disagreement between the asymptotic value for $\eta(k)$ and $\pi/2$ is interpreted at this stage as a consequence of quantum fluctuations. 
Finally, note that, if we start close to $\pi/2$, for $\eta(0)=\pi/2-0.01$ for instance, the flow converge for very larges values of $\bar{w}(0)$, typically $\bar{w}(0) \sim -10$, and the initial value for $\bar{q}(0)$ is very small. For these initial conditions, the critical value is  $\bar{w}(0) \sim -13$, and the convergence is resorted for $-\bar{q}(0)\geq 0.0005$. 
\medskip

To summary:

\begin{itemize}
    \item On Figure \ref{Figsum1} we show the behavior of different couplings for $\eta(0)=0.9$ just above and below the critical value $\bar{q}_c(0)\approx-0.0975$. At this stage, note that 1) the critical value depends weakly on the initial conditions, but divergences can be recovered for higher values, revealing the complicate structure of the phase space. We return on a numerical construction of it in a future work, here we are aiming to prove a concept only. 
\item On Figure \ref{Figsum2} we show the behavior of $v(k)$ and $q(k)$ for the non singular trajectories showed on Figure \ref{Figsum1}. Couplings converge toward a finite value, and we have to notice that the asymptotic value for $q(k)$ is far enough from the perturbation theory result of the section \ref{2PI}, pointing out maybe the non-perturbative nature of the transition.

\item On Figure \ref{Figsum2} we show the behavior for convergent trajectories, illustrating the fact that asymptotic averaging cosine is almost zero i.e. typically $\cos(\eta) \approx 0.02$ for the figure on left. We expect that the cosine does not vanishes exactly because of quantum fluctuations. 
\end{itemize}

\begin{figure}
\begin{center}
\includegraphics[scale=0.55]{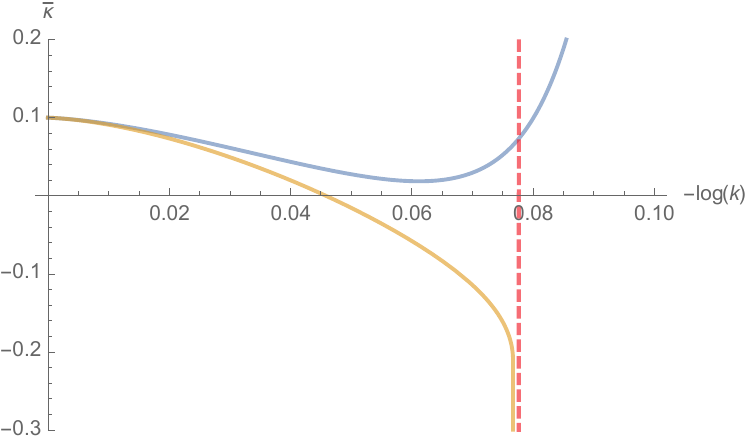}\quad \includegraphics[scale=0.55]{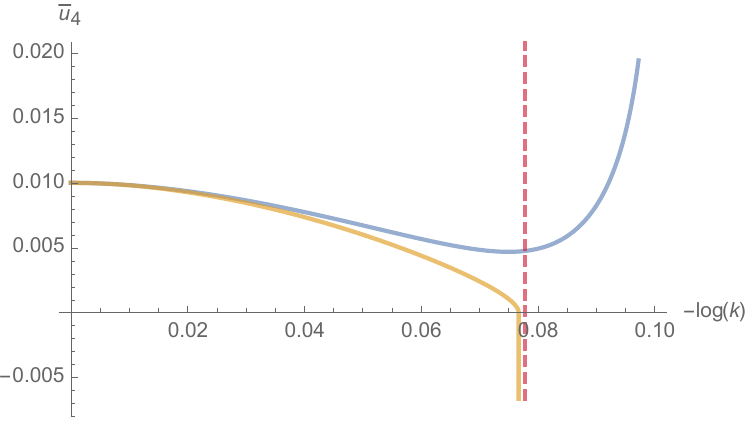}
\end{center}
\caption{Behavior of the RG flow for $\bar{w}(0)=-0.00001$ for $\bar{v}(0)=0$, $\eta(0)=0.9$ and $\bar{q}(0)=-0.097$ (yellow curve) and $\bar{q}(0)=-0.099$ (blue curve).}\label{Figsum1}
\end{figure}

\begin{figure}
\begin{center}
\includegraphics[scale=0.55]{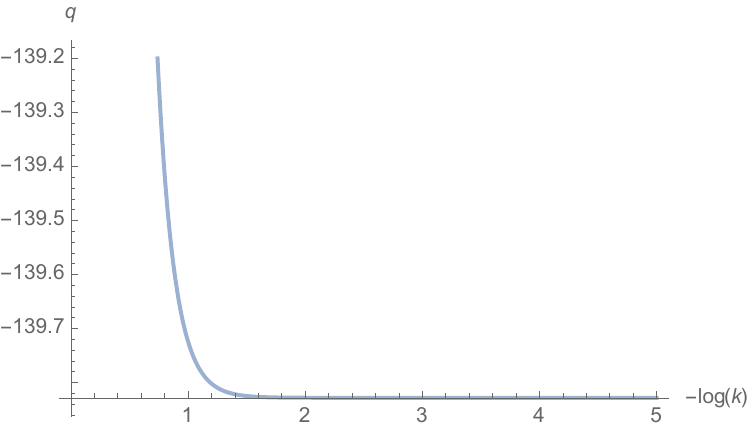}\quad \includegraphics[scale=0.55]{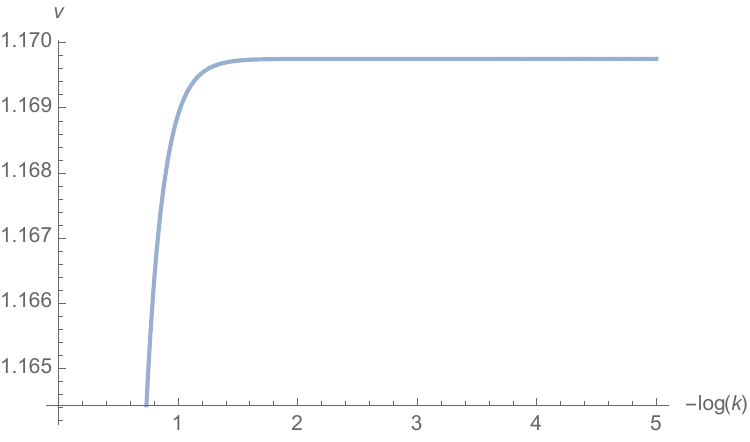}
\end{center}
\caption{Evolution of couplings $q(k)$ and $v(k)$.}\label{Figsum2}
\end{figure}

\begin{figure}
\begin{center}
\includegraphics[scale=0.53]{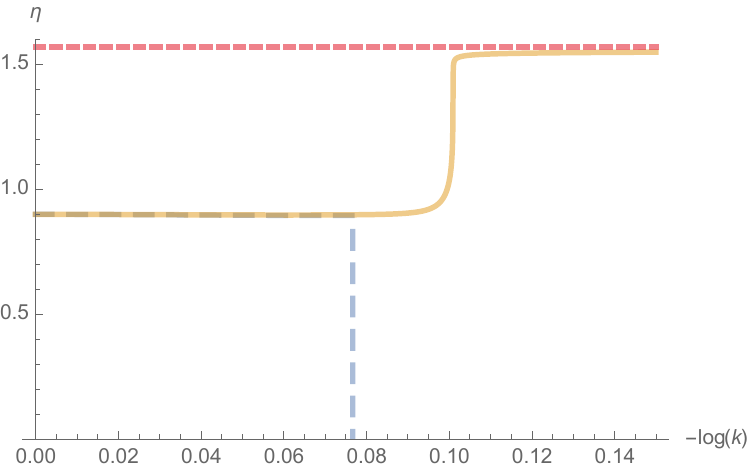}\quad \includegraphics[scale=0.58]{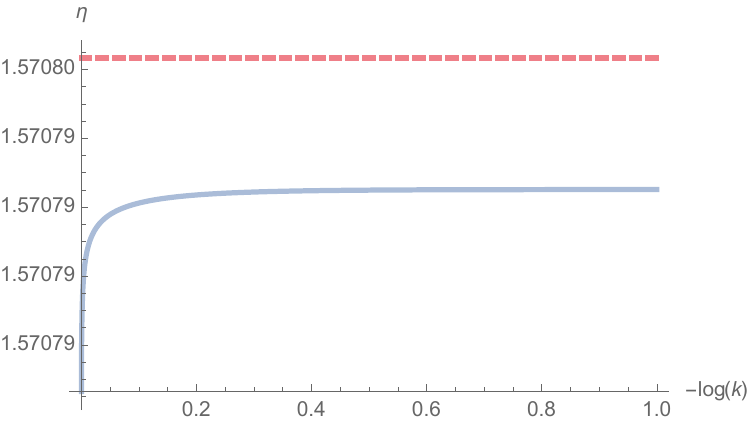}
\end{center}
\caption{Evolution of the effective angle $\eta(k)$. On left, for the initial conditions considered above in Figure \ref{Figsum1} for $\bar{q}=-0.097$ (blue dashed curve) and $\bar{q}(0)=-0.099$ (yellow curve). On the right, behavior of $\eta(k)$ for initial conditions $\eta(0)=\pi/2-0.001$ and $\bar{q}(0)=-0.0005$. Dashed red curve is for the value $\pi/2$.}\label{Figsum3}
\end{figure}

\begin{figure}
\begin{center}
\includegraphics[scale=0.53]{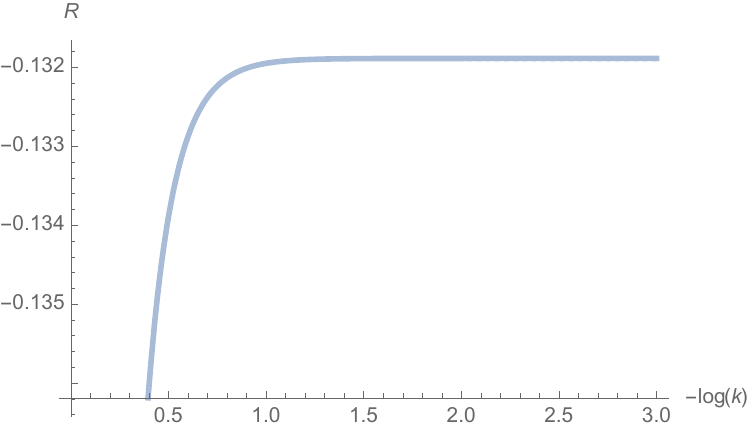}\quad \includegraphics[scale=0.58]{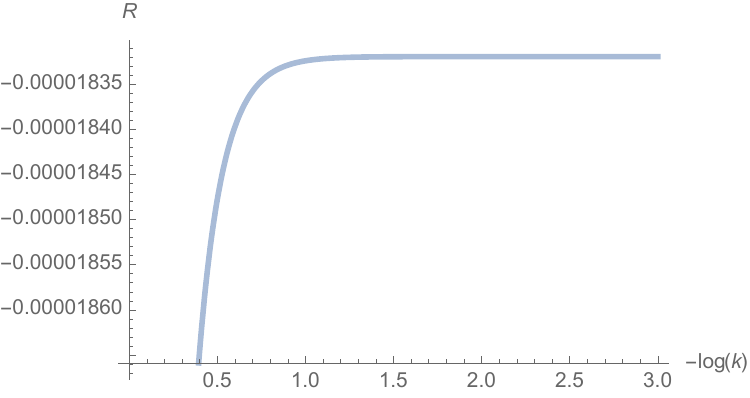}
\end{center}
\caption{Behavior of $R(k)$ for the initial condition $\eta(0)=0.9$ (on left) and for $\eta(0)=\pi/2-0.01$ (on right).}\label{figR}
\end{figure}

To conclude, our hypothesis concerning the relationship between the intensity of the overlapping and the vanishing of the mean cosine seems to be well verified by the initial conditions, but contradicted by the experiment concerning the asymptotic values, since we find $q(\infty)\sim -139.8$ for $\eta(0) = 0.9$ (see Figure \ref{Figsum2}) and $q(\infty)=-3430$ for $\eta(0)=\pi/2-0.01$. Note that the notation $\infty$ here designates the asymptotic value (the deep infrared). It should be noted, however, that these values have no intrinsic meaning, but must be compared to the typical scale of fluctuations in the IR, given by the mass scale. The effective mass of the fluctuations is of the order of $\sim u_4 \kappa$, and the relevant quantity is therefore the ratio $R(k):=q(k)/(u_4 \kappa)$, which is dimensionless. The results are reported in Figure \ref{figR}, and support our hypothesis: the asymptotic value of the ratio $R(\infty)$ is much smaller when the mean of the cosine is small. Numerically we have $\cos(\eta(\infty))\approx 0.6$ for $\eta(0)=0.9$ and $\cos(\eta(\infty))\approx 3.82\times 10^{-6}$ for $\eta(0)=\pi/2-0.01$.

\section{Enhanced vertex expansion and equilibrium ergodicity breaking}\label{sectionenhanced}

In this part, we will study the flow in the symmetric phase through vertex expansion. In the symmetric phase, the mean value of the field could be set as $M=0$. Here we take the case of $p=3$ with local and bi-local $2$-, $4$-, and $6$-point interaction. 
\medskip 

In this section, we will consider the equivalent truncation we considered in the previous section, but in the symmetric phase. The same truncation has been considered moreover in our previous work \cite{lahoche2024Ward}. The kinetic action is assumed of the form \eqref{ansatz2PI} with $q_{\alpha\beta}=-q$, and we consider the sextic truncation:

\begin{align}
\nonumber \Gamma_k[\{\textbf{M}_\alpha \}]&=\frac{1}{2}\int dt \sum_{i,\alpha,\beta} M_{i,\alpha}(t)[(-\partial_t^2+p_\mu^2+u_2)\delta_{\alpha\beta}-q]M_{i,\beta}(t)\\\nonumber
&+\sum_{n=2}^3\int dt \sum_{\mu,\alpha} \frac{(2\pi)^{n-1}u_{2n}}{(2n)!N^{n-1}}\, \bigg(\sum_\mu M_{\mu,\alpha}^2(t) \bigg)^n\\\nonumber 
&+\int dt \sum_{\mu,\alpha} \frac{(2\pi)v_{4,1}}{4N}\,\sum_{\alpha,\beta} \bigg(\sum_{i} M_{i,\alpha}(t)M_{i,\beta} \bigg)^2\\
&+\frac{(2\pi)\tilde{u}_6}{6!N^2}\int dt dt^\prime \sum_{\alpha,\beta}\, \bigg(\sum_\mu M_{\mu,\alpha}(t) M_{\mu,\beta}(t^\prime) \bigg)^3\,,\label{truncationGamma2}
\end{align}
We recall the explicit expression of the effective $2$-point function:
\begin{equation}
G_{k,\alpha\beta}(\omega)= \underbrace{\vcenter{\hbox{\includegraphics[scale=1]{locPot.pdf}}}}_{\frac{\delta_{\alpha\beta}}{\omega^2+u_2(k)+R_k(\omega)}}\,+\,\underbrace{\vcenter{\hbox{\includegraphics[scale=1]{NlocPot.pdf}}}}_{\frac{q}{(\omega^2+u_2(k)+R_k(\omega))(\omega^2+u_2(k)-n q+R_k(\omega))}}\,,
\end{equation}
and we will denote respectively as $G_{k,\alpha\beta}^{(0)}(\omega)$ and $G_{k,\alpha\beta}^{(1)}(\omega)$ the two components. The interaction part of the effective average action moreover reads, graphically:
\begin{equation}
\Gamma_{k,\text{int}}\,=\,\vcenter{\hbox{\includegraphics[scale=0.8]{V4.pdf}}}\,+\,\vcenter{\hbox{\includegraphics[scale=0.8]{V6.pdf}}}
\,+\,\vcenter{\hbox{\includegraphics[scale=0.8]{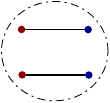}}}\,+\,\vcenter{\hbox{\includegraphics[scale=0.8]{V33.pdf}}}\,.
\end{equation}
Using the graphical conventions we defined in the previous section, we obtain the flow equations:

\begin{align}
\dot{u}_2\,=\,\vcenter{\hbox{\includegraphics[scale=0.7]{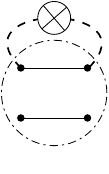}}}\,+\, \vcenter{\hbox{\includegraphics[scale=0.7]{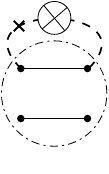}}}\,+\, \vcenter{\hbox{\includegraphics[scale=0.7]{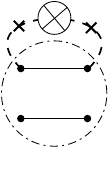}}}\,+\, \vcenter{\hbox{\includegraphics[scale=0.7]{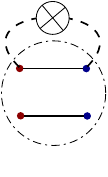}}}\,,
\end{align}

\begin{equation}
\dot{q}^\prime\,=\, \vcenter{\hbox{\includegraphics[scale=0.7]{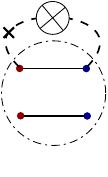}}}\,+\,\vcenter{\hbox{\includegraphics[scale=0.7]{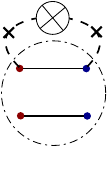}}}
\end{equation}

\begin{align}
\nonumber\dot{u}_4&\,=\,\vcenter{\hbox{\includegraphics[scale=0.7]{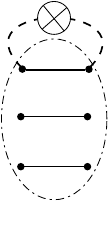}}}\,+\,\vcenter{\hbox{\includegraphics[scale=0.7]{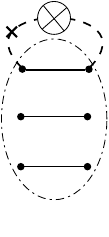}}}\,+\, \vcenter{\hbox{\includegraphics[scale=0.7]{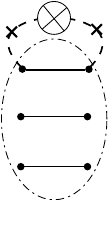}}}\,+\,\vcenter{\hbox{\includegraphics[scale=0.7]{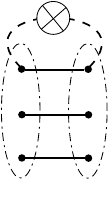}}}\,+\,\vcenter{\hbox{\includegraphics[scale=0.7]{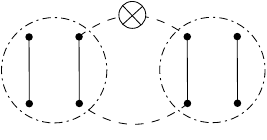}}}\,+\,\vcenter{\hbox{\includegraphics[scale=0.7]{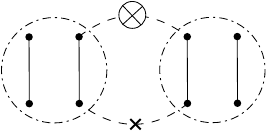}}}\\
&\,+\,\vcenter{\hbox{\includegraphics[scale=0.7]{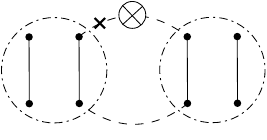}}}\,+\,\vcenter{\hbox{\includegraphics[scale=0.7]{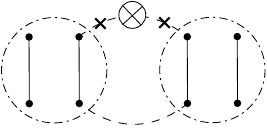}}}\,+\,\vcenter{\hbox{\includegraphics[scale=0.7]{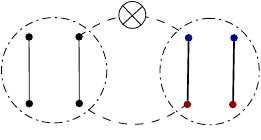}}}\,,
\end{align}

\begin{align}
\dot{v}_{4,1}&\,=\,\vcenter{\hbox{\includegraphics[scale=0.7]{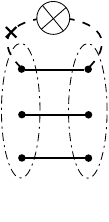}}}\,+\,\vcenter{\hbox{\includegraphics[scale=0.7]{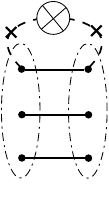}}}\,+\,\vcenter{\hbox{\includegraphics[scale=0.7]{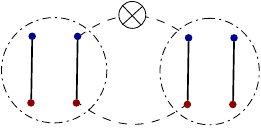}}}\,,
\end{align}

\begin{align}
\dot{w}_{6,1}&\,=\,0\,,
\end{align}

\begin{align}
\nonumber \dot{u}_{6}&\,=\,\vcenter{\hbox{\includegraphics[scale=0.7]{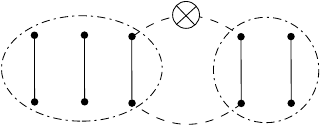}}}\,+\,\vcenter{\hbox{\includegraphics[scale=0.7]{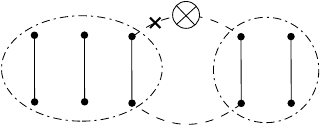}}}\,+\,\vcenter{\hbox{\includegraphics[scale=0.7]{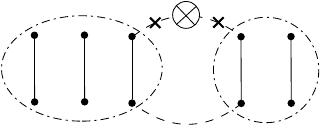}}}\\\nonumber
&\,+\,\vcenter{\hbox{\includegraphics[scale=0.7]{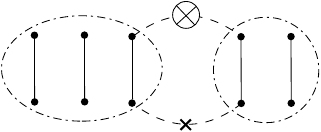}}}\,+\,\vcenter{\hbox{\includegraphics[scale=0.7]{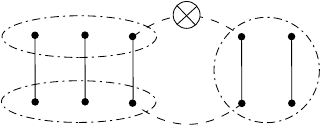}}}\,+\,\vcenter{\hbox{\includegraphics[scale=0.7]{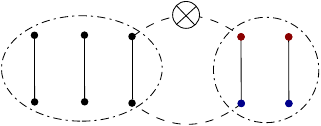}}}\\\nonumber
&\,+\,\vcenter{\hbox{\includegraphics[scale=0.6]{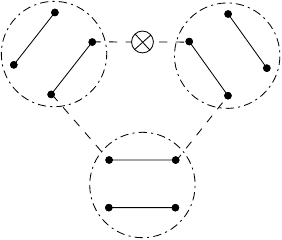}}}\,+\,\vcenter{\hbox{\includegraphics[scale=0.6]{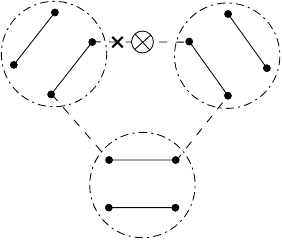}}}\,+\,\vcenter{\hbox{\includegraphics[scale=0.6]{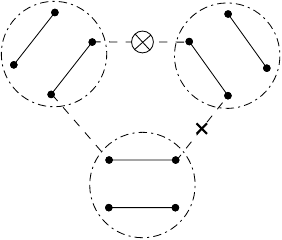}}}\\
&\,+\,\vcenter{\hbox{\includegraphics[scale=0.6]{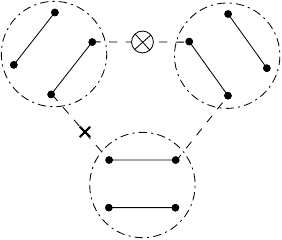}}}\,+\,\vcenter{\hbox{\includegraphics[scale=0.6]{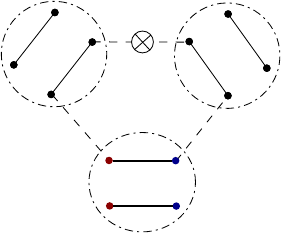}}}\,.
\end{align}

The computation of the different diagrams require the integral:
\begin{align}
I_{n,p}:=  \int \frac{d\omega}{2 \pi}\,\dot{R}_k(\omega)  (G_{k}^{(0)})^p (G_{k}^{(1)})^{n-p} \,,
\end{align}
which can be computed explicitly:
\begin{equation}
I_{n,p}(k,u_2,q)=  \frac{1}{2\pi} \frac{4 k^3}{(k^2+u_2)^p(k^2+u_2-n q)^{n-p}}\,,
\end{equation}
and the flow equations writes explicitly as:
\begin{equation}
\dot{\bar{u}}_2=-2\bar{u}_2-\bar{u}_4 (\bar{I}_{2,2}+2\bar{I}_{2,1}+n\bar{I}_{2,0})-\frac{2\bar{v}_{4,1}}{\pi }\frac{1}{(1+\bar{u}_2)^{2}}\,,
\end{equation}
\begin{equation}
\dot{\bar{q}}^\prime=-2 \bar{q}^\prime - \bar{v}_{4,1} (2\bar{I}_{2,1}+n\bar{I}_{2,0})
\end{equation}
\begin{align}
\nonumber \dot{\bar{u}}_4=&-3\bar{u}_4 - \frac{2\bar{\tilde{u}}_6 }{\pi}\frac{1}{(1+\bar{u}_2)^2}-2\bar{u}_6 (\bar{I}_{2,2}+2\bar{I}_{2,1}+n\bar{I}_{2,0})+32 \frac{\bar{v}_{4,1} \bar{u}_4}{\pi} \frac{1}{(1+\bar{u}_2)^{3}} \\
&+8\bar{u}_4^2 (\bar{I}_{3,3}+3\bar{I}_{3,2}+n\bar{I}_{3,1})\,.
\end{align}
\begin{align}
\dot{\bar{v}}_{4,1}&=-3{\bar{v}}_{4,1}-\frac{2\bar{\tilde{u}}_6 }{\pi(1+\bar{u}_2)^2}\left[\frac{2\bar{q}}{(1+\bar{u}_2+n\bar{q})}+\frac{n(\bar{q})^2}{(1+\bar{u}_2+n \bar{q})^2}\right]+\frac{16\bar{v}_{4,1}^2}{\pi} \frac{1}{(1+\bar{u}_2)^{3}}
\end{align}
\begin{align}
\nonumber\dot{\bar{{u}}}_6&=-4\, {\bar{{u}}}_6+6\bar{u}_4\bar{u}_6 (\bar{I}_{3,3}+3\bar{I}_{3,2}+n\bar{I}_{3,1})+\frac{6}{\pi} \frac{\bar{u}_4\bar{\tilde{u}}_6}{(1+\bar{u}_2)^{3}}+\frac{12}{\pi} \frac{\bar{v}_{4,1}\bar{u}_6}{(1+\bar{u}_2)^{3}}\\
&-24\bar{u}_4^3 (\bar{I}_{4,4}+4\bar{I}_{4,3})-\frac{144}{ \pi} \frac{\bar{u}^2_4\bar{v}_{4,1}}{(1+\bar{u}_2)^{4}}\,,
\end{align}
and:
\begin{equation}
\dot{\bar{\tilde{u}}}_6=\, -5{\bar{\tilde{u}}}_6\,.
\end{equation}
where $\bar{I}_{n,p}(\bar{u}_2,\bar{q}):=I_{n,p}(1,\bar{u}_2,\bar{q})$. 

\begin{figure}
\begin{center}
\includegraphics[scale=0.5]{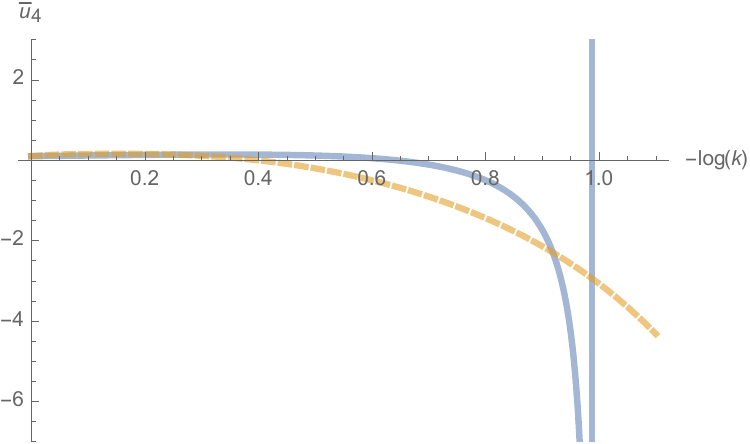}\qquad \includegraphics[scale=0.5]{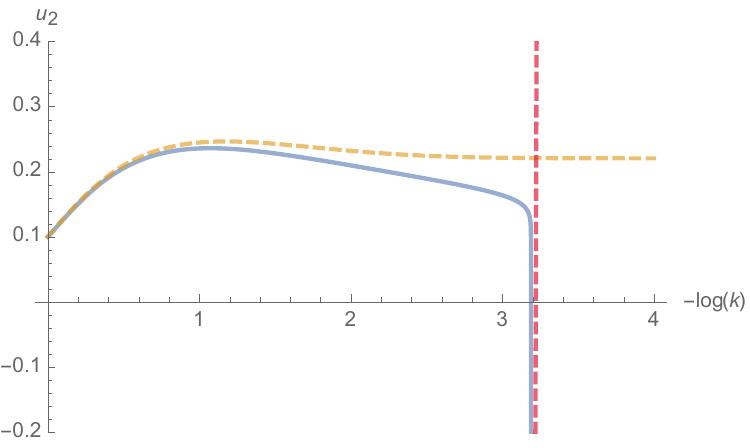}
\end{center}
\caption{Behavior of the RG flow for $q=0$ (blue curve) and $q\neq 0$ (dashed curve).}\label{figflowImproved}
\end{figure}

\begin{figure}
\begin{center}
\includegraphics[scale=0.5]{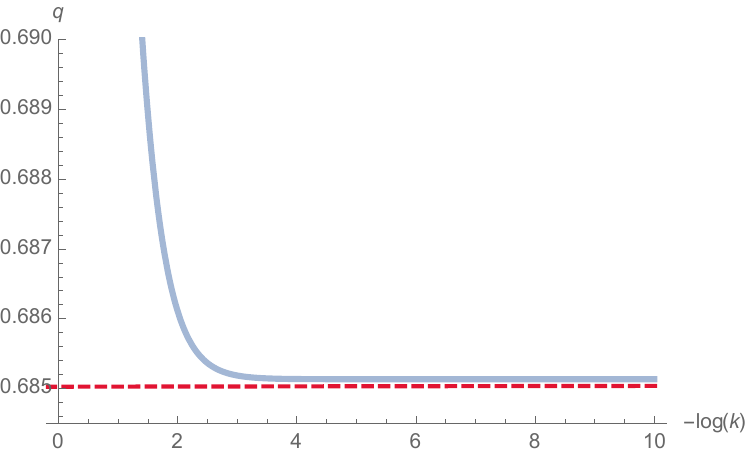}\quad \includegraphics[scale=0.5]{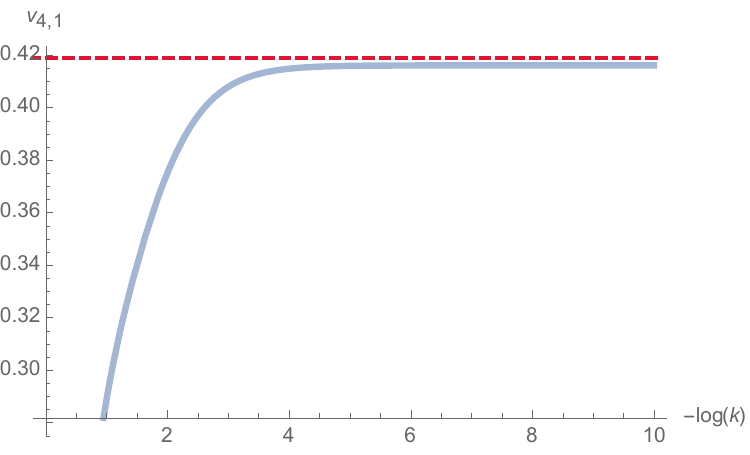}
\end{center}
\caption{Behavior of $q(k)$ and $v_{4,1}(k)$ for $q(0)=0.7$.}\label{figflowImproved2}
\end{figure}

The results are summarized on the Figures \ref{figflowImproved} and \ref{figflowImproved2}. We considered the RG flow for $n=2$, and the initial conditions:
\begin{equation}
S_0:=(\bar{u}_2(0)=\bar{u}_4(0)=\bar{u}_6(0)=0.1)\,,
\end{equation}
respectively for $\bar{q}(0)=0$ (blue curve for Figure \ref{figflowImproved}),
$\bar{q}(0)=0.7$ (yellow dashed curve on left) and $\bar{q}(0)=0.01$ (yellow dashed curve on right for Figure \ref{figflowImproved}). In all cases we set $\bar{v}_{4,1}(0)=0$. The results are again similar which what we obtained in the previous section in the broken phase; they moreover are quite similar also with the results obtained in \cite{lahoche2024Ward} and the reader may consult this reference for an extended discussion. The main features are the following:

\begin{enumerate}
    \item A non-vanishing initial value for $q$ cancel the finite scale singularity for some (but not all) initial conditions for other couplings. 
    \item The coupling $q$ converge toward a finite value (scaling regime) in the deep IR. 
\end{enumerate}

The last point evokes a phase transition, a convergence of the initial instability towards a non-zero vacuum. Finally, it should be noted that there is indeed a threshold from which the divergence disappears, when $q(0)$ is large enough, but that divergences can reappear for larger values. This seems to reveal part of the complexity of the phase space, as well as the limits of our approximations, which neglect a significant number of essential effects. This had also been observed in our previous work. It should be noted at this stage that taking into account other effects, dynamic rather than static ergodicity breaks, notably breaking the invariance by translation in time are likely to play a role, either complementary or superimposed on the effects that we have studied here. The reader is referred to our recent preliminary work on this subject \cite{achitouv2024timetranslationinvariancesymmetrybreaking}.

\section{Conclusion and Outlooks}\label{sec6}

In this paper we investigated the behavior of the perturbative RG flow of a $N$-dimensional quantum particle moving on a rugged landscape materialized with a Gaussian random tensor. This work follows \cite{lahoche2024Ward,lahoche2024largetimeeffectivekinetics,achitouv2024timetranslationinvariancesymmetrybreaking}, where we considered the RG for a $2+p$ quantum spin glass dynamics, from a coarse-graining over the matrix like disorder spectrum, which becomes deterministic in the large $N$ limit. The corresponding field theory looks at a standard 4D field theory in the IR, and our main results can be summarized as follows:
\begin{enumerate}
    \item Finite scale singularities arises from the naive truncation scheme based on large $N$ perturbation theory arguments, as the rank $p$ disorder is strong enough. 
    \item The effective potential for $2$-points (or more) interactions forbidden by perturbation theory has a second stable minimum (adding from the vanishing vacuum).
    \item Taking into account such an interaction avoids finite scale singularities, seeming to indicate that dynamics is dominated by meta-stables state for long range physics.
\end{enumerate}

In this paper, we considered a model with a unique disorder, of rank $p$, and constructed a coarse-graining on the frequencies. This is a field theory in dimension $1$, whose interactions are all relevant according to the dimensional analysis, which complicates the construction of approximations \cite{synatschke2009flow}, in particular for a non-local theory \cite{lahoche2021functional}. However, and by different approximations, we were able to gather evidence showing a strange similarity with the conclusions obtained in the previous case, based on a coarse-graining on the Wigner spectrum. In particular, we were able to observe in this case also the presence of singularities with a finite time scale, also analogous to those usually associated with the Larkin scale \cite{Tarjus,Tarjus2,dupuis2020bose}. Moreover, again, these singularities can be suppressed by interactions breaking time translation symmetry for the $2$-point functions, and/or imply correlations between replica. These interactions forbidden by the large $N$ perturbation theory correspond to meta-stable states dominating the long time dynamics. Beyond the disturbing similarity between the two approaches considered in our work, the results of this article shed light on the underlying physics, and allow us to glimpse a numerical reconstruction of the phase space of the system, by studying the phase transitions along the trajectories. This reconstruction, however, does not yet elucidate the different mechanisms at the origin of the appearance of its correlations, and in particular the way in which the quantum tunneling effect competes with disorder.

\medskip 

\noindent
\section*{Acknowledgment:} Vincent Lahoche and the authors warmly thank Mrs. Dalila Derdar for these very stimulating and inspiring exchanges from the first stages of the realization of this work. Without these highly stimulating exchanges, this work would not have been the same.

\pagebreak

\appendix

\section{Closed equations for self energy in a few words}\label{App1}

Let us review shortly the derivation of the so-called closed equation for self energy arising in the large $N$ limit for $O(N)$ models, including non-local vertices. A derivation was proposed in our previous work \cite{lahoche2024largetimeeffectivekinetics}, we propose here another way. The reader can also consult the references \cite{Lahoche:2021tyc} on this topic. 
\medskip

Let us start by the case $p=2$. We discussed the structure of leading order vacuum graphs in subsection \ref{EVE}, and we recalled the standard result that they look as planar (unrooted) trees in the LVR. A leading order contribution to the self energy can be obtained by deleting one of the leaf, which can be hooked to a red of blue edge. Two examples of typical rooted trees we obtain by this way are pictured on Figure \ref{fig2points1}. As explained in section \ref{EVE}, the rooted trees hooked to the root vertex $v_0$ are components of the expansion series of the self energy $\Sigma$, and because the corners are nothing but bare propagator edges $C$, what we reconstruct by summing all the diagrams is nothing but the Dyson series for the effective $2$-point function. Then, formally, the two allowed configurations for the root lead to, graphically\footnote{We assume that signs and numerical factors are included in the diagrams.}:

\begin{equation}
\Sigma= \, \vcenter{\hbox{\includegraphics[scale=0.8]{V41eff1.pdf}}}\,+\,  \vcenter{\hbox{\includegraphics[scale=0.8]{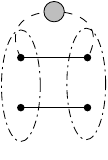}}}\,.
\end{equation}

For the case $p=3$, and including local sextic interactions in the bare action, we have to distinguish three cases, depending if the deleted leaf is hooked with a black, blue or red edge root edge:

\begin{align}
\nonumber \Sigma&=\, \Bigg\{\cdots\vcenter{\hbox{\includegraphics[scale=0.4]{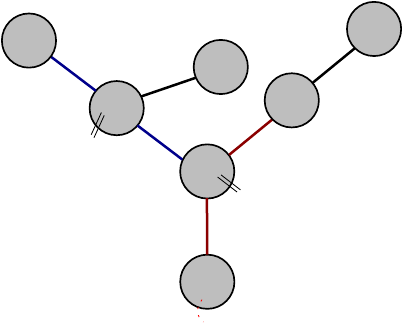}}}\cdots \Bigg\}\,+\,  \Bigg\{\cdots \vcenter{\hbox{\includegraphics[scale=0.4]{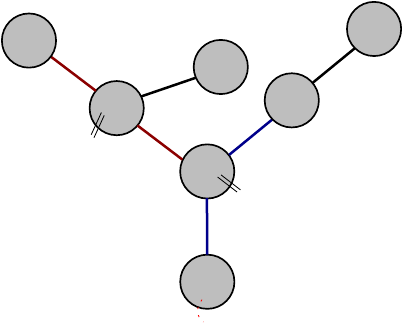}}}\cdots \Bigg\}\\
&\,+\, \Bigg\{\cdots  \vcenter{\hbox{\includegraphics[scale=0.4]{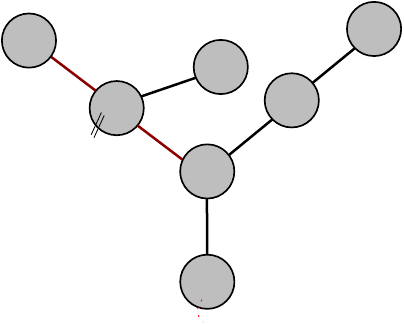}}}\cdots \Bigg\}\,, 
\end{align}

When the root edge is black, we recover the same situation as before and the external edges are hooked to a quartic vertex. Where the rooted edge is red or blue, the external edges are hooked to a sextic vertex, and a moment of reflection shows that:

\begin{equation}
\Sigma=\, \vcenter{\hbox{\includegraphics[scale=0.8]{V41eff1.pdf}}}\,+\,  \vcenter{\hbox{\includegraphics[scale=0.8]{V61eff1.pdf}}}\,+\, \vcenter{\hbox{\includegraphics[scale=0.8]{V62eff1.pdf}}}\,.
\end{equation}

\begin{figure}
\begin{center}
\includegraphics[scale=0.6]{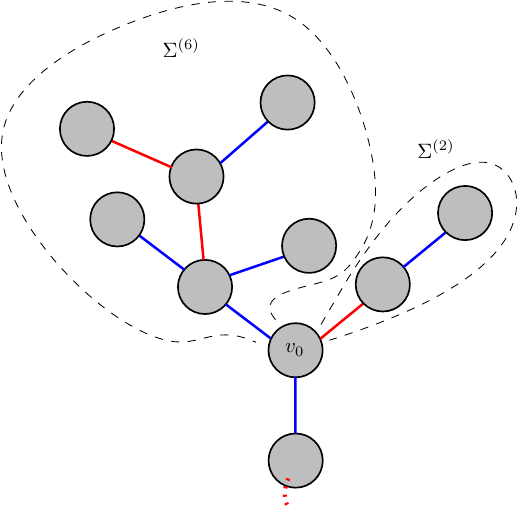}\qquad \includegraphics[scale=0.6]{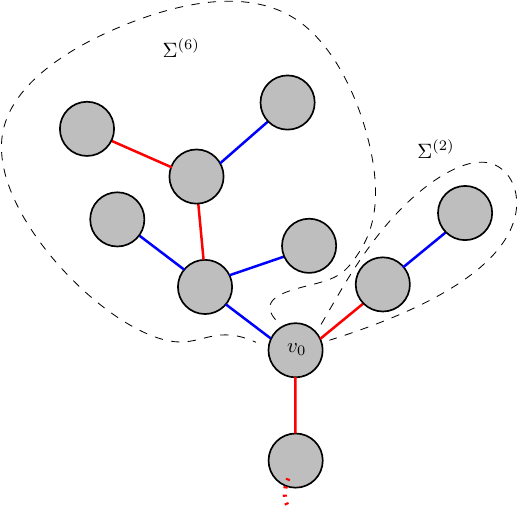}
\end{center}
\caption{The two typical rooted trees for the quartic theory.}\label{fig2points1}
\end{figure}

For the case $p=3$, and including local sextic interactions in the bare action, we have to distinguish three cases, depending if the deleted leaf is hooked with a black, blue or red edge root edge:

\begin{align}
\nonumber \Sigma&=\, \Bigg\{\cdots\vcenter{\hbox{\includegraphics[scale=0.4]{OneLoop6pts1.pdf}}}\cdots \Bigg\}\,+\,  \Bigg\{\cdots \vcenter{\hbox{\includegraphics[scale=0.4]{OneLoop6pts2.pdf}}}\cdots \Bigg\}\\
&\,+\, \Bigg\{\cdots  \vcenter{\hbox{\includegraphics[scale=0.4]{OneLoop6pts3.pdf}}}\cdots \Bigg\}\,, 
\end{align}

When the root edge is black, we recover the same situation as before and the external edges are hooked to a quartic vertex. Where the rooted edge is red or blue, the external edges are hooked to a sextic vertex, and a moment of reflection shows that:

\begin{equation}
\Sigma=\, \vcenter{\hbox{\includegraphics[scale=0.8]{V41eff1.pdf}}}\,+\,  \vcenter{\hbox{\includegraphics[scale=0.8]{V61eff1.pdf}}}\,+\, \vcenter{\hbox{\includegraphics[scale=0.8]{V62eff1.pdf}}}\,.
\end{equation}

\begin{figure}
\begin{center}
\includegraphics[scale=0.6]{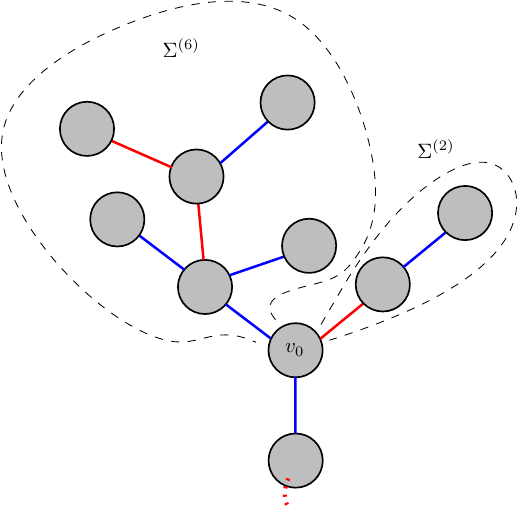}\qquad \includegraphics[scale=0.6]{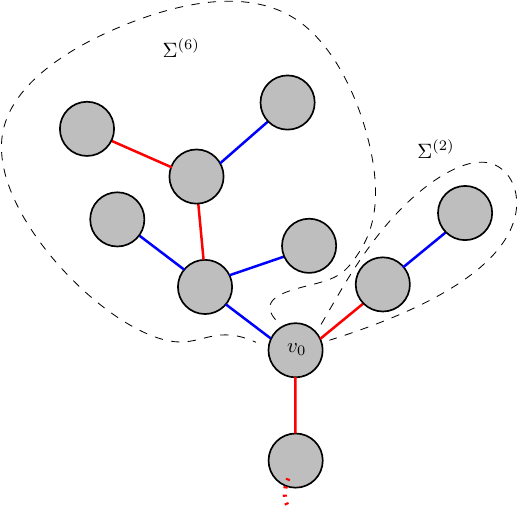}
\end{center}
\caption{The two typical rooted trees for the quartic theory.}\label{fig2points1}
\end{figure}

\printbibliography[heading=bibintoc]
\end{document}